\documentclass[journal]{new-aiaa} 
\usepackage[utf8]{inputenc}
\usepackage{textcomp}

\usepackage{graphicx}
\usepackage{rotating}
\usepackage{amsmath}
\usepackage[version=4]{mhchem}
\usepackage{siunitx}
\usepackage{longtable,tabularx}

\usepackage{ar} 
\usepackage{subfig} 
\usepackage{color}

\usepackage{multirow}
\usepackage{bm}
\usepackage{caption}
\DeclareCaptionFont{xbf}{\bfseries\boldmath}
\newcommand\degrees[1]{\ensuremath{#1^\circ}}
\newcommand\Rey{\mbox{\textit{Re}}}  
\newcommand\etal{\mbox{\textit{et al.}}} 

\title{The Unsteady Lift Produced by a Flat-Plate Wing Translating Past Finite Obstacles}

\author{Karan A. Shah\footnote{Former M.S.\ Student, Department of Mechanical \& Aerospace Engineering, 211 Bell Hall, Buffalo NY 14260.} and Matthew J. Ringuette\footnote{Associate Professor, Department of Mechanical \& Aerospace Engineering, 211 Bell Hall, Buffalo NY 14260.}}
\affil{University at Buffalo, the State University of New York, Buffalo, NY 14260 USA}

\begin{document}

\maketitle

\begin{abstract}

The unsteady lift of a high-angle-of-attack, flat-plate wing encountering finite-length obstacles is studied using towing-tank force measurements. The wing translates from rest and interacts with a rectangular channel, ceiling, or ground obstacle. Variations with angle of attack, obstacle length, mid-chord height to the obstacle, and starting distance between the wing leading edge (LE) and obstacle (typically 1 chord) are examined. For channels, as the gap height decreases, circulatory-lift peaks attributed to leading-edge vortices (LEVs) are the largest, and from the second peak onward occur earliest. This is likely from wing blockage enhancing the flow speed. The lift reduces while exiting a channel, and is lowest afterward if exiting during a lift peak. For ceilings, the first circulatory-lift maximum increases for smaller LE-to-ceiling gaps, but for gaps of 0.5 chords or less, subsequent peaks are below the no-obstacle case yet still earlier. For grounds, with lower wing height the first circulatory-lift peak is larger but the second peak’s behavior varies with angle of attack, and the lift decreases near the ground end. Grounds affect peak timing the least, indicating a reduced influence on the LEV. Changing the starting distance to a channel alters the lift, likely from different LEV timing.

\end{abstract}

\section{Introduction}\label{sec:intro}
\begin{figure}
	\centering
	\includegraphics[width=0.99\textwidth,keepaspectratio]{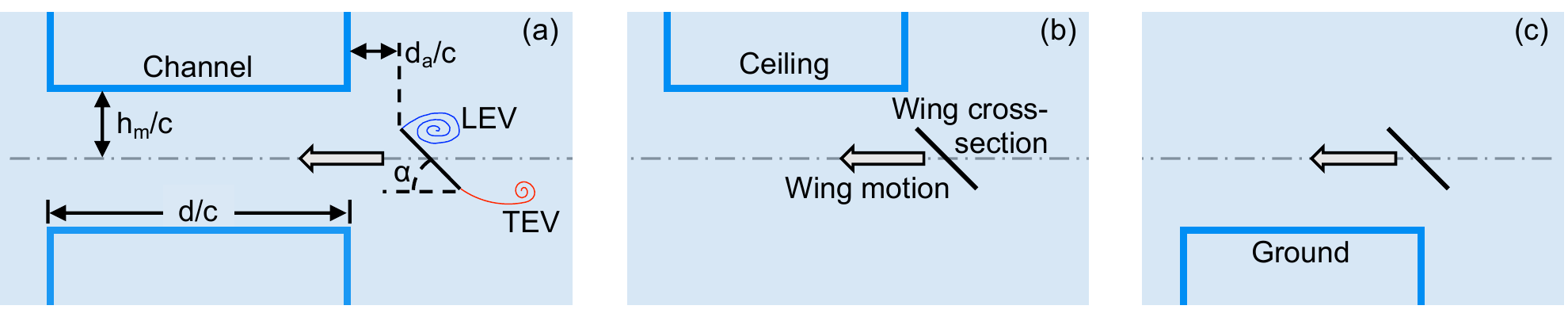}
	\caption{Schematic of the obstacles in a wing cross-sectional plane; a centerline marks the wing mid-chord path. A cartoon of the wing LEV and TEV flow is given for illustration. (a) Channel obstacle; (b) ceiling; (c) ground. The parameters labeled for the channel apply to all obstacles.}\label{fig:obstacle_schematic}
\end{figure}

\lettrine{T}{he} goal of this paper is to understand the changes in the unsteady lift force on a translating, high-angle-of-attack wing as it moves past or through finite-sized obstacles. This work applies to small uninhabited air vehicles (UAVs) operating in complex environments, such as near irregular ground or ceiling geometries, or in confined spaces including air ducts \cite{Ol2008}. For simplicity, the obstacles chosen are finite-length rectangular channels, ceilings, and grounds, with a flat-plate wing; Fig.\ \ref{fig:obstacle_schematic} gives a schematic. During maneuvers the angle of attack, $\alpha$, or effective angle of attack, $\alpha_\mathrm{eff}$, may become high and cause separation and vortex formation at the leading edge (LE) with unsteady forces, called dynamic stall \cite{McCroskey1982,AVT202}. Here $\alpha$ is fixed at various large values, to incorporate leading-edge vortex (LEV) generation but simplify the kinematics; trailing-edge vortices (TEVs) also form (Fig. \ref{fig:obstacle_schematic}). The following background covers research on high-$\alpha$ foils and wings interacting with solid boundaries in a Reynolds number (\Rey) range of $\mathcal{O}(10^2\mbox{--}10^4)$ relevant to small UAVs \cite{Mueller2003}.

Foils and wings at large $\alpha$ with no obstacles present have been studied extensively, as reviewed in Refs.\ \cite{AVT202,Stevens2017,Eldredge2019}. Topics include translating, pitching, plunging, and rotating motions, LEV and TEV dynamics, circulatory and non-circulatory (added-mass) forces, aspect ratio (\AR) effects, and low-order modeling. Here the focus is boundary interactions.

Ground effect (GE) occurs when the height from the ground to the wing trailing edge (TE), $h_\mathrm{TE}$, is within about one span or chord length, $c$ \cite{Rozhdestvensky2006,Cui2010,Vogt2012}. For low $\alpha$ and steady motion, GE causes a pressure rise on the airfoil's lower side from decreased velocity due to flow blockage \cite{Vogt2012,Qu2015}, called the ram effect \cite{Rozhdestvensky2006}. For finite wings, GE also reduces the tip vortex (TV) downwash, increasing the lift and decreasing the induced drag \cite{Rozhdestvensky2006,Cui2010}. With lower height the TVs shift outward and weaken \cite{Han2005}, raising the effective \AR\ \cite{Bleischwitz2017,Bleischwitz2018}. High-$\alpha$ GE for infinite or large grounds is covered next.

Quinn \etal\ \cite{Quinn2014} used experiments and a potential-flow model to examine a periodically-pitching foil near a ground in a freestream. Cases with GE have an upward-tilted vortex-dipole wake, instead of a reverse von K\'{a}rm\'{a}n vortex street, due to ground-image-vorticity interactions. The lift is positive/negative at smaller/larger heights from the ground, respectively. In between, they found an “equilibrium location” with 40\% higher thrust. Kurt \etal\ \cite{Kurt2019} showed that this is from the balance of circulatory lift (<0) from the asymmetric wake, quasi-steady circulatory lift (>0), and added-mass force (<0). Dai \etal\ \cite{Dai2016} simulated periodically-plunging, self-propelled rigid and flexible foils near an infinite ground and found increasing cruising speed with smaller height. Further work on scaling and \AR\ effects has continued \cite{Mivehchi2021,Zhong2021}. 

Flapping-wing GE has also been investigated. Gao and Lu \cite{Gao2008} studied a 2D flapping foil in hover (periodic horizontal heaving, with pitching at stroke reversal) over an infinite ground computationally. They found 3 mean-force trends with dimensionless distance from the ground to mid-chord, $h_\mathrm{m}/c$: for $h_\mathrm{m}/c<1.5$, a lift coefficient ($C_\mathrm{L}$) increase with decreasing $h_\mathrm{m}/c$, with $C_\mathrm{L}$ above the no-ground case; for $1.5<h_\mathrm{m}/c<3.5$, $C_\mathrm{L}$ below the no-ground value with a minimum; and for $h_\mathrm{m}/c>3.5$, an eventual restoration of the no-ground level. This results from changes in vortex interactions at wing stroke-reversal, e.g., the LEV from the prior stroke strengthening the current one in the $h_\mathrm{m}/c<1.5$ regime. Lu \etal\ \cite{Lu2014} experimentally investigated a nominally-2D hovering foil in GE. They similarly found a mean-$C_\mathrm{L}$ reduction as $h_\mathrm{m}/c$ is lowered to $\sim$2, then a large rise with nearness to the ground, caused, respectively, by a less-intense LEV and enhanced wake-capture (stroke-reversal wing-flow interaction). Close to the ground, the TEV cannot travel downward and at stroke reversal it pairs with the shedding LEV to cause a jet toward the wing pressure side, increasing lift. Zheng \etal\ \cite{Zheng2019} used simulations to examine GE for a 2D hovering foil, also showing an initial decrease in mean-$C_\mathrm{L}$ with smaller $h_\mathrm{m}/c$, then increase for $h_\mathrm{m}/c<2.5$. In the latter case, the ground's blockage effect strengthens the LEV and TEV and alters the LEV structure. This improves the LEV-TEV jet on the lower side at stroke reversal (as found by Lu \etal\ \cite{Lu2014}), and overall the pressure changes on the upper and lower foil sides give comparable lift increases.

Considering 3D flows, Su \etal\ \cite{Su2013} simulated a bird in ground effect to a minimum height of $\sim$1 semi-span, showing a rise in lift to a maximum of 47\% with reduced height, mainly from higher pressure below the wings during the downstroke. Kim \etal\ \cite{Kim2014} investigated hovering hummingbirds, finding lower power expended with smaller proximity to the ground, primarily from reduced downwash and also from root-vortex upwash (``fountain effect''). Lu \etal\ \cite{Lu2016} studied 3D flapping wings, and showed the same 3 regimes as Refs.\ \cite{Gao2008,Lu2014}, but smaller effects likely from reduced wake capture; the fountain effect was observed in all ground cases. Closest to the ground, they found smaller downwash with lower height; this and the fountain effect will increase lift from a higher $\alpha_\mathrm{eff}$. Van Truong \etal\ \cite{VanTruong2013} used a dynamically-scaled experiment of a flapping beetle hind wing to study GE during take-off by changing the stroke plane (tilt) and ground height for different runs, and found that when the wing is closest to the ground during the first two strokes and the downstroke phase, the LEV becomes larger than the no-ground case.

High-$\alpha$ GE studies without flapping are covered here. Qu \etal\ \cite{Qu2015} used CFD to analyze an airfoil at fixed $\alpha$ from \degrees{-4} to \degrees{20} in a freestream over a matched moving ground. For larger $\alpha$, e.g.\ \degrees{18}, the lift first lowers then rises again with smaller $h_\mathrm{TE}/c$, but is always below the no-ground case, then drops substantially at $h_\mathrm{TE}/c=0.05$. The latter behavior is from the separation region growing toward the LE with lower $h_\mathrm{TE}/c$ and becoming larger versus no ground, and a weaker pressure increase below the airfoil. Bleischwitz \etal\ \cite{Bleischwitz2017,Bleischwitz2018} measured forces and performed PIV in a wind tunnel with a moving ground matching the freestream, using $\AR=2$ flat-plate and membrane wings for $\alpha=\degrees{10}\mbox{--}\degrees{25}$. For rigid wings they found that the mean $C_\mathrm{L}$ increases with lower $h_\mathrm{TE}/c$, and can be much higher than the no-ground case for $h_\mathrm{TE}/c<0.5$. The flow separates at the LE in this $\alpha$ range, and as $h_\mathrm{TE}/c$ decreases the separation region grows, from the blockage below the wing redirecting more flow over the upper surface via upwash. With greater $\alpha$ and smaller $h_\mathrm{TE}/c$, the near-wing separated region is a mainly-steady recirculating flow that does not add further lift \cite{Bleischwitz2017}. However, there is slowed flow and thus higher pressure below the wing (ram effect) that creates substantial lift in GE \cite{Bleischwitz2017}, and may be the primary mechanism in extreme GE ($h_\mathrm{TE}/c=0.1c$) \cite{Bleischwitz2017,Bleischwitz2018}. Also, in GE the trailing-edge flow is more aligned with the ground than the chord \cite{Bleischwitz2017}, and the TVs shift outward and upward, and enlarge and break down \cite{Bleischwitz2018}. Adhikari \etal\ \cite{Adhikari2022} tested perching flat-plate wings with a plunge toward a large ground coupled with deceleration, also with pitch-up. The pitching raises the initial peak $C_\mathrm{L}$, related mainly to a higher $\alpha_\mathrm{eff}$ and non-circulatory force, but with a faster post-peak decline versus pure plunge, likely from enhanced LEV detachment and TEV downwash.

Meng \cite{Meng2019} studied ceiling interactions for simulated hovering wings with $\alpha=\degrees{40}$ at mid-stroke; the ceiling extends over the whole domain. A monotonic increase in stroke-averaged lift with decreasing gap height was shown. The ceiling acts as a mirror-image LEV to enhance the LE-flow speed, and creates an upwash that raises $\alpha_\mathrm{eff}$; both give greater lift. Meng \etal\ \cite{Meng2020} added a freestream to emulate forward flight, which yields a smaller $\alpha_\mathrm{eff}$ and reduces the ceiling effect.

Regarding foils and wings in channels, with top and bottom walls parallel to the freestream or wing motion, work has been done on wind-tunnel blockage and corrections, e.g., Refs. \cite{Maskell1963,Barlow1999}. Abernathy \cite{Abernathy1962} studied nominally-2D flat plates with $\alpha=\degrees{30}$--\degrees{90}, using free-streamline theory and wind-tunnel experiments, for test-section heights from $2h_\mathrm{m}/c=3.5$ to $14$. The mean height of the separated wake is $\sim$$\sqrt{2}c\sin{\alpha}$ in all cases. Ota and Okamoto \cite{Ota1990} used a discrete vortex method (DVM) to model 2D high-$\alpha$ flat plates in a wind tunnel, and showed the normal force and Strouhal number ($St$) grow with 0--0.40 blockage ratio (plate frontal-projected height divided by test-section height), consistent with Abernathy \cite{Abernathy1962}. Ota \etal\ \cite{Ota1994} used these DVM results to develop blockage corrections, improving them via empirical data. Yeung \cite{Yeung2008} re-scaled the pressure coefficient to be invariant with $2h_\mathrm{m}$, then defined a new wake height and $St$ used in a free-streamline theory to accurately predict the base pressure and drag of flat plates and cylinders for various blockage ratios. Recently, Zhou \etal\ \cite{Zhou2019} used computations and experiments to examine blockage ratios up to 0.365 with $\alpha=\degrees{30}$--\degrees{80} for airfoils in wind tunnels. By varying the blockage ratio for different fixed $\alpha$, then changing $\alpha$ with different fixed blockage, they showed that $St$ increases with blockage as in prior studies, and grows with $\alpha$ especially above $\sim$\degrees{40}. The LEV and TEV formation and spread are confined by the walls for higher blockage, and lose swirl downstream. Zhou \etal\ \cite{Zhou2019} also developed a new $St$ correction accounting for blockage and $\alpha$ effects.  

Jeong \etal\ \cite{Jeong2021} used simulations to study self-propelled flexible thin fins in harmonic plunge with no walls, a ground wall, and a $2c$-high channel. For a single fin, the narrow channel causes the alternating-sign wake vortices to be close to the centerline, and redirects much of the fin-induced vertical velocity to give the largest rearward jet flow of all cases; the vortex circulation is also highest. The channel-confined vortex pattern is similar to that of Zhou \etal\ \cite{Zhou2019}.

Considering quadrotor UAVs, Gao \etal\ \cite{Gao2019} and Carter \etal\ \cite{Carter2021} showed experimentally that the lift increases monotonically with decreasing distance to ceilings or grounds, also consistent with studies of larger UAVs cited therein. Carter \etal\ \cite{Carter2021} further found that the GE lift is higher than the ceiling-interaction value for a given distance.

There is much less research on interactions with finite or semi-infinite obstacles. Wang and Yeung \cite{Wang2016} examined a hovering (flapping) 2D foil over a semi-infinite platform numerically. When the wing is close vertically but farther horizontally, the shed TEV induces an opposite-sign vortex at the platform tip and the pair create a downward jet that increases lift; adverse interactions occur if the wing is nearer horizontally. Yin \etal\ \cite{Yin2019} studied 2D hovering near circular obstacles computationally. In GE, if the obstacle diameter is $\leq1c$ the Gao and Lu \cite{Gao2008} trends switch. A small obstacle near the TE yields a weaker LEV, less wake-capture, interrupted TEV pairing, and lower lift. The ceiling-proximity trend follows Meng \cite{Meng2019}, with more lift for larger obstacles. Cai \etal\ \cite{Cai2023} studied UAV propellers near finite ceilings and grounds: areas equal to the rotor disk act as infinite, and power effects add with ceiling area.

Zhi \etal\ \cite{Zhi2022} computationally studied GE with periodic water waves traveling below an airfoil at $\alpha=0$--\degrees{4}. For $h_\mathrm{TE}/c\geq0.4$, there is no appreciable difference between results with a wave-shaped ground and water waves, and the lift oscillates with the wave frequency but shifted earlier. The ground shape dictates the pressure gradient below the foil, e.g., a divergent foil-ground gap creates a pressure increase; $h_\mathrm{TE}/c$ modulates this. Water interactions occur for $h_\mathrm{TE}/c\leq 0.2$.

Despite this progress, there remains a lack of data on high-$\alpha$ wings encountering finite obstacles. Here the kinematics are simplified to pure translation at fixed $\alpha$ with no flapping, and this work is unique in its objective to understand the unsteady lift of a wing approaching a single finite-sized obstacle, interacting with it, then passing it. As shown in Fig.\ \ref{fig:obstacle_schematic}, the obstacles tested are rectangular channels, ceilings, and grounds, with their length and proximity to the wing varied; the wing starting (approach) distance to a channel is also examined. This first study is intended to characterize the influence of these parameters and identify interesting lift-force phenomena. The experimental techniques are presented first, next results for each obstacle type, a comparison of types, the approach-distance tests, then conclusions.

\vspace{-5 pt}
\section{Experimental Setup and Methods}

\begin{figure}
	\centering
	\includegraphics[width=0.75\textwidth,keepaspectratio]{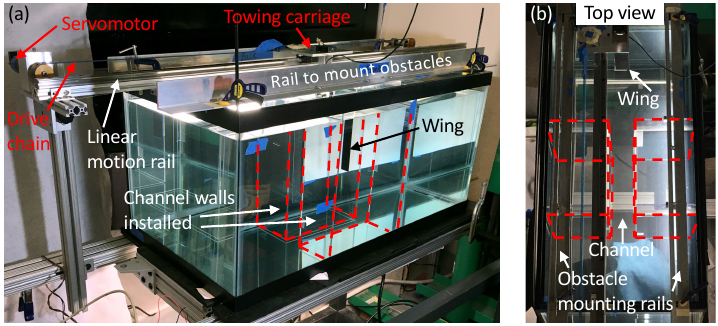}
	\caption{Water towing tank. (a) Overall photo showing the drive system, wing, and an installed $6c$-long channel obstacle (the submerged portions of the channel walls are outlined by dashed red lines). (b) Overhead view of the tank, highlighting the channel geometry and mounting rails for the lateral channel walls.}\label{fig:tank}
\end{figure}

\vspace{-5 pt}
\subsection{Tank Facility, Wing Model, and Obstacles} 
The experiments are done in a 90~cm $\times$ 45~cm $\times$ 40~cm glass towing tank with water as the working fluid (Fig.\ \ref{fig:tank}a). The tank rests on extruded-aluminum support beams, elevated 85~cm on a steel frame. The drive system above it consists of a linear guide rail attached to an extruded-aluminum beam, supported at each end by further aluminum framing. A bearing pillow block slides on the rail, and an aluminum carriage plate mounted above the block overhangs to one side. The wing-model assembly attaches below this so the wing is submerged vertically and centered laterally in the tank. The carriage is moved via a chain-sprocket system, with a sprocket near each end of the rail and a near-zero-backlash drive chain (WM Berg 32GCF, polyurethane with 2 embedded stainless-steel cables) around the sprockets that connects to the carriage. One sprocket is driven by a DC servomotor (Faulhaber 3863H024CR) with a 20:1 gearhead and optical encoder with 2,000 counts per revolution in quadrature, the other is an idler. Velocity programs are created with Faulhaber Motion Manager and uploaded to a motion controller (Faulhaber MCDC 3006 S), which uses the encoder feedback. The wing position is retained within $\sim$6 microns or $\sim$0.01\%$c$ from run to run, and the encoder shows a velocity error under 5\% between the executed and programmed motion. Before each run, the wing is moved back slightly behind the starting position, then forward to it, to remove any small drive-chain slack; the wait between runs (Sec.\ \ref{sec:force_setup}) begins after this. 

\begin{figure}
	\centering
	\includegraphics[width=0.55\textwidth,keepaspectratio]{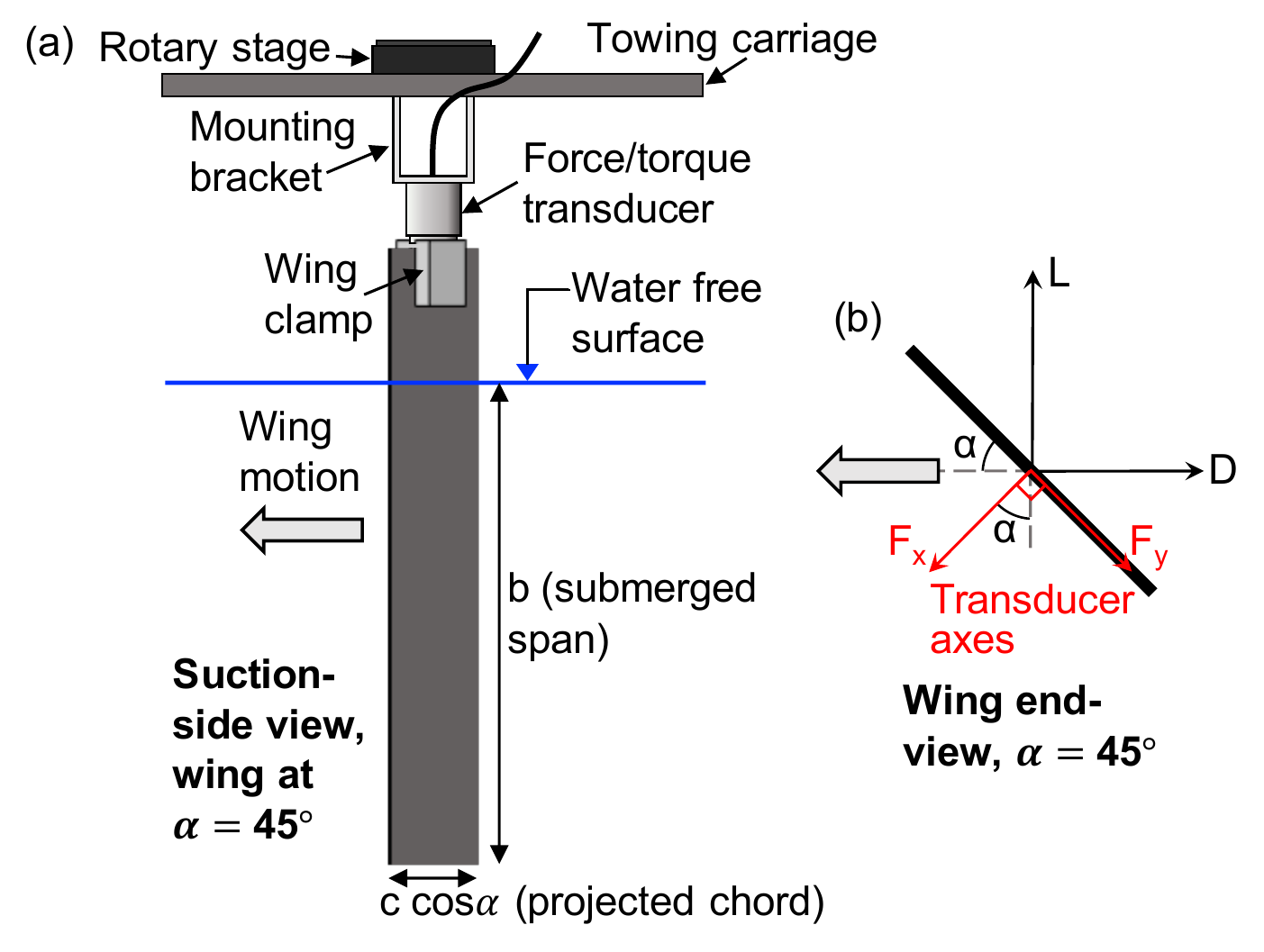}
	\vspace{-5pt}
	\caption{Experimental wing model. (a) Schematic of the assembly viewed from the wing suction side. (b) Coordinate systems for the transducer and the lift and drag in the wing-frame, end-view.}\label{fig:wing}
\end{figure}

The rectangular-planform, flat-plate wing is made from a 1.6~mm thick carbon-fiber composite sheet, with $c=44.5$~mm, a physical $227$~mm span, and square edges. Figure \ref{fig:wing}a gives the wing-model assembly: a manual Thorlabs rotary stage is fixed above the carriage, attached below this through a cutout is an aluminum box-section to mount the force transducer and accommodate its cable, and the wing is screwed to the transducer's sensing end through an aluminum clamp epoxied to the wing. The assembly is designed to be light and rigid. The wing $\alpha$ is set and fixed via the rotary stage, which has a \degrees{0.33} resolution and estimated $\pm\degrees{0.33}$ uncertainty. To obtain the $\alpha=\degrees{0}$ reference, which can be slightly different than the stage's \degrees{0} setting if the rotational alignment of the attached wing is imperfect, transducer tests are done with the plate at different $\alpha$ near the \degrees{0} reading until the lift is as close to 0 as possible within the transducer uncertainty. Until the wing is unmounted, this is taken as $\alpha=\degrees{0}$. The vertically-oriented wing has a submerged aspect ratio of $\AR=b/c=4$, where $b$ is the span below water; Sec.\ \ref{sec:params} discusses the free-surface condition. The wing tip is $4.1c$ from the tank bottom to mitigate any wall-interference effects there. At the maximum $\alpha=\degrees{60}$, the forward-projected, submerged wing area is 4.2\% of the filled tank cross-sectional area, so blockage effects are small.

The obstacles are made from vertical walls placed inside the tank (Fig.\ \ref{fig:tank}a,b, dashed red lines), which extend from above the free surface to the tank bottom. Each obstacle consists of 1 or 2 box shapes, e.g., a ceiling has a lateral (cross-stream) wall where it begins, streamwise ceiling wall, and lateral end wall, with the tank glass serving as the opposite (outer) streamwise wall of the box. A channel (Fig.\ \ref{fig:tank}b) has 2 such boxes in a mirror-image configuration about the tank centerline (wing mid-chord path), and the wing enters and exits through it. The obstacles are constructed from flat plastic sheets: 2.78~mm thick acrylic for the streamwise walls (channel, ceiling, and ground surfaces) and 4.76~mm thick polycarbonate for the lateral walls at the start/end of each obstacle. Aluminum clamps above the tank grip the tops of the lateral start/end walls and hold them rigidly in-place, and these can be positioned as desired along two streamwise, slotted aluminum rails on either side of the linear-motion rail (Fig.\ \ref{fig:tank}b). The ends of the streamwise walls are attached orthogonal and flush to those of the lateral walls via Scotch brand heavy-duty, clear packaging tape run smoothly over the joint; the tape is affixed when the walls are dry and remains indefinitely underwater. This thin tape is applied along a depth encompassing the wing length for the corners at the start of each obstacle and at two positions along the end corners. Any gaps between the joints are small enough that no significant flow within them is possible. Care is taken to ensure that the mounted obstacle walls are vertical in the tank, mutually orthogonal, aligned with the wing geometry and motion, and have consistent positioning and dimensions across the tested cases, e.g., the same distance between the streamwise channel walls for different channel lengths. Interchangeable walls are used to vary the obstacle dimensions. Due to the silicone bead at the tank-floor perimeter, the cross-stream obstacle walls stand off slightly from the tank side walls; this is acceptable as the small gap is away from the flow of interest and allows water to fill into the obstacle boxes.

\vspace{-5 pt}
\subsection{Force Measurements}\label{sec:force_setup}
A 6-axis ATI Nano 17 IP68 force/torque sensor measures the time-varying lift and drag (Fig.\ \ref{fig:wing}a); the focus here is lift. A PC with a 16-bit National Instruments data acquisition card (NI 6036E) acquires the sensor signals, operated via LabVIEW software, with a 1~kHz sampling frequency. In all cases, $N=10$ runs are acquired with the wing in water and 10 in air, the latter serve as dynamic-tare runs for subtracting off the inertial forces of the wing model. A time of 8 min or more is allowed between each run for the flow to settle, meaning the force signals return to the quiescent-tank values.

The force data are processed in MATLAB. For each case, the $N$ water and $N$ air runs are first aligned, respectively, in time via a correlation. The start time ($t=0$) for each aligned set is found by averaging then low-pass filtering the $N$ runs, then obtaining the time when the $y$-torque first departs from the mean wing-at-rest level by 2 standard deviations. This ensures the start time is not affected by noise before the wing moves, and this torque is used as its response to the wing motion is highest. These operations simply align the raw data, and crop them prior to $t=0$, before the main processing.

A strike test is used to find the vibrational frequencies of the towing-tank setup with no wing motion; the transducer data are evaluated by estimating the power spectral density (PSD) in MATLAB. The frequencies during wing motion corresponding to mechanical vibration, electrical noise, and fluid-dynamic forces are examined by comparing the PSD for air and water runs, and the strike test. Based on this, a third-order low-pass Butterworth filter with a 5~Hz cutoff frequency is applied to each of the aligned and cropped raw $N$ runs for water and air, using MATLAB's ``filtfilt'' command. This mitigates the majority of vibrational and electrical noise while preserving the fluid-dynamic forces. The filtered water and air data are averaged over their $N$ runs, then the mean air curve is subtracted off the mean water result.

The ATI sensor employs the SI-25-0.25 calibration, and precision weights are used in static tests to verify its accuracy for the range of measured forces ($\sim$0.1--0.3~N). For the transducer $F_\mathrm{x}$-axis this bias error is within 2\%, and within 0.5\% for $F_\mathrm{y}$, but it raises to $\sim$4\% for $F_\mathrm{x}$ when the force reduces to $\sim$0.1~N, which is about the smallest value experienced after startup in some runs. Figure \ref{fig:wing}b shows the wing-aligned transducer axes in relation to the lift and drag directions. This paper focuses on the lift coefficient, $C_\mathrm{L}=2L/\rho U^2S$, where $L$ is the lift force, $\rho$ is the fluid density, $U$ is the constant wing velocity after the startup acceleration, and $S=cb$ is the wing area. The random uncertainty in the mean lift coefficient, $\delta\overline{C}_\mathrm{L}(t)$, is calculated for each case from the precision error of the mean due to the variations over $N$ runs for the water and air data at every time step as follows (the last equation is the error propagation for the lift direction):
\vspace{-7 pt}
\begin{gather*}
\overline{C}_\mathrm{(x,y),water/air}=2(F_\mathrm{x,water/air},F_\mathrm{y,water/air})/\rho U^2S,\\
\delta\overline{C}_\mathrm{(x,y),water/air}(t)=t_{\nu,0.95}\sigma_\mathrm{(x,y),water/air}(t)/\sqrt{N},\\
\overline{C}_\mathrm{L,water/air}=-\cos(\alpha)\overline{C}_\mathrm{x,water/air}-\sin(\alpha)\overline{C}_\mathrm{y,water/air},\\
\delta\overline{C}_\mathrm{L,water/air}(t)=\sqrt{\cos^2(\alpha)\left(\delta\overline{C}_\mathrm{x,water/air}(t)\right)^2+\sin^2(\alpha)\left(\delta\overline{C}_\mathrm{y,water/air}(t)\right)^2}.
\end{gather*}
Here $t_{\nu,0.95}$ is the Student's $t$-distribution for a confidence level of 95\% and $\nu=N-1$ degrees of freedom, and $\sigma$ is the sample standard deviation. For a given case, where the air curve is subtracted from the water data, the precision error is expressed as: $\overline{C}_\mathrm{L}(t)\pm\delta\overline{C}_\mathrm{L}(t)=(\overline{C}_\mathrm{L,water}(t)\pm\delta\overline{C}_\mathrm{L,water}(t))-(\overline{C}_\mathrm{L,air}(t)\pm\delta\overline{C}_\mathrm{L,air}(t))$.

\begin{figure}
	\centering
	\subfloat{
	\hspace{1pt}\includegraphics[height=0.27\textheight,keepaspectratio]{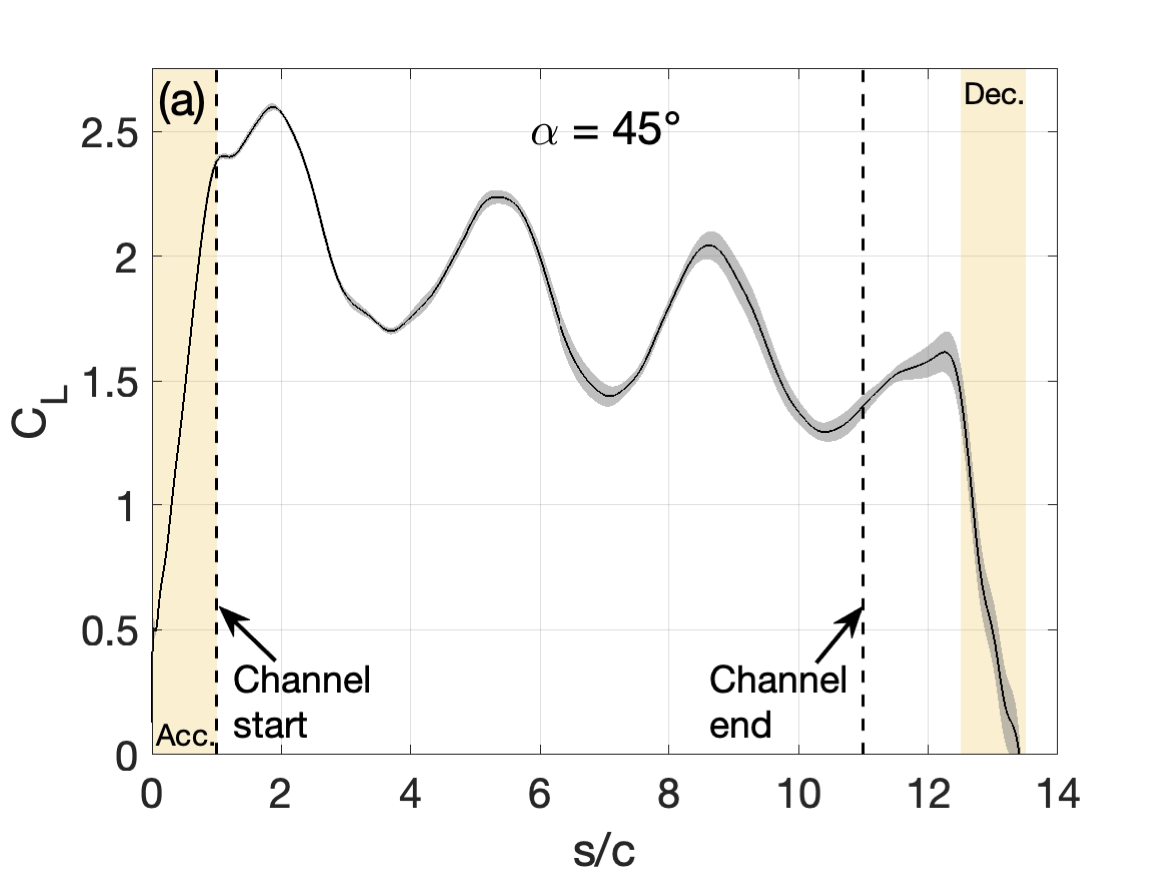}\label{fig:error_45}}
	\subfloat{
	\includegraphics[height=0.27\textheight,keepaspectratio]{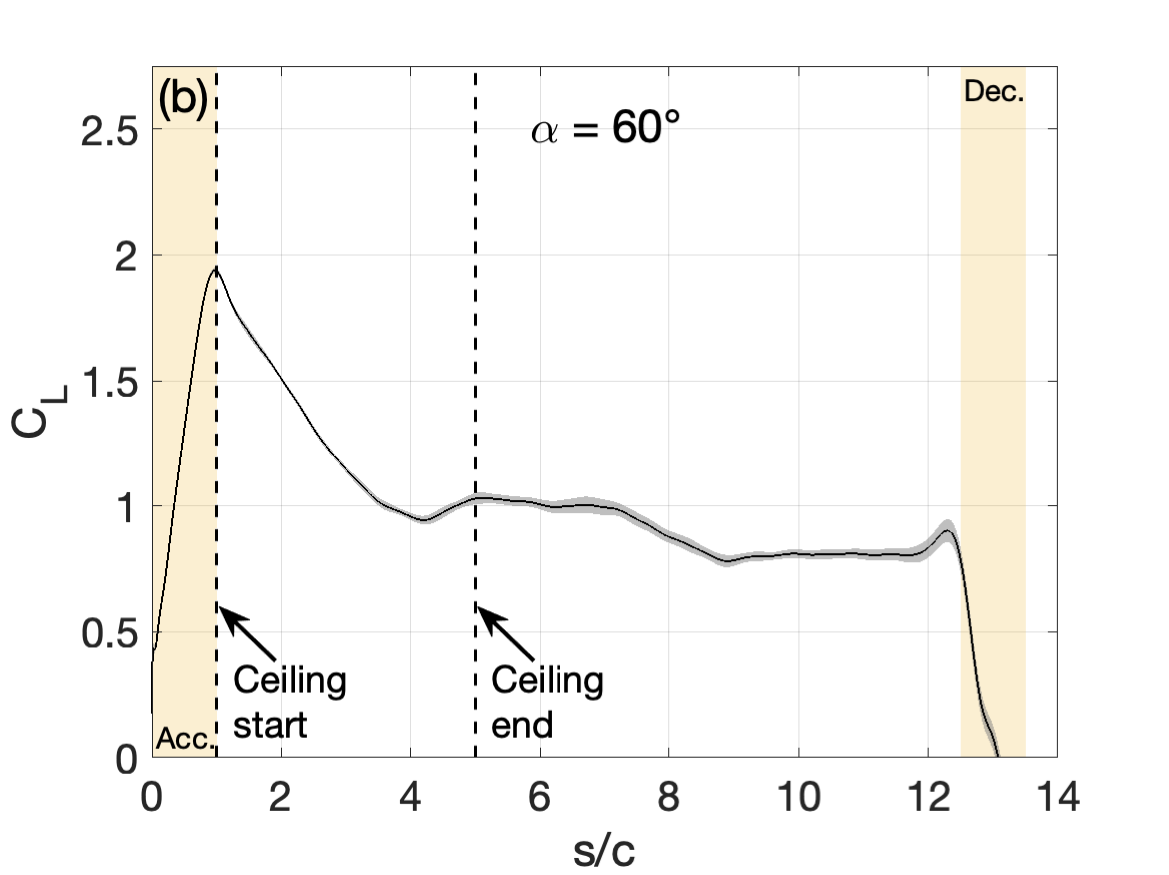}\label{fig:error_60}}
	\vspace{-4pt}
	\caption{The $C_\mathrm{L}$ vs.\ $s/c$ averaged over 10 runs with a gray precision-error band: (a) case with the largest overall force signal ($\alpha=\degrees{45}$, $10c$-long channel, $h_\mathrm{m}/c=0.865$, $d_\mathrm{a}/c=1$); (b) case having the lowest overall force ($\alpha=\degrees{60}$, $4c$-long ceiling, $h_\mathrm{m}/c=0.865$, $d_\mathrm{a}/c=1$). Beige marks the wing acceleration/deceleration regions.
	}\label{fig:error}
\end{figure}

Figure \ref{fig:error} shows two representative cases: panel (a) having the highest overall force signal, for an $\alpha=\degrees{45}$ wing passing through the narrowest $d/c=10$-long channel, and panel (b) with the lowest, for $\alpha=\degrees{60}$ and a $d/c=4$-long ceiling. For both, the height from mid-chord to the obstacle is $h_\mathrm{m}/c=0.865$, the approach distance is $d_\mathrm{a}/c=1$ (Fig.\ \ref{fig:obstacle_schematic}), and the motion is acceleration from rest over $1c$ to a constant velocity with $\Rey=7,000$, then deceleration to rest over the final $1c$, and $13.5c$ total travel (Fig.\ \ref{fig:motion}). Gray bands show the precision-error bounds, the overbar for the mean $C_\mathrm{L}$ is dropped from here on, and $s/c$ is chords traveled. In panel (a), the run-to-run $C_\mathrm{L}$ variability increases after $s/c\approx4$, and is greatest at the $C_\mathrm{L}$ peaks and minima, which should be mainly related to LEV formation and shedding (Sec.\ \ref{sec:chan_H}); a likely contributor is per-run changes in the unsteady vortex flow. The error is also larger after the wing exits the channel at $s/c=11$, possibly from variability in the exiting flow interactions, and as deceleration starts ($s/c=12.5$) when mechanical vibrations will be prominent. For case (b), the error band becomes larger also beyond $s/c\approx4$ until deceleration begins. Over the full wing motion, the time-averaged precision error bounds on $C_\mathrm{L}$ are $\pm3.16\%$ ($\pm0.05$) and $\pm2.75\%$ ($\pm0.04$) for panels (a) and (b), respectively. Overall, the measurement uncertainty is considered reasonable.

\vspace{-5 pt}
\subsection{Experimental Parameters}\label{sec:params}

The parameters tested are given here. For the vertically-oriented wing with submerged $\AR=4$, the free surface can be treated as a reflected (symmetry) boundary condition for low Froude number ($Fr$) and if the free-surface depressions from connecting vortices are small versus the span \cite{Kim2011,Wibawa2012}, giving an effective $\AR_\mathrm{eff}=8$. Here $Fr=U/\sqrt{gc}=0.25$ for all cases, where $g$ is the acceleration due to gravity. For $Fr$ closer to 1, free-surface waves may inhibit the vortex shedding near the water surface \cite{Vlachos2008}. Videos of the free surface indicate only small wave-motion deflections, less than $\sim$1.5~mm (<1\% span). The low-pressure cores from the growing and shedding LEVs and TEVs create free-surface dimple deformations, showing the normal vortex connections needed for the symmetry condition. The resulting free-surface curvature will produce some surface-parallel vorticity \cite{Lang2000}, so the reflected condition is not perfect. However, these deflections are at most $\sim$5\% of the wing span and therefore small, so the symmetry-plane assumption is reasonable.

Chowdhury and Ringuette \cite{Chowdhury2021} tested the symmetry assumption by comparing their $C_\mathrm{L}$ measurements for vertically-submerged, towed $\AR=1$ and 2 ($\AR_\mathrm{eff}=2$ and 4) flat-plate wings to the fully-submerged, horizontally-oriented, towed $\AR=2$ and $4$ data from joint Air Force Research Lab and University of Maryland measurements \cite{Stevens2017}. For that comparison, \Rey\ ranged from 12,000 to 20,000, higher than the present 7,000, and the vertically-submerged wings of Chowdhury and Ringuette \cite{Chowdhury2021} had $Fr=0.17$--0.25, which overlaps that here. They found excellent agreement for $\AR=4$, with some deviation for $\AR=2$ that is comparable to the variability of the experiments and simulations in Ref.\ \cite{Stevens2017}. This supports the reflected boundary as a good assumption if free-surface effects are small, therefore $\AR_\mathrm{eff}=8$ is assumed here. This $\AR_\mathrm{eff}$ ensures an inboard portion of LEV and TEV shedding not suppressed by TV downwash effects, whereas outboard between this and the TV there will likely be a wake region of more complex ``braid-like'' \cite{Zhang2020} vortices with streamwise components \cite{Zhang2020,Burtsev2022,Ribeiro2023,Pandi2023}. For $\AR_\mathrm{eff}=8$, the torques also stay within the ATI sensor's limits.

Figure \ref{fig:obstacle_schematic} shows a schematic of the rectangular channels, ceilings, and grounds tested, which are simplifications of obstacles a UAV may interact with. The view is the wing cross-sectional plane, with the span orthogonal to the page. The channels are symmetric about the wing's streamwise path at mid-chord, while the ceilings and grounds are above and below this, respectively. As mentioned earlier, $h_\mathrm{m}/c$ is the dimensionless orthogonal distance or height from the mid-chord path to the streamwise obstacle surface, making the wall-to-wall channel gap $2h_\mathrm{m}/c$; the streamwise obstacle length is $d/c$. The ``approach distance,'' $d_\mathrm{a}/c$, is the streamwise length between the wing LE $t=0$ position and obstacle start, i.e., the chords traveled ($s/c$) value at which the LE first reaches the obstacle. The wing starting position is adjusted to maintain the desired $d_\mathrm{a}/c$ as $\alpha$ is changed. One obstacle is tested per case, so obstacles in series are not considered.

\begin{table*}
\centering
\caption{Parameter values tested, 60 total cases; NO is the no-obstacle case. For all experiments, $\Rey=7{,}000$.}\label{tab:parameters}
\label{tab:cases}
\begin{tabular}{lcccc}
\hline\noalign{\smallskip}
\hline\noalign{\smallskip}
\textbf{Obstacle type} & \textbf{Obstacle dist.\ from mid-chord} & \textbf{Obstacle length} & \textbf{Approach dist.} & \textbf{Angle of attack}\\
& \bm{$(h_\mathrm{m}/c)$} & \bm{$(d/c)$} & \bm{$(d_\mathrm{a}/c)$} & \bm{$(\alpha)$}\\\hline
\multirow{2}{8em}{Channels, ceilings, \& grounds}& 0.865 & 2, 4, 6, 10 & \multirow{2}{0.6em}{1} & \multirow{2}{5.2em}{\degrees{30}, \degrees{45}, \degrees{60}}\\
& 1.35& 4, 10& & \cr
\hline\noalign{\smallskip}
Channels & 0.865 & 6 & 4 & \degrees{30}, \degrees{45}, \degrees{60}\\
\hline\noalign{\smallskip}
NO & & & & \degrees{30}, \degrees{45}, \degrees{60}\\
\noalign{\smallskip}\hline
\noalign{\smallskip}\hline
\end{tabular}
\end{table*}

Table \ref{tab:parameters} gives the obstacle parameters and wing $\alpha$ tested. The no-obstacle case is called NO following Yin \etal\ \cite{Yin2019}, and the distance to each tank side wall gives $h_\mathrm{m}/c=5.06$. From preliminary tests, for the obstacles $h_\mathrm{m}/c=0.865$ and 1.35 are chosen, which yield significant force changes; these values include the slight cross-stream wall offsets from the tank side walls due to the silicone seal around the bottom. The streamwise $d/c$ range is $2$--$10$, with 2 being closer to the wing scale, 4 and 6 encompassing at least one LEV formation and shedding cycle as indicated by NO data, and 10 ensuring the wing-obstacle encounter occurs for at least two cycles \cite{Taira2009,AVT202,Mancini2015,Stevens2017,Mulleners2017}. To study $\alpha$ effects for the same $h_\mathrm{m}/c$, fixed $\alpha$ values of \degrees{30}, \degrees{45}, and \degrees{60} are tested. All exhibit vortex formation and shedding in the NO case \cite{Taira2009,AVT202,Mancini2015,Zhang2020}. For nearly all tests $d_\mathrm{a}/c=1$, but to examine approach-distance effects, a channel case is repeated with $d_\mathrm{a}/c=4$.

\begin{figure}
\vspace{-10pt}
	\centering
	\subfloat{
	\hspace{1pt}\includegraphics[height=0.27\textheight,keepaspectratio]{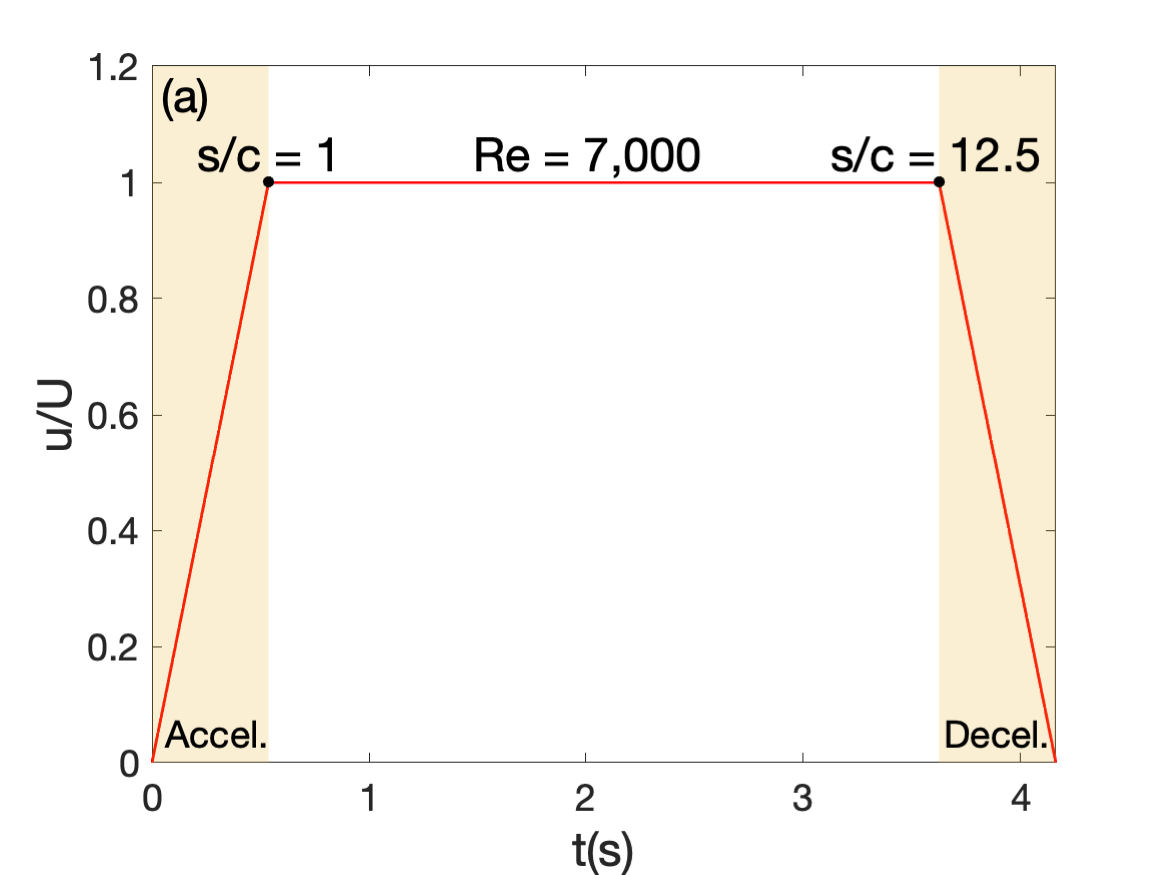}\label{fig:motion_t}}
	\subfloat{
	\includegraphics[height=0.27\textheight,keepaspectratio]{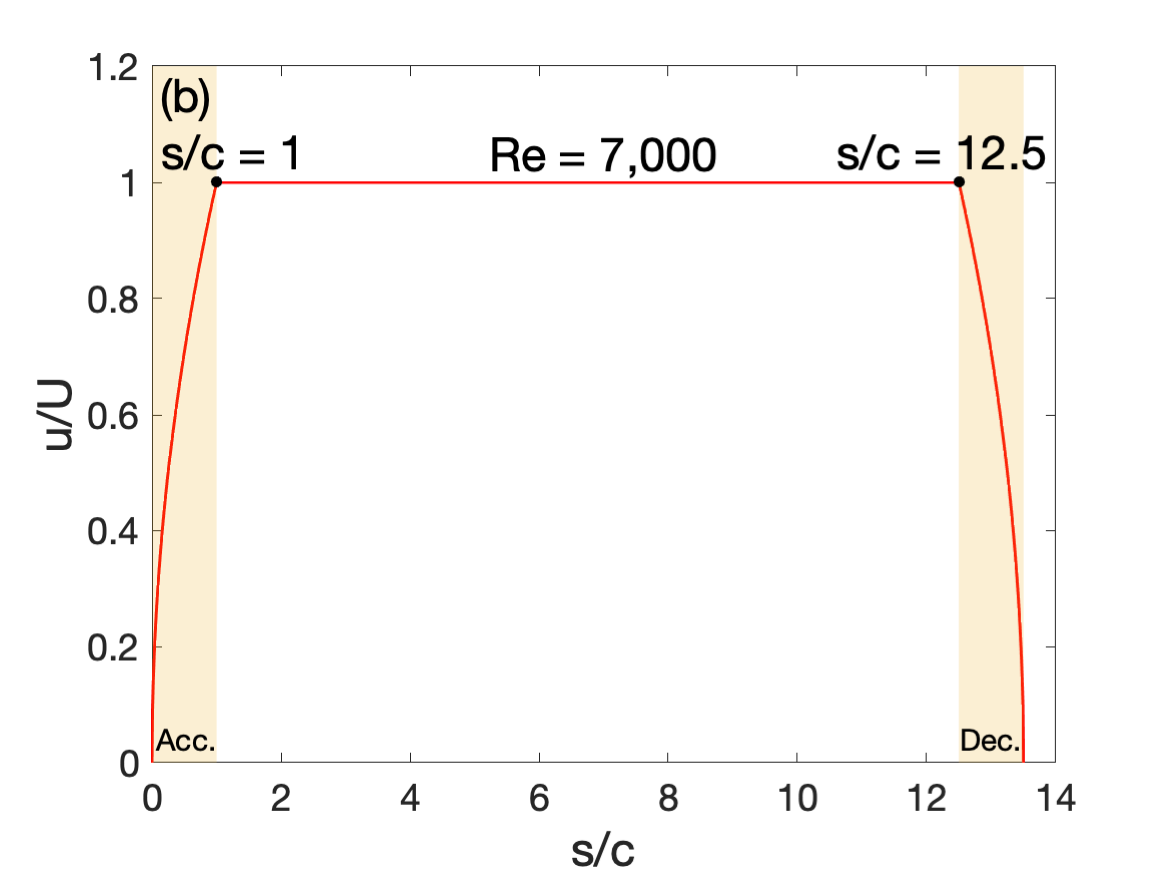}\label{fig:motion_soc}}
	\vspace{-4pt}
	\caption{Wing motion program: (a) dimensionless velocity versus time, with chords traveled ($s/c$) labeled; (b) versus chords traveled. Acceleration and deceleration portions are colored in beige.}\label{fig:motion}
\end{figure}

The wing's streamwise velocity program $u$ is trapezoidal, with constant acceleration over a $1c$ distance to a constant velocity $U$ with $\Rey=7{,}000$ (Fig.\ \ref{fig:motion_t}); Fig.\ \ref{fig:motion_soc} gives $u/U$ versus $s/c$. The total travel is $13.5c$, with constant deceleration to rest for the last $1c$, and $1.5c$ of constant-$U$ motion past the longest $d/c=10$ obstacle. Wings translating from rest with acceleration for $1c$ have been studied previously, e.g., Refs.\ \cite{AVT202,Mancini2015,Stevens2017,Mulleners2017}, and we refer to this to relate force peaks and vortex formation. The chosen \Rey\ is in the small UAV range \cite{Mueller2003} and gives the highest forces while keeping $Fr$ low.

\vspace{-5 pt}
\section{Results}

The results are organized by obstacle type, examining changes in the height to the wing $h_\mathrm{m}/c$, then length $d/c$. Next all types are compared, then the influence of approach distance, $d_\mathrm{a}/c$, is covered. The obstacle interactions can modify the magnitude and timing of the wing $C_\mathrm{L}$ peaks from the NO data. Based on this and the translating-wing NO work cited above, we comment on how the obstacles might alter the vortex dynamics. Acquiring flow data is a topic for future work.

\vspace{-5 pt}
\subsection{Effect of Channel Gap Height}\label{sec:chan_H}

\begin{figure}[t!]
	\centering
	\subfloat{
		\hspace{-5pt}\includegraphics[height=0.175\textheight,keepaspectratio]{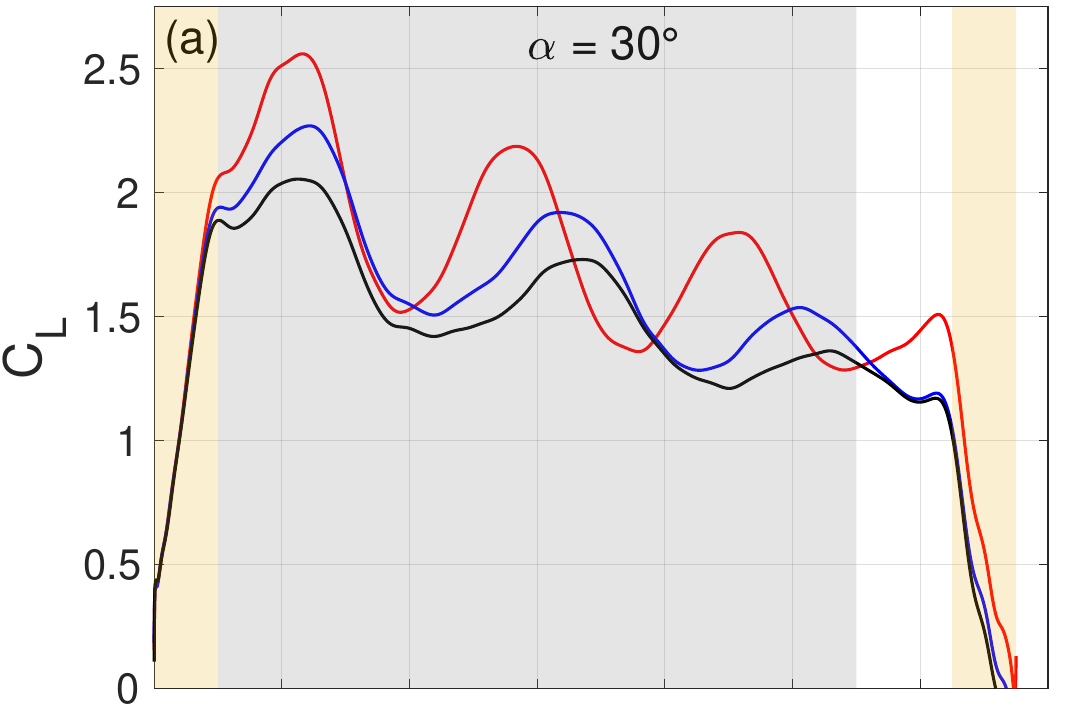}\label{fig:chan_H_10_30deg}}
	\subfloat{
		\includegraphics[height=0.175\textheight,keepaspectratio]{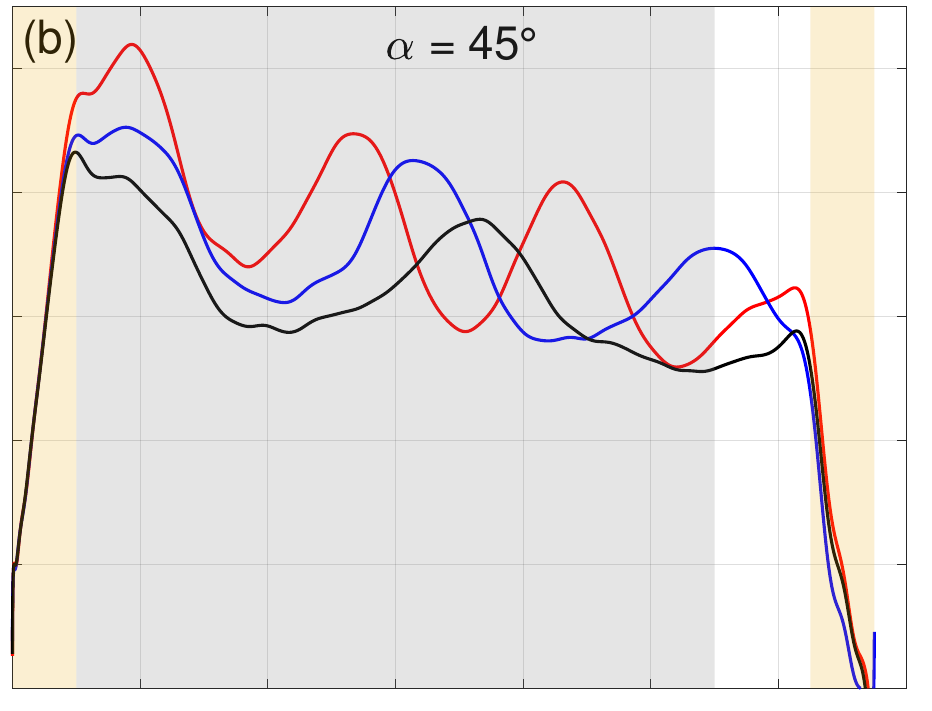}\label{fig:chan_H_10_45deg}}
	\subfloat{
		\includegraphics[height=0.175\textheight,keepaspectratio]{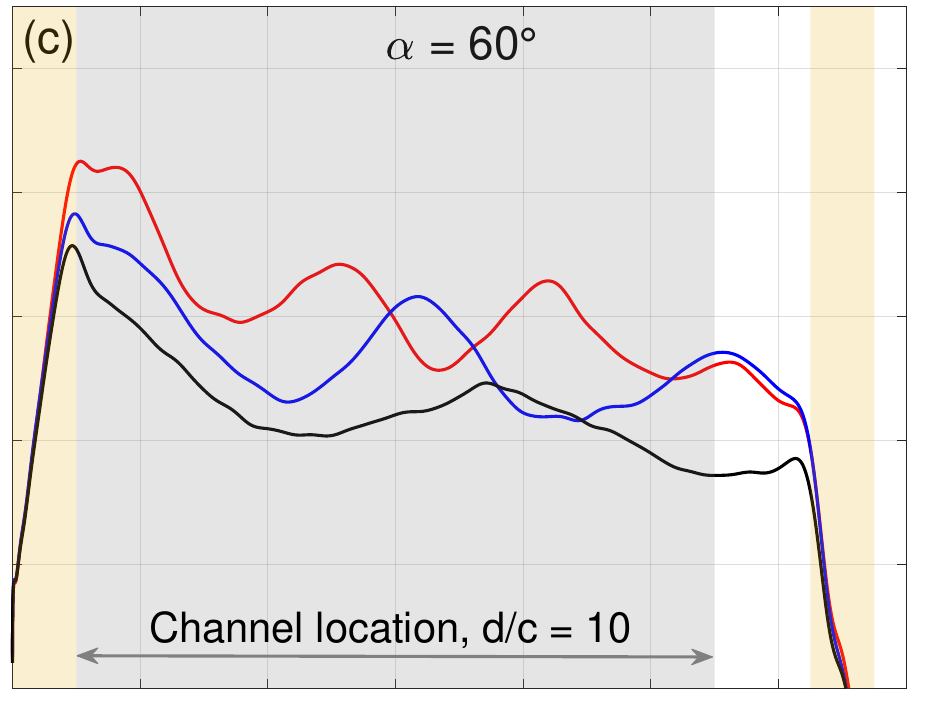}\label{fig:chan_H_10_60deg}}
	\\ \vspace{-10pt}
	\subfloat{
		\hspace{-5pt}\includegraphics[height=0.19914\textheight,keepaspectratio]{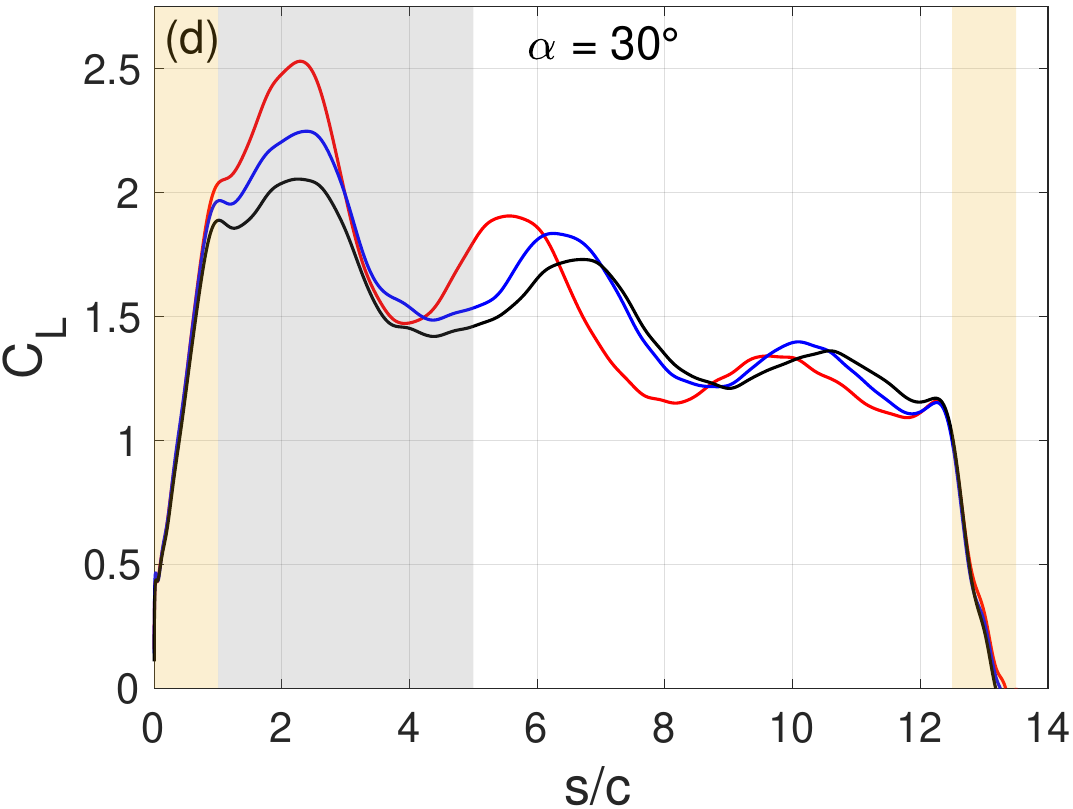}\label{fig:chan_H_04_30deg}}
	\subfloat{
		\includegraphics[height=0.19914\textheight,keepaspectratio]{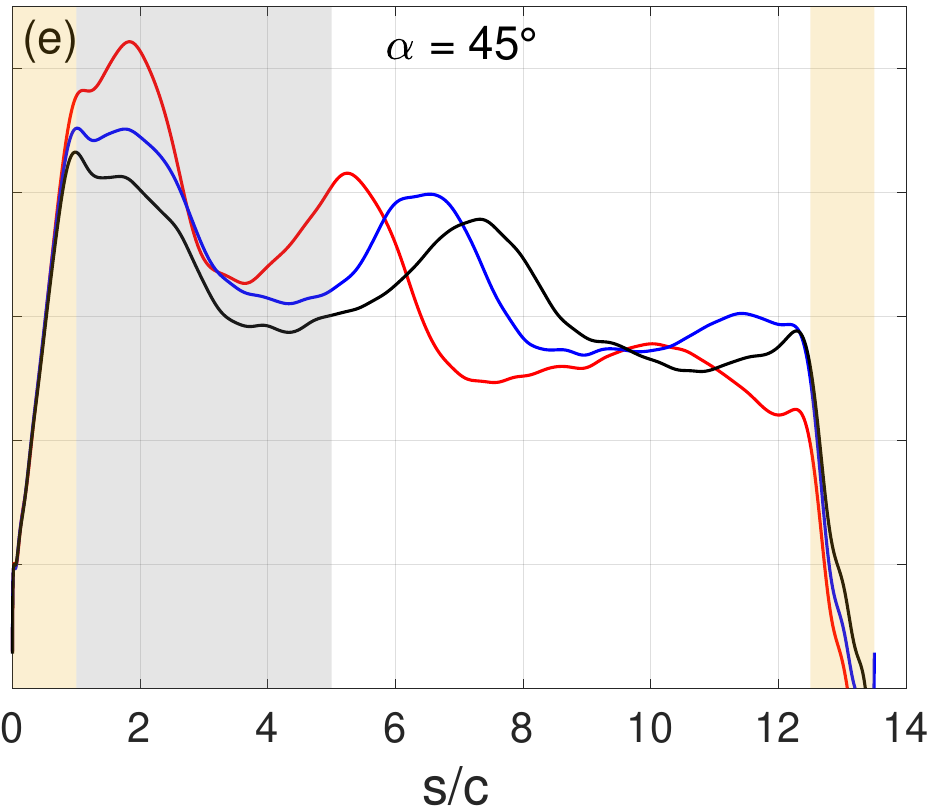}\label{fig:chan_H_04_45deg}}
	\subfloat{
		\includegraphics[height=0.19914\textheight,keepaspectratio]{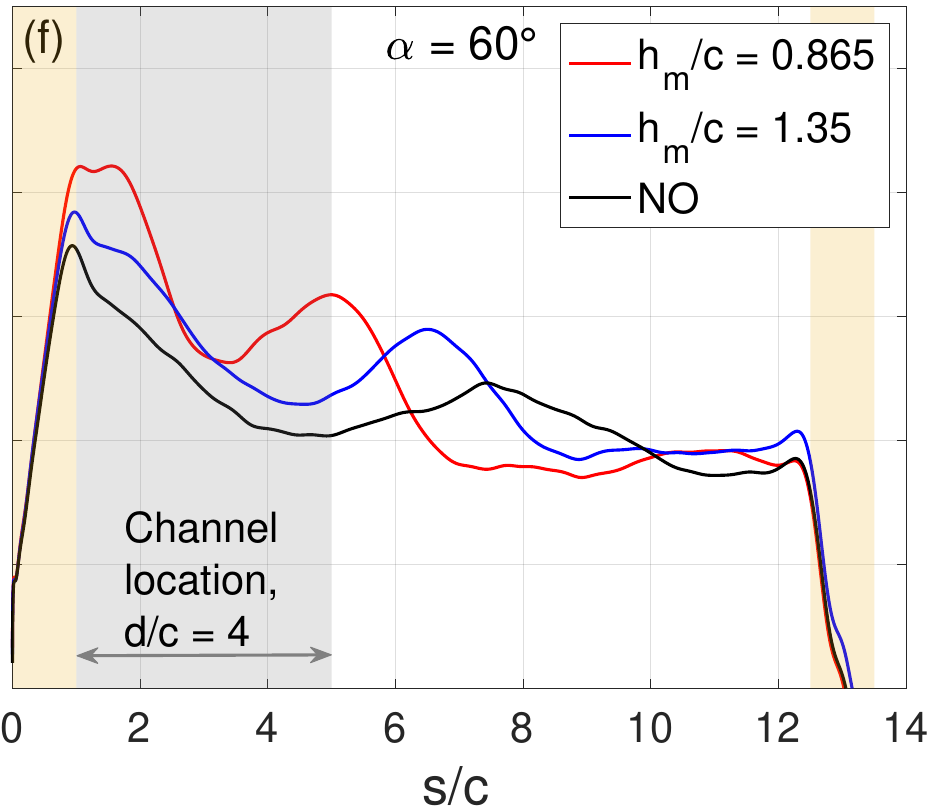}\label{fig:chan_H_04_60deg}}
	\vspace{-4pt}
	\caption{
	Channel gap-height ($h_\mathrm{m}/c$) variations; $d_\mathrm{a}/c=1$. Top row: $d/c=10$, bottom row: $d/c=4$.
	}\label{fig:chan_H}
\end{figure}

Figure \ref{fig:chan_H} shows $C_\mathrm{L}$ versus $s/c$ for channels of fixed $d/c$ but different $h_\mathrm{m}/c=0.865$ and 1.35; the vertical gap (with respect to the wing coordinates) between the streamwise channel walls is $2h_\mathrm{m}/c=1.73$ and 2.70. The $\alpha=\degrees{30}$, \degrees{45}, and \degrees{60} cases are in separate sub-plots, and the top and bottom rows show $d/c=10$ and 4, respectively; the NO data are also given (black curve). In all cases, $d_\mathrm{a}/c=1$, i.e., the wing LE reaches the channel start at $s/c=1$. Beige regions indicate the $s/c$ ranges for wing acceleration and deceleration, and the gray patch shows the streamwise channel extent. 

The $\alpha=\degrees{45}$, NO case shows typical $C_\mathrm{L}$ features (Fig.\ \ref{fig:chan_H_10_45deg}). The first $C_\mathrm{L}$ peak near $s/c=1$ where acceleration ceases is from added-mass (non-circulatory) force, proportional to the acceleration, and a contribution related to the growing and evolving vortex (LEV, TEV, and TV) flow and bound circulation, called circulatory force \cite{AVT202,PittFord2014,Mancini2015,Stevens2017,Eldredge2019}; ``vortex-induced force'' is also used, in force and moment partitioning \cite{Menon2021}. After $s/c=1$, only the circulatory force remains and the $C_\mathrm{L}$ peaks and minima are associated mainly with formation and shedding of LEV-signed circulation and TEVs \cite{Dickinson1993,AVT202,Mancini2015,Stevens2017,Mulleners2017,Eldredge2019}. In $\AR=4$, $\alpha=\degrees{45}$ studies, the LEV during the first peak is more coherent, while the second is a weaker recirculating flow \cite{Mancini2015,Stevens2017}. The current $\alpha=\degrees{45}$, NO case has two circulatory-$C_\mathrm{L}$ peaks then a third $C_\mathrm{L}$ rise halted at $s/c=12.5$ by wing deceleration, during which there is a $C_\mathrm{L}<0$ added-mass force. Successive circulatory-$C_\mathrm{L}$ peak heights decay, especially visible in the $\alpha=\degrees{30}$, NO case (Fig.\ \ref{fig:chan_H_10_30deg}). Mulleners \etal\ \cite{Mulleners2017} showed that this coincides with an aft shift in the LEV and TEV interactions with time. For the present NO cases, the $C_\mathrm{L}$ curves for each $\alpha$ are similar to prior $\AR=4$ data \cite{AVT202,Mancini2015}, but with higher values likely from the larger $\AR_\mathrm{eff}=8$. However, here the second circulatory-$C_\mathrm{L}$ peak is greater for $\alpha=\degrees{45}$ versus \degrees{30}, opposite the $\AR=4$ trend, probably due to the higher $\AR$ \cite{Taira2009}.

For the channel cases with $d/c=10$ (Fig.\ \ref{fig:chan_H}, top row), for all $\alpha$ there are clear $C_\mathrm{L}$ trends with narrowing $h_\mathrm{m}/c$, from the NO case to 1.35 to 0.865: the $C_\mathrm{L}$ peaks become larger, and after the first circulatory peak they occur earlier at progressively-smaller $s/c$. The $C_\mathrm{L}$ maximum when acceleration ceases ($s/c=1$) is also larger with smaller $h_\mathrm{m}/c$, likely from an increase in the circulatory-force contribution as the wing flow begins to interact with the channel entrance. For $\alpha=\degrees{45}$ and \degrees{60}, all channel cases have at least an additional $C_\mathrm{L}$ peak before deceleration, likely indicating a further LEV formation/shedding cycle. These trends imply that the flow over the wing evolves relatively faster for smaller $h_\mathrm{m}/c$.

The wing's blockage effect in the channel will increase the velocity over it, since the area for the flow to pass around the wing decreases versus the NO case. For example, at $\alpha=\degrees{45}$ the blockage defined by the forward-projected wing area divided by the cross-sectional channel-gap area (from the free surface to the tank bottom) is 12.9\% and 20.2\% for $h_\mathrm{m}/c=1.35$ and 0.865, respectively, but 3.5\% for the NO reference. Higher blockage will raise $C_\mathrm{L}$ and enhance the flow evolution, consistent with longer-time studies that show larger $C_\mathrm{L}$ and $St$ with greater blockage \cite{Abernathy1962,Ota1990,Zhou2019}.

Ignoring viscous effects, the potential-flow method of images can give an alternate explanation \cite{Meng2019,Quinn2014,Katz2001}. Assuming 2D flow and an infinite channel length, the upper and lower channel walls can be modeled using mirror-image (opposite-sign) LEVs and TEVs at equal distances from each wall but on the other side, to satisfy the no-flow-through condition; infinite sets of image vortices are required \cite{Katz2001}. Considering the near-LE region, the image-LEV above the wing, on the other side of the ceiling, will enhance the growth and advection of the real LEV via the Biot-Savart law; similarly for the TEV and its image. This will yield a higher $C_\mathrm{L}$ and earlier formation/shedding peaks, as found here.

For $d/c=10$, all cases show a second circulatory-$C_\mathrm{L}$ peak well past the channel entrance, and the timing trends with $\alpha$ depend on $h_\mathrm{m}/c$. For the NO results, the peaks are located at about $s/c=6.7$, 7.3, and 7.4 for $\alpha=\degrees{30}$, \degrees{45}, and \degrees{60}, respectively, increasing with larger $\alpha$. For $h_\mathrm{m}/c=1.35$, the peak timing is similar at $s/c=6.4$, 6.3, and 6.4, respectively. With $h_\mathrm{m}/c=0.865$, the trend reverses from the NO data, with $s/c=5.7$, 5.3, and 5.1, respectively, decreasing with greater $\alpha$. This is likely from the larger blockage with higher $\alpha$, which is highest for $h_\mathrm{m}/c=0.865$. The blockage percentages for $\alpha=\degrees{30}$, \degrees{45}, and \degrees{60} are 14.3\%, 20.2\%, and 24.7\%, respectively, so the flow over the wing should evolve relatively faster with larger $\alpha$. Note that this second peak always occurs earlier in a channel than the NO case.

Figure \ref{fig:chan_H}, second row shows variations with $h_\mathrm{m}/c$ for the shorter $d/c=4$ channel. The $C_\mathrm{L}$ curves are close to those of $d/c=10$ up to the first circulatory peak, but before the wing reaches the channel exit ($s/c=5$) they become lower. This happens earliest for $h_\mathrm{m}/c=0.865$ versus 1.35, from the greater channel influence. The second circulatory maximum, occurring near or after the exit for $d/c=4$, is therefore reduced compared to $d/c=10$, but still higher than the NO case and the peak timing remains similar to that of $d/c=10$. As with $d/c=10$, this $C_\mathrm{L}$ peak is larger for $h_\mathrm{m}/c=0.865$ than 1.35. However, after this second peak the $C_\mathrm{L}$ for $h_\mathrm{m}/c=0.865$ falls to a minimum below the other cases. For $\alpha=\degrees{30}$ and \degrees{45}, the third circulatory-$C_\mathrm{L}$ maximum for $d/c=4$ is much smaller than for $d/c=10$, closer to the NO level, and with the narrowest $h_\mathrm{m}/c=0.865$ channel its timing is later (relaxed) versus $d/c=10$. At $\alpha=\degrees{60}$, both $C_\mathrm{L}$ curves for $d/c=4$ eventually level off at low values with no third circulatory peak unlike $d/c=10$. Overall, the $d/c=4$ data show a lift loss as the wing exits the channel; for $h_\mathrm{m}/c=0.865$ the lift is worse at times than the NO data.

\vspace{-5 pt}
\subsection{Variations with Channel Length}\label{sec:chan_L}

\begin{figure}[t!]
	\centering
	\subfloat{
		\hspace{-5pt}\includegraphics[height=0.175\textheight,keepaspectratio]{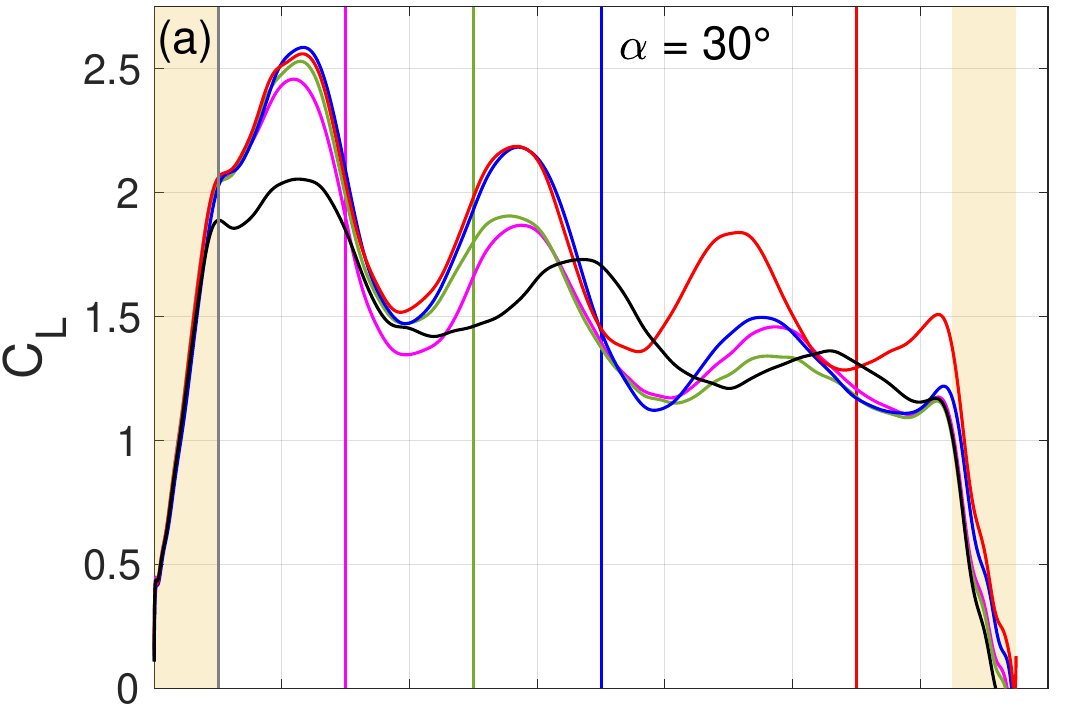}\label{fig:chan_L_0p865_30deg}}
	\subfloat{
		\includegraphics[height=0.175\textheight,keepaspectratio]{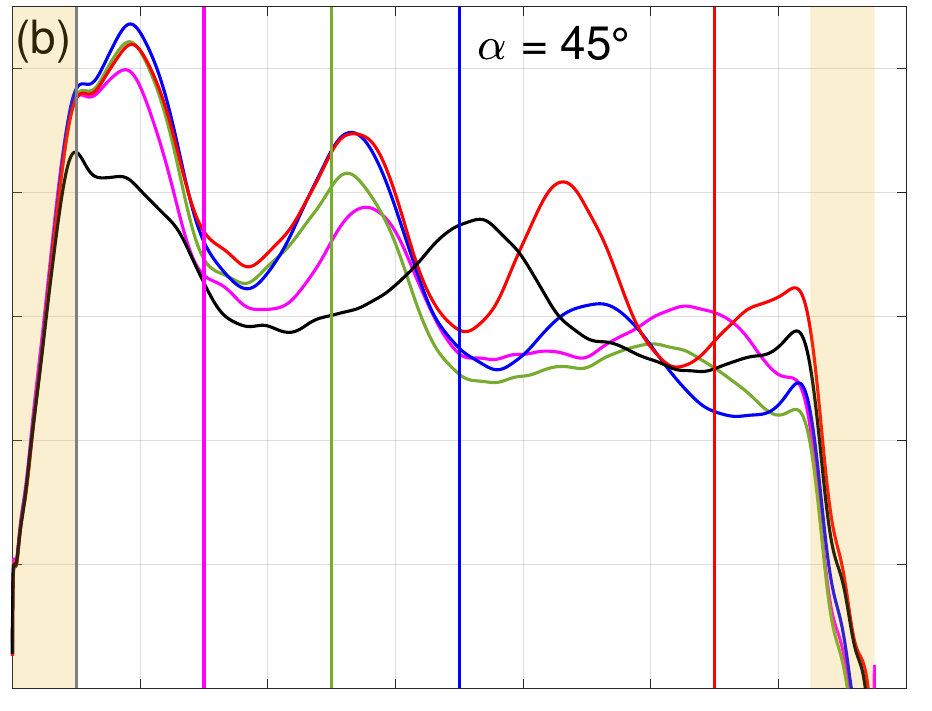}\label{fig:chan_L_0p865_45deg}}
	\subfloat{
		\includegraphics[height=0.175\textheight,keepaspectratio]{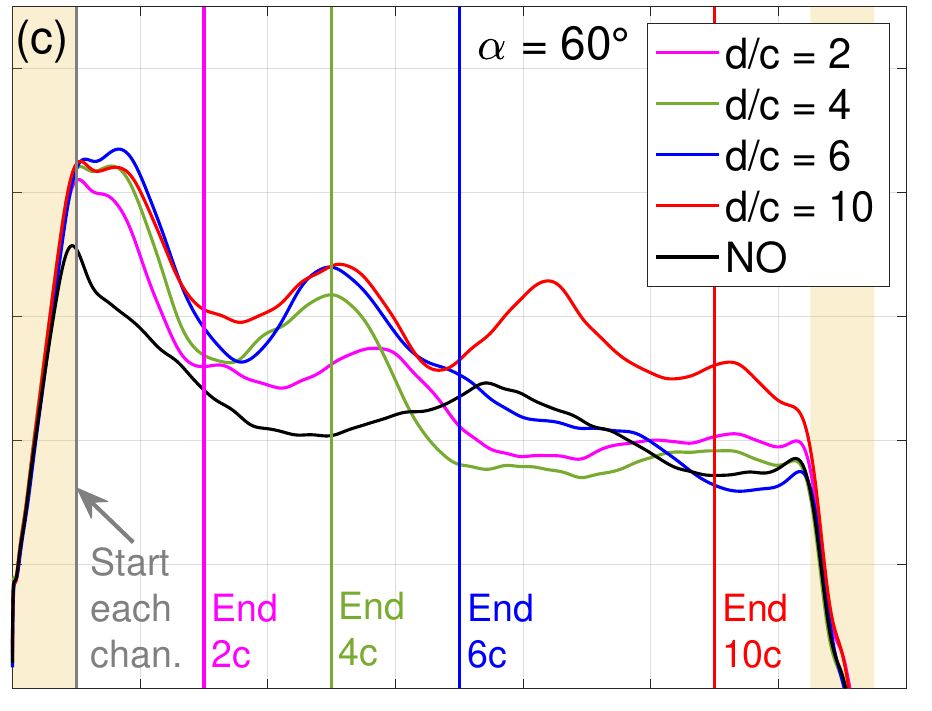}\label{chan_L_0p865_60deg}}
	\\ \vspace{-10pt}
	\subfloat{
		\hspace{-5pt}\includegraphics[height=0.19914\textheight,keepaspectratio]{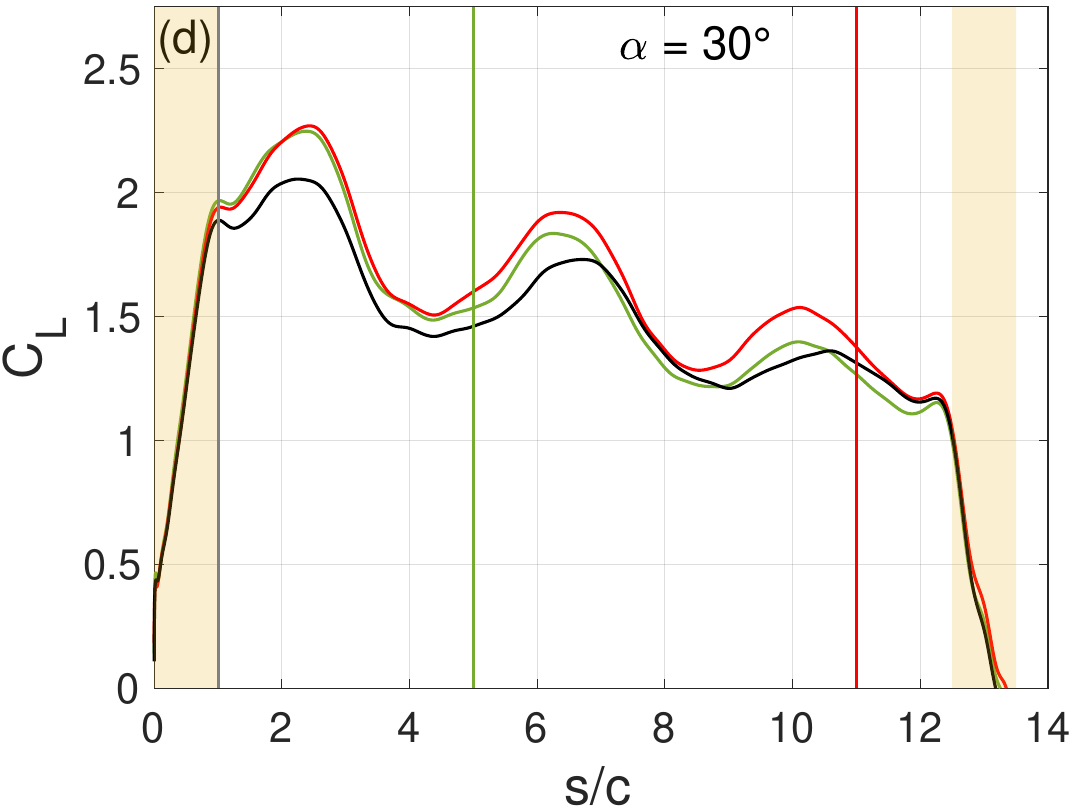}\label{fig:chan_L_1p35_30deg}}
	\subfloat{
		\includegraphics[height=0.19914\textheight,keepaspectratio]{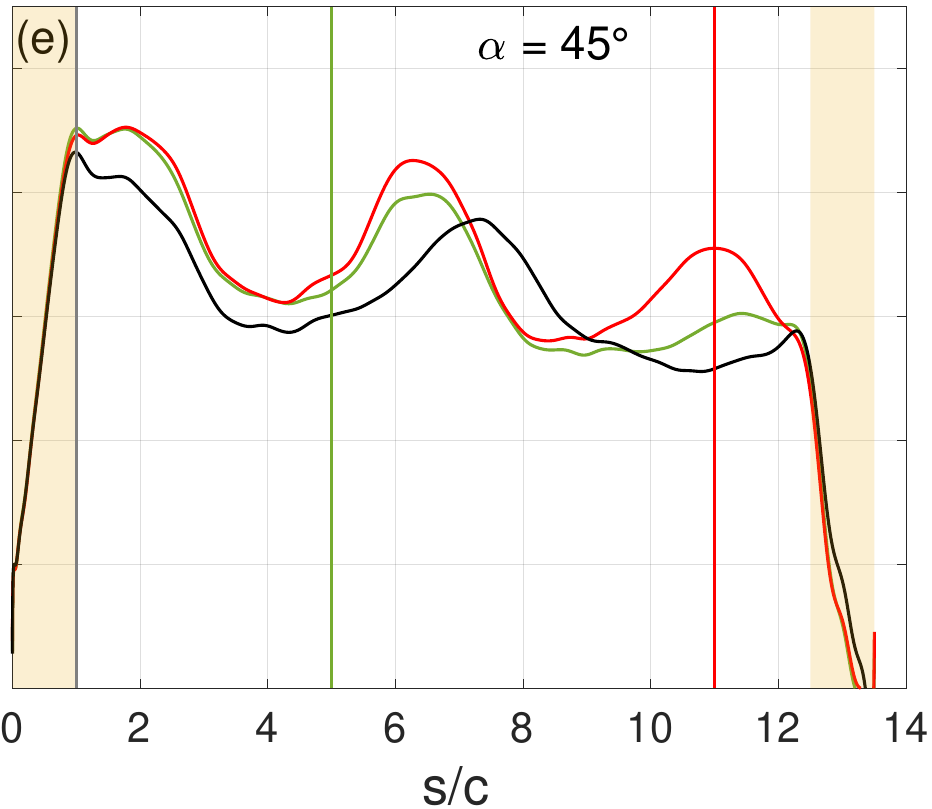}\label{fig:chan_L_1p35_45deg}}
	\subfloat{
		\includegraphics[height=0.19914\textheight,keepaspectratio]{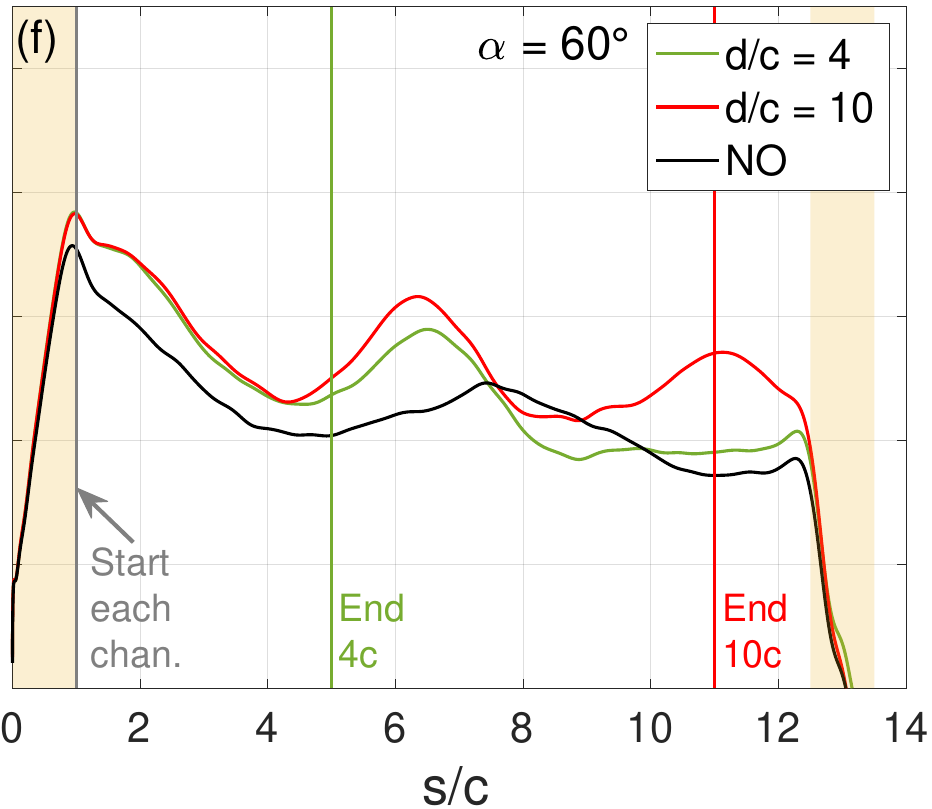}\label{fig:chan_L_1p35_60deg}}
	\vspace{-4pt}
	\caption{
	Channel length ($d/c$) variations; $d_\mathrm{a}/c=1$. Top row: $h_\mathrm{m}/c=0.865$, bottom row: $h_\mathrm{m}/c=1.35$.
	}\label{fig:chan_L}
\end{figure}

The $C_\mathrm{L}$ changes with channel length from $d/c=2$ to 10 are shown in Fig.\ \ref{fig:chan_L}; the top row is for $h_\mathrm{m}/c=0.865$, bottom is 1.35 for $d/c=4$ and 10. Different colors indicate each $d/c$, all channels start at $s/c=1$ (vertical gray line), and each ending position is marked by a color-matched vertical line. In Fig.\ \ref{fig:chan_L}, top row, for $d/c=4$ and 10 the first circulatory-$C_\mathrm{L}$ maximum after the peak at $s/c\approx1$ is very similar in magnitude at a given $\alpha$. However, for $d/c=6$ (blue), in the original data for each $\alpha$ the first circulatory-$C_\mathrm{L}$ peak is slightly higher than for $d/c=4$ and 10, which is unexpected since $d/c=6$ is between these; also, for $\alpha=\degrees{45}$ the peak at $s/c\approx1$ is somewhat larger. To verify these data the measurements were retaken, which occurred after the setup had been disassembled and reassembled. Great care was taken in setting all parameters. For $\alpha=\degrees{30}$ and $\degrees{45}$, the correlation-based alignment was incorrect for 1 of the 10 new water runs, although visually these are comparable to the others; to be conservative, these were omitted rather than doing a manual alignment. The new data are similar to the originals within the uncertainty, and it was decided to average the prior and new runs for each $\alpha$; these results are shown for $d/c=6$ in Fig.\ \ref{fig:chan_L}. With these new runs, the only appreciable difference is that for $\alpha=\degrees{30}$, the first circulatory-$C_\mathrm{L}$ peak is slightly closer to that of $d/c=4$ and 10; for $\alpha=\degrees{45}$ and \degrees{60}, the peak difference remains. The cause is unknown and future flow measurements will examine it.

For $h_\mathrm{m}/c=0.865$ and $d/c=2$ (Fig.\ \ref{fig:chan_L}, top row), the first circulatory-$C_\mathrm{L}$ peak is lower than for the other channels, likely because the channel exit is only $\sim$$1c$ away from the wing LE when the peak occurs. As the induced flow from the moving wing passes through the exit, it will slow because there is a factor of 5.85 area increase from the channel exit to the main tank. This should reduce the pressure on the front of the wing, decrease the oncoming flow speed in the wing frame, and weaken the separated shear layers over the LE and TE, affecting LEV and TEV formation and lowering $C_\mathrm{L}$.

For $h_\mathrm{m}/c=0.865$ channels with $d/c<10$, all $C_\mathrm{L}$ peaks occurring after a channel exit are lower than their $d/c=10$ counterparts (Fig.\ \ref{fig:chan_L}, top row). This makes sense, as the $d/c=10$ channel should sustain $C_\mathrm{L}$ enhancement the longest. In all cases except $\alpha=\degrees{30}$ and \degrees{60} for $d/c=6$, the departure below the $d/c=10$ data is observed within $1c$--$3c$ of travel before the channel exit. This is expected for an incompressible flow and might be used in future studies to help sense the exit. The exit for $d/c=2$ and 6, at $s/c=3$ and 7, respectively, is reached while the wing $C_\mathrm{L}$ decreases, roughly $1c$ prior to a $C_\mathrm{L}$ minimum between peaks in almost all cases. For $d/c=6$, $\alpha=\degrees{60}$ no nearby minimum exists, but $C_\mathrm{L}$ reduces while exiting. Conversely, with $d/c=4$ the exit ($s/c=5$) occurs for $\alpha=\degrees{30}$ during a $C_\mathrm{L}$ increase less than $1c$ before a maximum, and at $\alpha=\degrees{45}$ and \degrees{60} the exit is nearly aligned with a $C_\mathrm{L}$ peak. Exiting the channel when the circulatory-$C_\mathrm{L}$ is close to or at a peak ($d/c=4$ cases) should correspond to a large LEV near the wing, while exiting during a $C_\mathrm{L}$ decrease ($d/c=2$ and 6) should coincide with LEV shedding, based on prior NO studies \cite{Dickinson1993,AVT202,Mancini2015,Stevens2017,Mulleners2017,Eldredge2019}.

The $d/c=4$ cases have poor post-exit ($s/c>5$) lift. At \degrees{30}, $C_\mathrm{L}$ for $d/c=4$ is the lowest of the channel cases for $8.2<s/c<10.7$, and at \degrees{45} and \degrees{60} this is true for about $6.5<s/c<9.5$. At $\alpha=\degrees{30}$ and \degrees{45}, the third circulatory-$C_\mathrm{L}$ peak for $d/c=4$ is the lowest of the channel data. It is difficult to judge the post-exit $d/c=4$ lift against the NO case, as the feature timing differs, but the $C_\mathrm{L}$ minima for $d/c=4$ are comparable to or lower than the NO minima.

In summary, exiting the channel lowers the wing $C_\mathrm{L}$ versus the channel continuing, which often begins before the exit is reached. The post-exit $C_\mathrm{L}$ depends on the exit timing with respect to $C_\mathrm{L}$ peaks or decreases, reducing toward or below the NO value for the former, likely related to LEV formation and shedding. Flow experiments are required to examine this, including the effect of any vortices that form off the channel-exit corners from the wing-induced flow.

The shortest $d/c=2$ case (Fig.\ \ref{fig:chan_L}, top row) shows the sustained, post-exit channel influence for the more favorable exit condition, i.e., a $C_\mathrm{L}$ decrease. For each $\alpha$, after the $s/c=3$ exit any circulatory-$C_\mathrm{L}$ peaks are earlier and larger than the NO data, which persists $9.5c$ of travel past the exit until wing deceleration.

The second circulatory-$C_\mathrm{L}$ maximum for $d/c\ge4$ occurs at nearly the same $s/c$ for a given $\alpha$ (Fig.\ \ref{fig:chan_L}, top row), since the wing is in or near the channel. With increasing $\alpha$, this $s/c$ value is progressively earlier than the respective NO peak, likely from greater blockage at the same $h_\mathrm{m}/c$. For $d/c=2$, where this peak occurs after $2c$ traveled from the exit, for the higher-blockage $\alpha=\degrees{45}$ and \degrees{60} cases this maximum has a somewhat larger $s/c$ than for $d/c\ge4$. Meaning, there is a smaller timing shift from the NO peak with the wing being past the exit. At $\alpha=\degrees{30}$ and \degrees{45}, all channel cases show a third circulatory-$C_\mathrm{L}$ peak, found after the exit for $d/c\le6$. Its $s/c$ value increases toward the NO case's with decreasing $d/c$, even for $d/c=4$ with the lowest peak magnitude, but remains less than the NO's (at \degrees{45} there is no third NO peak).

For the greater $h_\mathrm{m}/c=1.35$ channel, Fig.\ \ref{fig:chan_L} (bottom row) shows $d/c=4$, 10, and NO results. The larger gap height has a smaller influence on $C_\mathrm{L}$, so for all $\alpha$ the second circulatory-$C_\mathrm{L}$ channel peak occurs later and is lower than for $h_\mathrm{m}/c=0.865$ (Fig.\ \ref{fig:chan_L}, top row). For $d/c=4$  (Fig.\ \ref{fig:chan_L}, bottom row), this later shift means the wing exits the channel shortly after a $C_\mathrm{L}$ minimum, rather than near or at a maximum as for $h_\mathrm{m}/c=0.865$, which reduces the adverse exit effect described above for the latter case. Therefore, with increased gap height the dependence on channel length $d/c$ is reduced, so for $h_\mathrm{m}/c=1.35$ the post-exit $C_\mathrm{L}$ for $d/c=4$ is closer to that of $d/c=10$ (past the second circulatory-$C_\mathrm{L}$ peak) compared to $h_\mathrm{m}/c=0.865$. Also, for the larger gap there is a smaller difference in the third circulatory-$C_\mathrm{L}$ peak's timing for $d/c=4$ and 10, at $\alpha=\degrees{30}$ and \degrees{45}. However, for all $\alpha$ and $h_\mathrm{m}/c=1.35$, the $C_\mathrm{L}$ for $d/c=4$ still lowers slightly below the $d/c=10$ case starting $\sim$$0.5c$ traveled before the exit, giving a ``warning'' as for $h_\mathrm{m}/c=0.865$.

\vspace{-5 pt}
\subsection{Effect of Vertical Distance to the Ceiling}\label{sec:ceil_H}

\begin{figure}[t!]
	\centering
	\subfloat{
		\hspace{-5pt}\includegraphics[height=0.175\textheight,keepaspectratio]{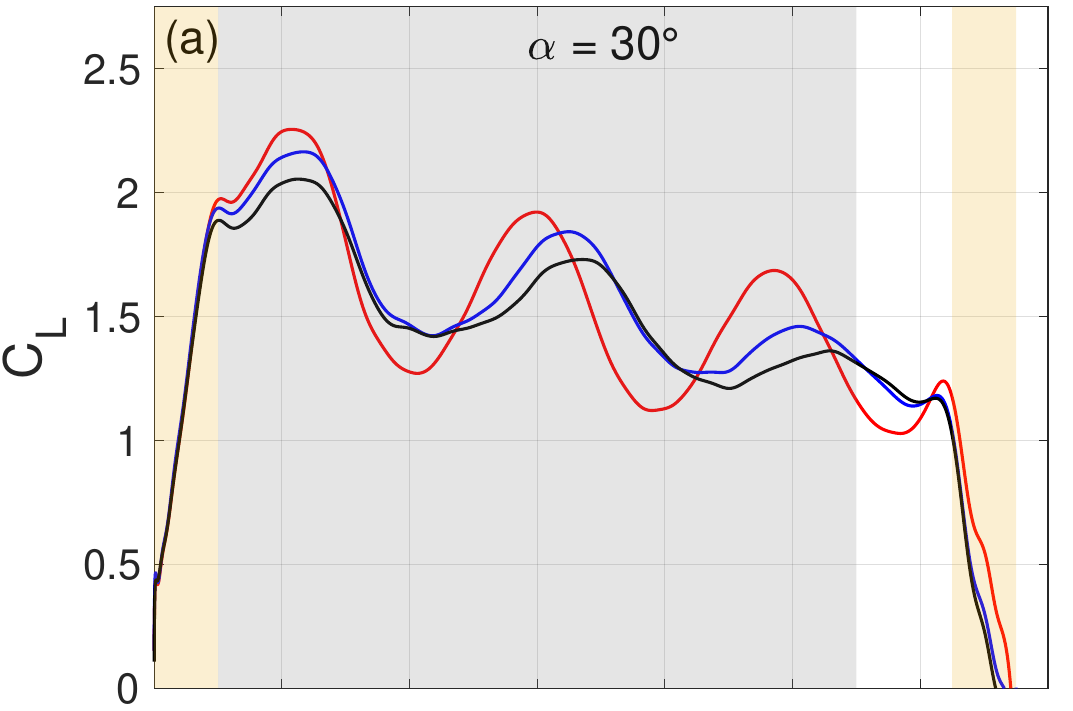}\label{fig:ceil_H_10_30deg}}
	\subfloat{
		\includegraphics[height=0.175\textheight,keepaspectratio]{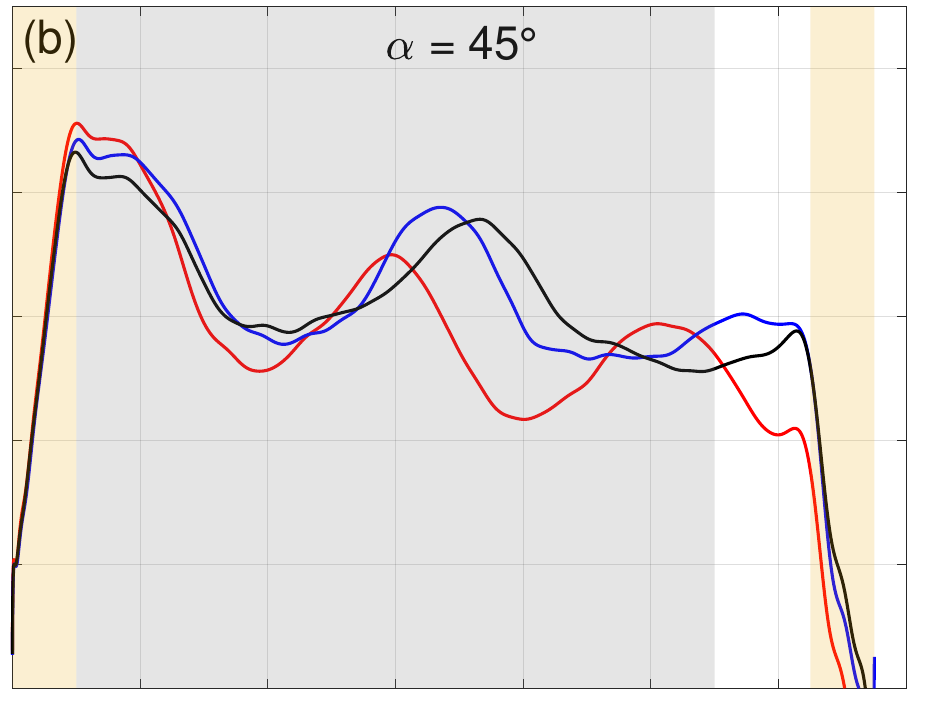}\label{fig:ceil_H_10_45deg}}
	\subfloat{
		\includegraphics[height=0.175\textheight,keepaspectratio]{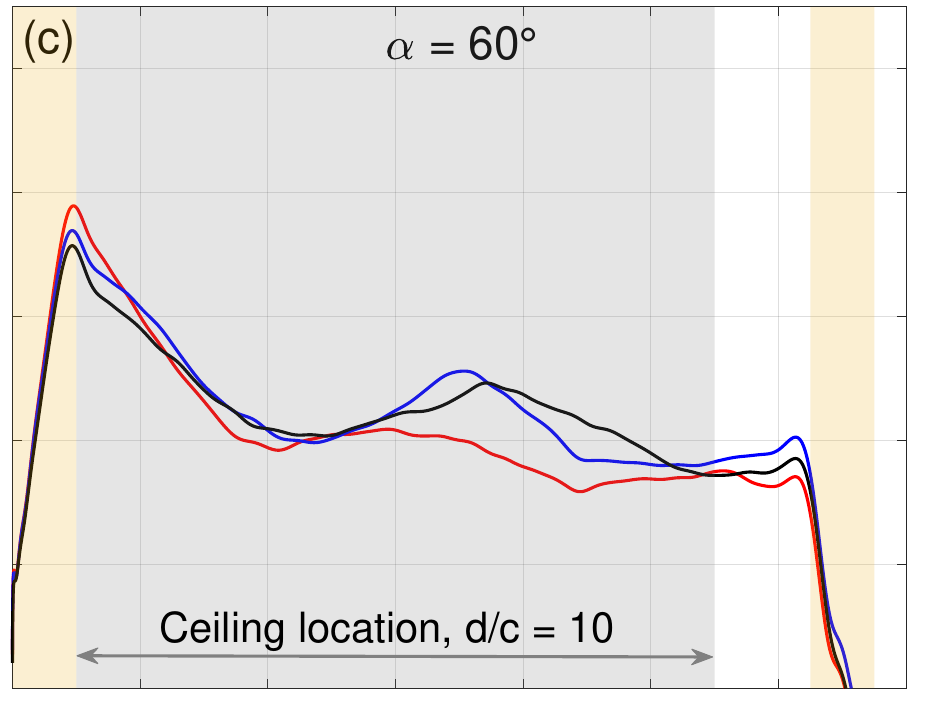}\label{fig:ceil_H_10_60deg}}
	\\ \vspace{-10pt}
	\subfloat{
		\hspace{-5pt}\includegraphics[height=0.19914\textheight,keepaspectratio]{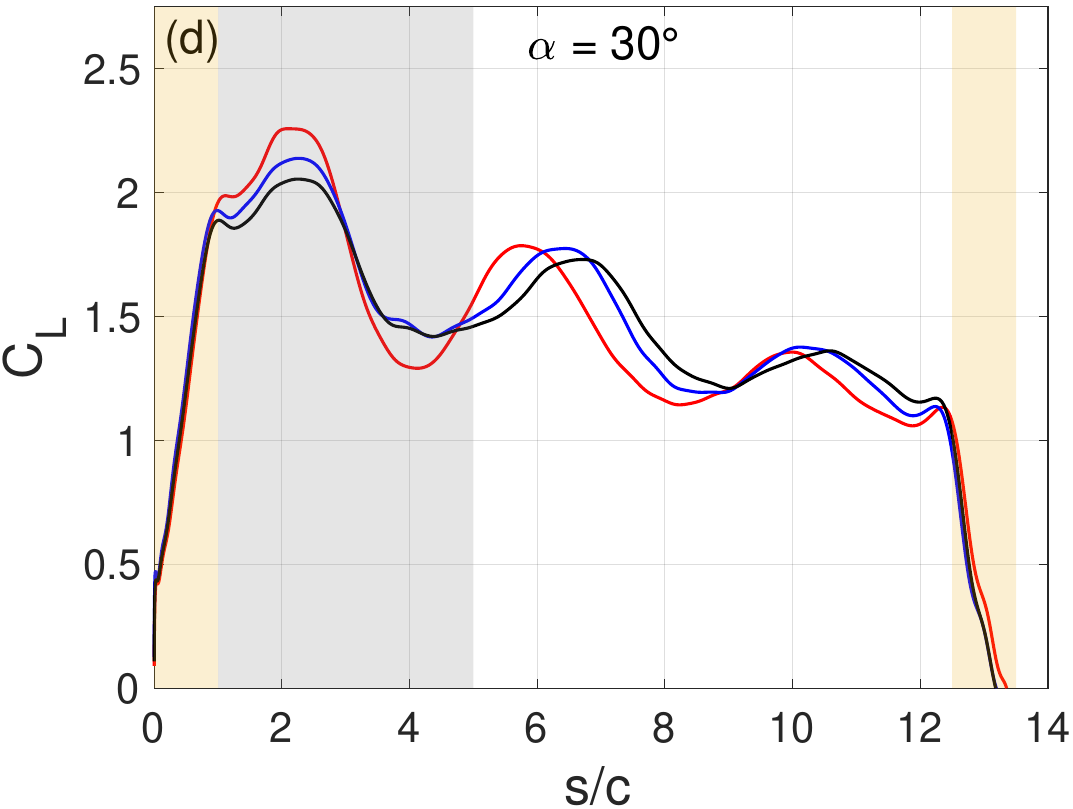}\label{fig:ceil_H_04_30deg}}
	\subfloat{
		\includegraphics[height=0.19914\textheight,keepaspectratio]{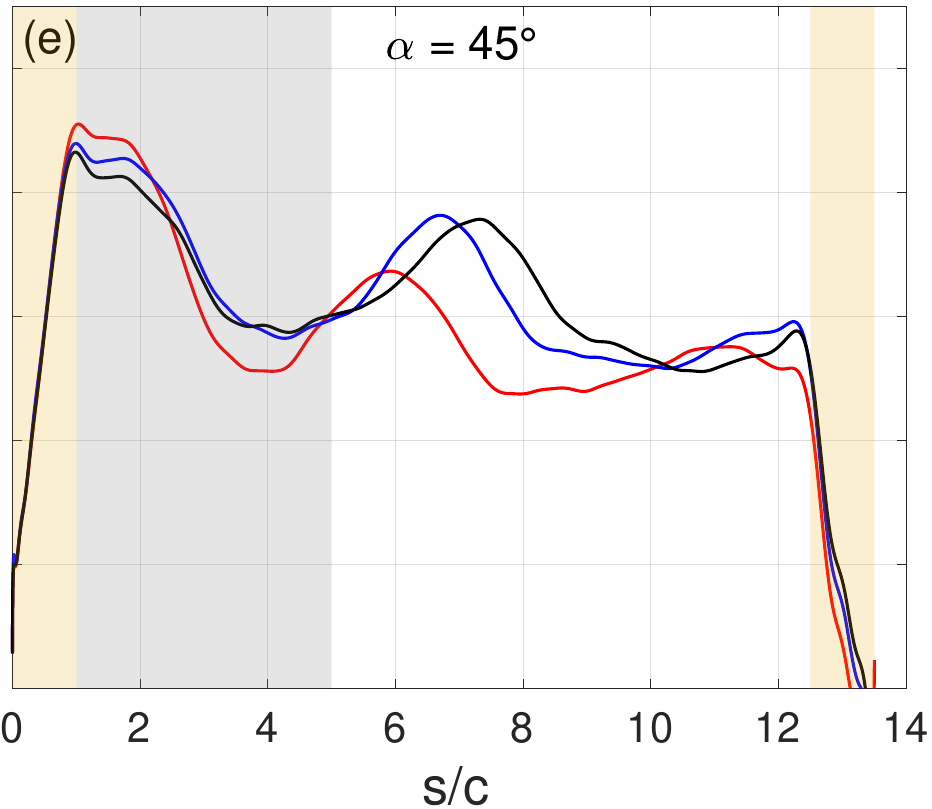}\label{fig:ceil_H_04_45deg}}
	\subfloat{
		\includegraphics[height=0.19914\textheight,keepaspectratio]{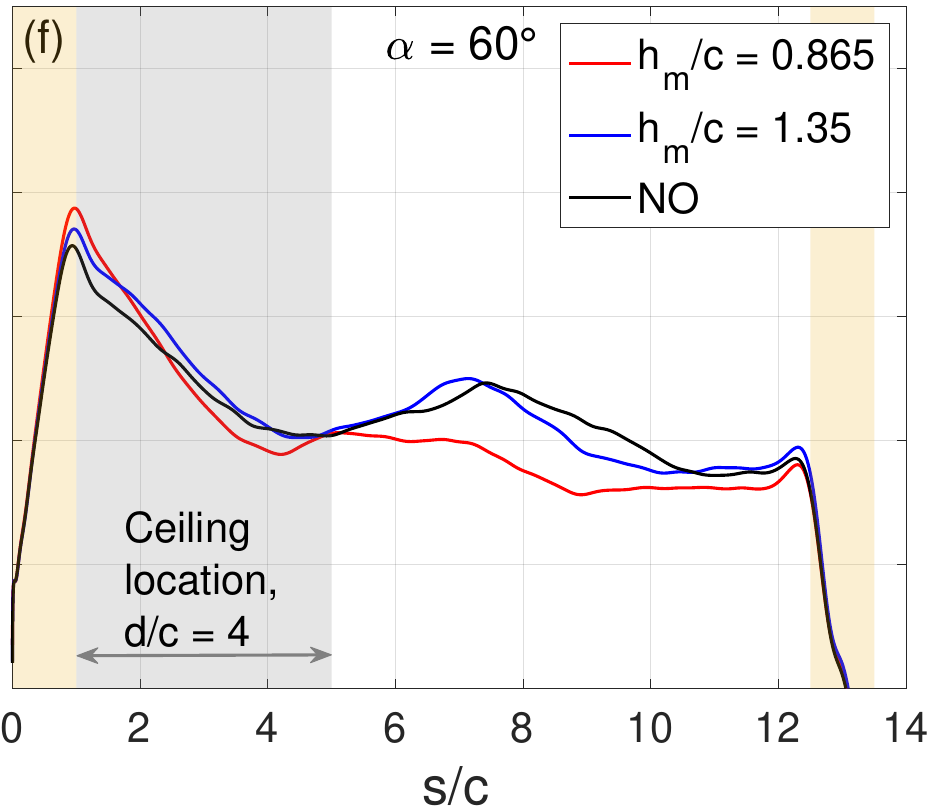}\label{fig:ceil_H_04_60deg}}
	\vspace{-4pt}
	\caption{
	Variations with vertical distance to the ceiling ($h_\mathrm{m}/c$); $d_\mathrm{a}/c=1$. Top row: $d/c=10$, bottom: $d/c=4$.
	}\label{fig:ceil_H}
\end{figure}

The effect of the vertical distance between the ceiling and mid-chord, $h_\mathrm{m}/c$, is shown in Fig.\ \ref{fig:ceil_H}; $d/c=10$ long ceilings (top row) are considered first. For $h_\mathrm{m}/c=0.865$ and $\alpha=\degrees{30}$, \degrees{45}, and \degrees{60}, the vertical distance or gap between the LE and streamwise ceiling surface is $h_\mathrm{LE}/c=0.62$, $0.51$, and $0.43$, respectively; with $h_\mathrm{m}/c=1.35$, it is $h_\mathrm{LE}/c=1.1$, $1.0$, and $0.92$. The gray patch in each sub-plot gives the streamwise ceiling position ($s/c=1$ to 11).

For $d/c=10$ and $h_\mathrm{m}/c=1.35$, the $C_\mathrm{L}$ peaks are always larger than the corresponding NO maxima, for all $\alpha$. Also, the second circulatory-$C_\mathrm{L}$ peak occurs at an earlier $s/c$ versus the NO data. A third circulatory maximum exists for $\alpha=\degrees{30}$ and \degrees{45}, timed before the third NO peak for \degrees{30}, but for \degrees{45} no such NO maximum forms within the run length. These ceiling results have the same trends as the $h_\mathrm{m}/c=1.35$ channel cases, but the latter have greater $C_\mathrm{L}$ peak magnitudes and, for $\alpha\ge\degrees{45}$, slightly larger timing differences with the NO data. At $\alpha=\degrees{30}$, the ceiling and channel $C_\mathrm{L}$ curves are somewhat closer. Note that for certain $\alpha=\degrees{60}$ cases (NO, all ceilings, and $h_\mathrm{m}/c=1.35$ channels and grounds), the first circulatory-$C_\mathrm{L}$ peak is not clearly distinct from the added-mass peak and so must occur very near to it.

With the ceiling closer ($h_\mathrm{m}/c=0.865$) and $d/c=10$ (Fig.\ \ref{fig:ceil_H}, top row), for $\alpha=\degrees{30}$ this yields the highest $C_\mathrm{L}$ maxima versus the NO and $h_\mathrm{m}/c=1.35$ cases, and earliest peak timing after the first circulatory peak. This trend also follows the channel data, but the channel-case peaks are larger and earlier (Fig.\ \ref{fig:chan_H}). For this $\alpha=\degrees{30}$ ceiling case, the $C_\mathrm{L}$ minima are lower than for the $h_\mathrm{m}/c=1.35$ and NO curves, which is not found for the channels. Interestingly, the peak-to-peak (maximum to subsequent minimum) changes in $C_\mathrm{L}$ for the three circulatory-$C_\mathrm{L}$ peaks are within $\sim$0.1 of the those for the channel at $\alpha=\degrees{30}$, so the ceiling-$C_\mathrm{L}$ oscillations are very similar but shifted lower.

The lift performance changes at $\alpha\ge\degrees{45}$ (Fig.\ \ref{fig:ceil_H}, top row, $d/c=10$). For $\alpha=\degrees{45}$ with $h_\mathrm{m}/c=0.865$, the first circulatory-$C_\mathrm{L}$ peak is the highest of the three curves, but the second is the smallest and the third is only comparable to that of $h_\mathrm{m}/c=1.35$. The $C_\mathrm{L}$ minima for $\alpha=\degrees{45}$ and $h_\mathrm{m}/c=0.865$ are lowest  versus the 1.35 and NO curves, as for $\alpha=\degrees{30}$, also influenced by the reduced second circulatory peak. At $\alpha=\degrees{60}$ and $h_\mathrm{m}/c=0.865$, the initial blended added-mass and circulatory-$C_\mathrm{L}$ maximum is highest, but the next peak near $s/c=6$ is lower and earlier than for the $h_\mathrm{m}/c=1.35$ and NO cases, similar to $\alpha=\degrees{45}$. Afterward, this $\alpha=\degrees{60}$ and $h_\mathrm{m}/c=0.865$ case has the poorest $C_\mathrm{L}$. 

Therefore, unlike the channel cases a smaller LE-to-ceiling gap can be detrimental for lift after the first circulatory-$C_\mathrm{L}$ peak. This occurs for $\alpha\ge\degrees{45}$ with $h_\mathrm{m}/c=0.865$, with gaps of $h_\mathrm{LE}/c=0.51$ and 0.43 at $\alpha=\degrees{45}$ and \degrees{60}, respectively. Flow fields are required to examine if the lift loss is from, e.g., flow restriction in this gap. Prior flapping-wing work on ceiling interactions \cite{Meng2019,Meng2020} showed only a monotonic $C_\mathrm{L}$ increase with decreasing $h_\mathrm{m}/c$, as found here for the first circulatory-$C_\mathrm{L}$ peak. However, the flapping-wing studies did not test $h_\mathrm{m}/c<1$, which leads to the adverse effect here, they use a rotational (waving) motion with a sinusoidal velocity program that generally yields one main circulatory-$C_\mathrm{L}$ peak per half-stroke for $\sim$$4.6c$--$9.8c$ of wingtip travel, and pitching alters $\alpha$ near the start and end of each half-stroke.

For $d/c=4$ length ceilings (Fig.\ \ref{fig:ceil_H}, bottom row), before the obstacle ends at $s/c=5$ the $C_\mathrm{L}$ curves for all cases are very similar to the $d/c=10$ results. For $\alpha=\degrees{30}$, where $h_\mathrm{LE}/c$ is largest, after the ceiling ends the $C_\mathrm{L}$ peak magnitudes for $h_\mathrm{m}/c=1.35$ and 0.865 diminish toward the NO values, below the $d/c=10$ levels. Despite this post-obstacle $C_\mathrm{L}$ reduction, the peak timing remains near that of $d/c=10$, so the earlier shift versus the NO data persists. The $\alpha=\degrees{45}$, $d/c=4$ with $h_\mathrm{m}/c=1.35$ case has similar trends, as do the channel data (Fig.\ \ref{fig:chan_H_04_30deg},e). As for channels, the timing shift makes it challenging to compare the $d/c=4$ and NO curves, to pinpoint a post-ceiling relaxation time. Conversely, for $\alpha=\degrees{45}$ and the smaller $h_\mathrm{m}/c=0.865$, the second circulatory-$C_\mathrm{L}$ peak is slightly lower than for $d/c=10$, reducing it further from the NO level rather than recovering, so this is a weakly-adverse ceiling-end effect. The next $C_\mathrm{L}$ minimum is, however, higher than for $h_\mathrm{m}/c=1.35$, then the $C_\mathrm{L}$ reaches the NO level $\sim$$5c$ after the ceiling end, and the third $C_\mathrm{L}$ peak is delayed versus $d/c=10$. For $\alpha=\degrees{60}$ with smaller $h_\mathrm{LE}/c$, the ceiling-end effect is weaker. The $h_\mathrm{m}/c=1.35$ curve nearly matches the $d/c=10$ case until $s/c=10$, then slightly lowers to follow the NO data so the ceiling influence on the magnitude subsides (Fig.\ \ref{fig:ceil_L_1p35_60deg}). With the ceiling closer ($h_\mathrm{m}/c=0.865$), the $d/c=4$ and 10 curves are also almost identical until $s/c\approx9$, then the former is somewhat below the $d/c=10$ and NO results, shifting it away from recovery.

To summarize the $d/c=4$ results, unlike for channels there is no $C_\mathrm{L}$ loss, or ``warning,'' ahead of the ceiling end, versus $d/c=10$. For larger LE-to-ceiling gaps there is a post-obstacle $C_\mathrm{L}$ reduction to the NO level, but the peak timing behaves as if the ceiling continues, for the available run length. For $\alpha=\degrees{30}$ this is true for both the $h_\mathrm{LE}/c=1.1$ and 0.62 gaps, for $\alpha=\degrees{45}$ this is found for $h_\mathrm{LE}/c=1.0$ ($h_\mathrm{m}/c=1.35$), and at $\alpha=\degrees{60}$ it occurs for $h_\mathrm{LE}/c=0.92$ ($h_\mathrm{m}/c=1.35$). The time at which the ceiling and NO curves start to show the most overlap is more delayed with higher $\alpha$. However, with the smaller $h_\mathrm{LE}/c=0.51$ and 0.43 gaps for $\alpha=\degrees{45}$ and \degrees{60}, respectively (both for $h_\mathrm{m}/c=0.865$), at $\alpha=\degrees{45}$ there is an initially-adverse post-exit $C_\mathrm{L}$ then a recovery, and at $\alpha=\degrees{60}$ the ceiling effect persists then for $s/c>9$ the $C_\mathrm{L}$ is somewhat below the NO level (adverse). There is some $\alpha$ dependence, but roughly for $h_\mathrm{LE}/c>0.6$ there is a gradual post-exit $C_\mathrm{L}$ decrease to the NO value. With $h_\mathrm{LE}/c<0.6$, there are slightly-adverse ceiling-end effects along with the detrimental ceiling influence described for $d/c=10$, yet for $\alpha=\degrees{45}$ there is an eventual recovery.

\vspace{-5 pt}
\subsection{Variations with Ceiling Length}\label{sec:ceil_L}

\begin{figure}[t!]
	\centering
	\subfloat{
		\hspace{-5pt}\includegraphics[height=0.175\textheight,keepaspectratio]{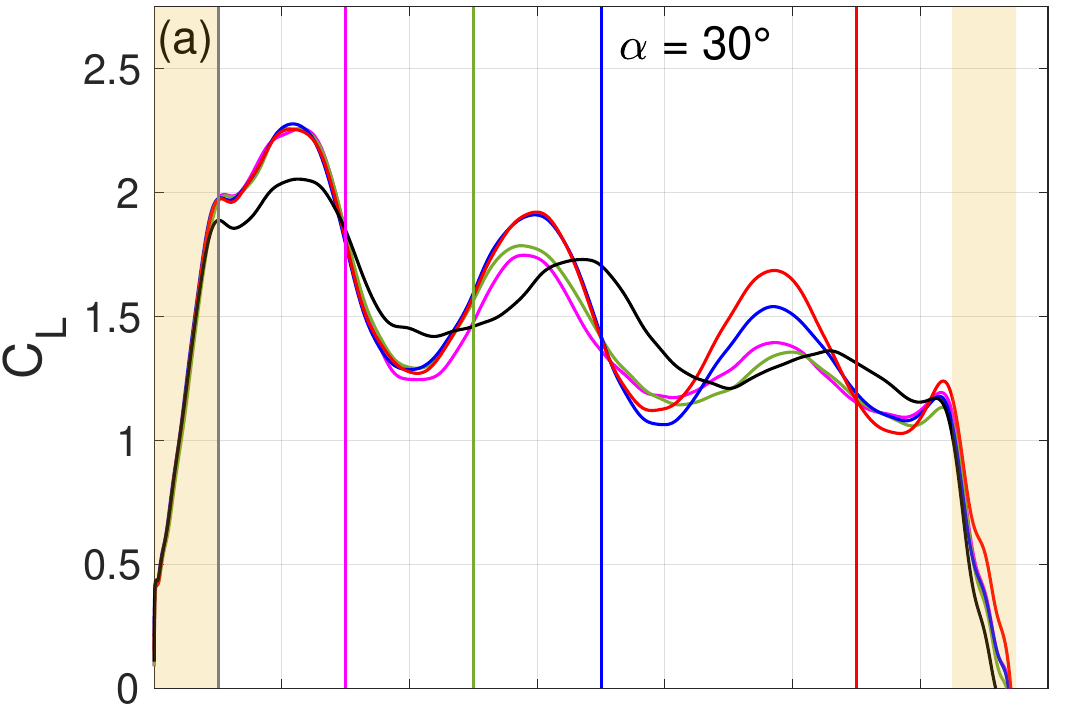}\label{fig:ceil_L_0p865_30deg}}
	\subfloat{
		\includegraphics[height=0.175\textheight,keepaspectratio]{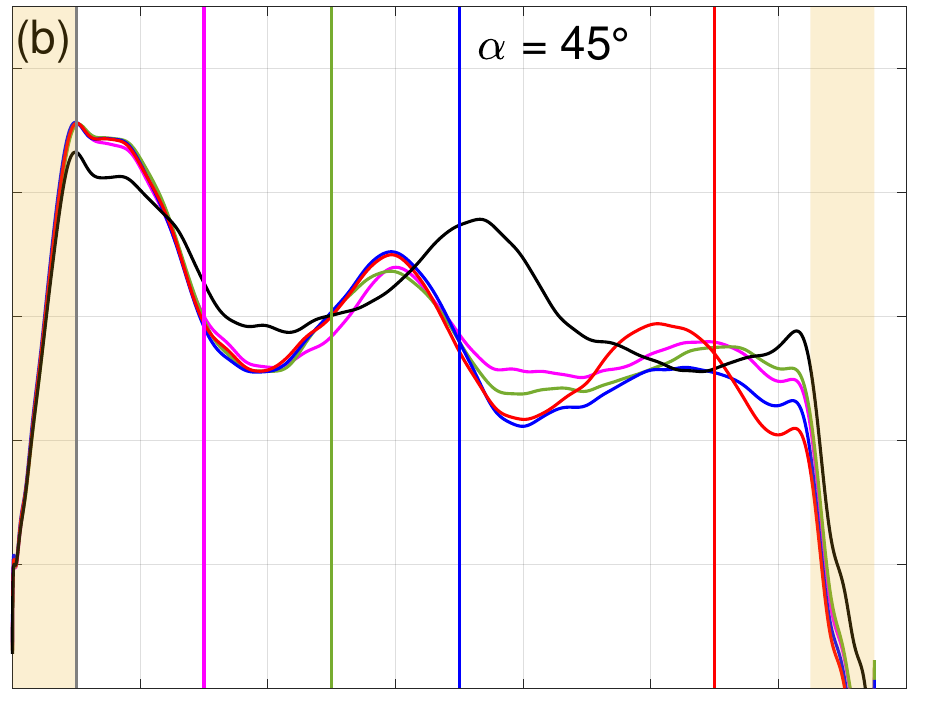}\label{fig:ceil_L_0p865_45deg}}
	\subfloat{
		\includegraphics[height=0.175\textheight,keepaspectratio]{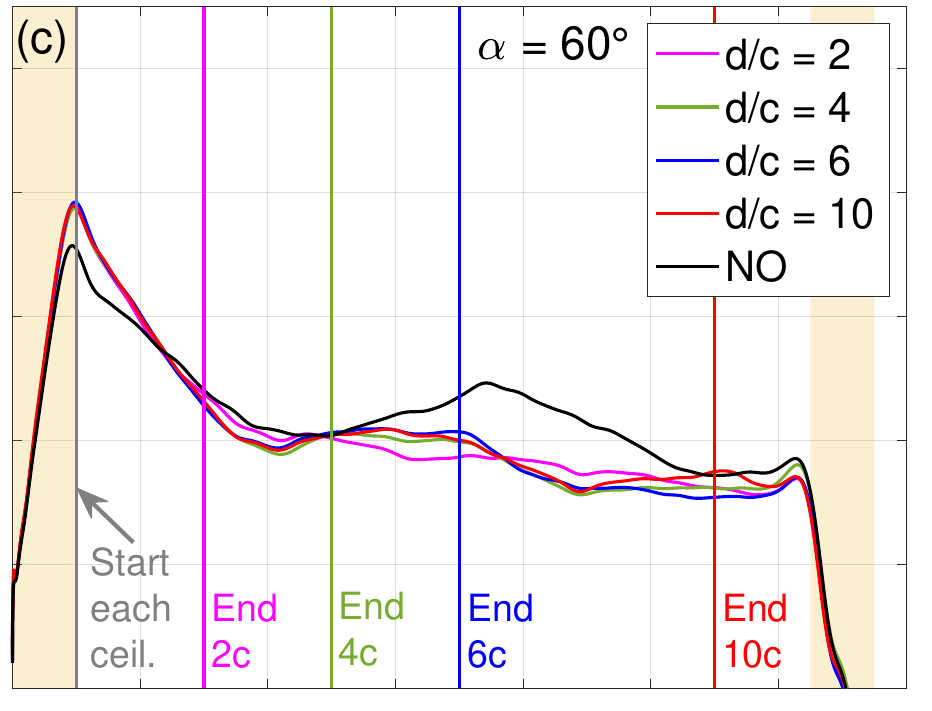}\label{ceil_L_0p865_60deg}}
	\\ \vspace{-10pt}
	\subfloat{
		\hspace{-5pt}\includegraphics[height=0.19914\textheight,keepaspectratio]{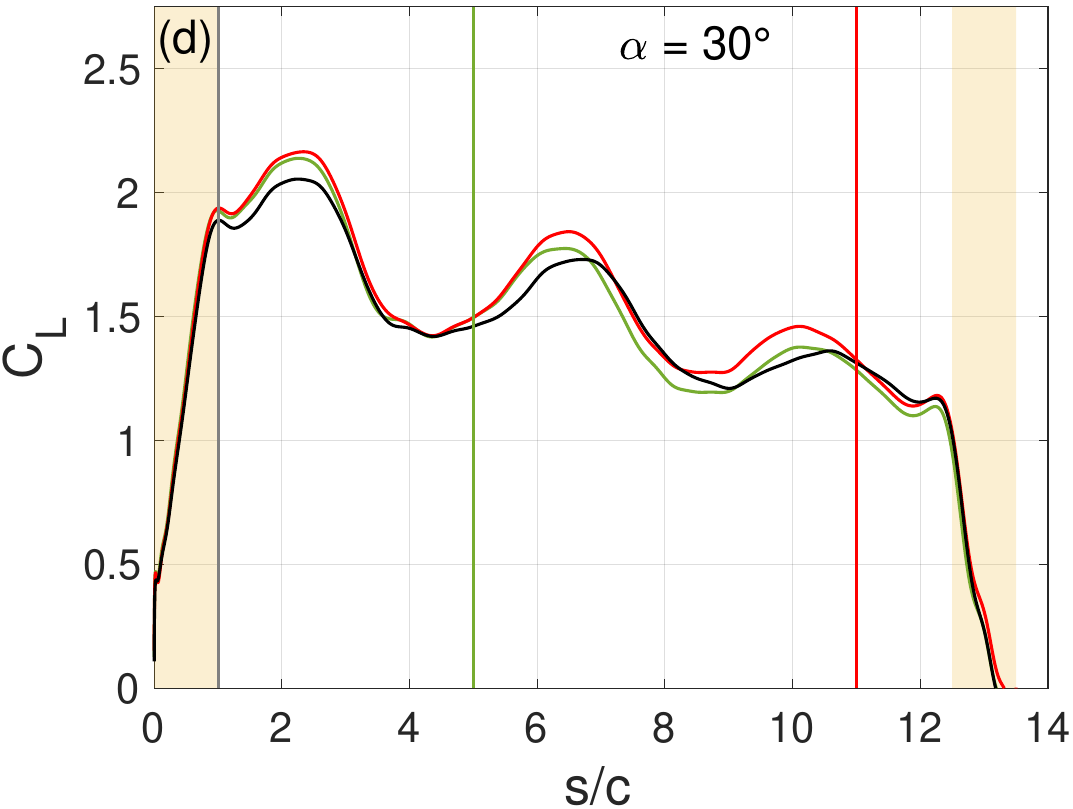}\label{fig:ceil_L_1p35_30deg}}
	\subfloat{
		\includegraphics[height=0.19914\textheight,keepaspectratio]{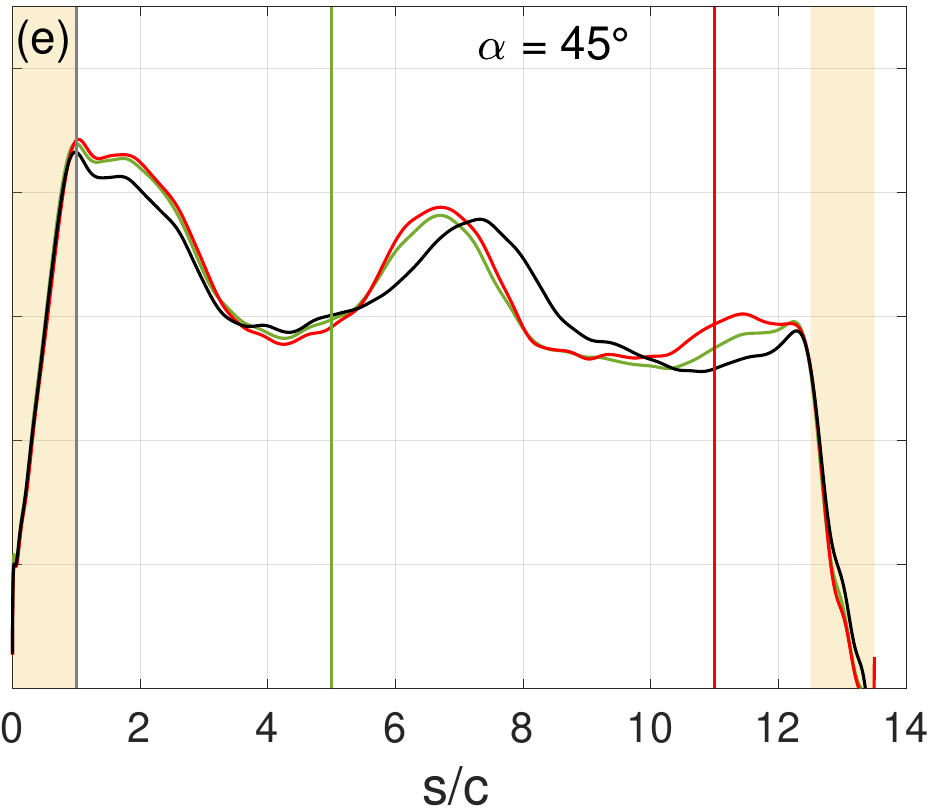}\label{fig:ceil_L_1p35_45deg}}
	\subfloat{
		\includegraphics[height=0.19914\textheight,keepaspectratio]{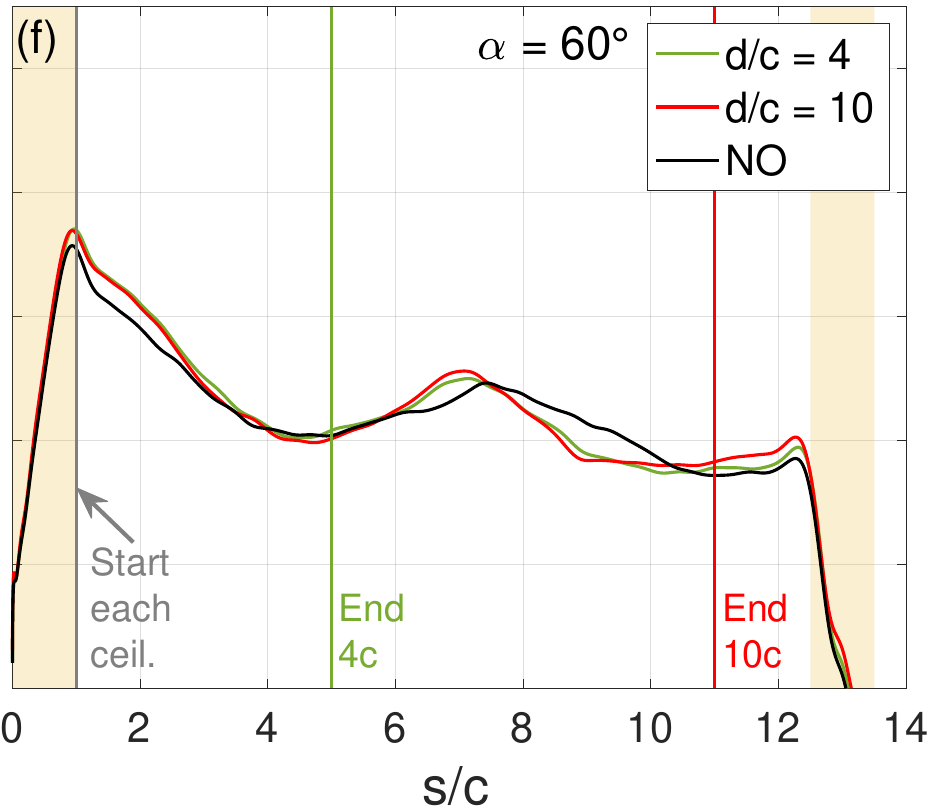}\label{fig:ceil_L_1p35_60deg}}
	\vspace{-4pt}
	\caption{
	Ceiling length ($d/c$) variations; $d_\mathrm{a}/c=1$. Top row: $h_\mathrm{m}/c=0.865$, bottom row: $h_\mathrm{m}/c=1.35$.
	}\label{fig:ceil_L}
\end{figure}

Changes with ceiling length are given in Fig.\ \ref{fig:ceil_L}: the top row shows $h_\mathrm{m}/c=0.865$ and $d/c=2$--10, the bottom presents $h_\mathrm{m}/c=1.35$ with $d/c=4$ and 10. Considering first $h_\mathrm{m}/c=0.865$, for $\alpha=\degrees{30}$ the ceiling-case trends are similar to the channel data (Fig.\ \ref{fig:chan_L_0p865_30deg}), but with generally lower and later $C_\mathrm{L}$ peaks, with the timing differences occurring from the second circulatory-$C_\mathrm{L}$ peak onward. Exceptions are the third circulatory maximum with $d/c=4$ and 6, where the more-pronounced channel-exit effects cause this peak to be slightly lower than the ceiling data.

For the $\alpha=\degrees{45}$ and \degrees{60} ceiling results with $h_\mathrm{m}/c=0.865$ (Fig.\ \ref{fig:ceil_L}, top row), there are greater differences with the channel cases, related to the poorer (ceiling) performance after the first circulatory peak described above. Ceiling-end effects are analyzed here by comparing the $d/c<10$ curves to the longest $d/c=10$ case. With $\alpha=\degrees{45}$, for $d/c=2$ and 4 the second circulatory-$C_\mathrm{L}$ peak, occurring after the ceiling ends, is somewhat below that of $d/c=6$ and 10 (Sec.\ \ref{sec:ceil_H}  discusses $d/c=4$). For all $d/c\leq10$, $\alpha=\degrees{45}$ cases, this peak has similarly-early timing and is lower than the corresponding NO maximum. Later the $C_\mathrm{L}$ for $d/c=2$ is larger and close to the post-peak NO level, indicating some lift recovery, then both the $d/c=2$ and 4 ceiling cases reach a similar third maximum value. For the $d/c=4$ channel, the exit influence is so detrimental that the post-exit $C_\mathrm{L}$ is almost as low as the ceiling result, despite a much higher second circulatory-$C_\mathrm{L}$ peak beforehand. Returning to ceilings, for $d/c=6$ its third-peak is the smallest of the cases, perhaps because the $d/c=2$ and 4 $C_\mathrm{L}$ have recovered further by this time, while $d/c=10$ is the highest. Unlike the $\alpha=\degrees{45}$ channel data, for the ceiling the timing of the obstacle end with respect to the $C_\mathrm{L}$ increasing or decreasing does not show a clear trend in subsequent $C_\mathrm{L}$ behavior. However, consistently the ceiling ending lowers the circulatory-$C_\mathrm{L}$ peak values versus $d/c=10$. As for $\alpha=\degrees{30}$, with $\alpha=\degrees{45}$ the variations with $d/c$ are generally less than for the channel. At $\alpha=\degrees{60}$, the $C_\mathrm{L}$ changes with $d/c$ are even smaller versus the channel, consistent with the $d/c=4$ discussion above (cf.\ Fig.\ \ref{fig:ceil_H_04_60deg}). The $d/c=2$ case has the greatest differences with $d/c=10$, with no large maxima or minima.

Figure \ref{fig:ceil_L}, bottom row, shows the ceiling-length study with wings shifted farther away ($h_\mathrm{m}/c=1.35$), for $d/c=4$ and 10. Per the $h_\mathrm{m}/c$ discussion (Sec.\ \ref{sec:ceil_H}), for all $\alpha$ these results are overall closer to the NO data. For $\alpha=\degrees{30}$ and \degrees{45}, the $C_\mathrm{L}$ trends versus $d/c$ follow the $h_\mathrm{m}/c=0.865$ results, but with smaller peak-magnitude changes; for channels the $C_\mathrm{L}$ variations with $d/c$ are also lower for larger $h_\mathrm{m}/c$. At $\alpha=\degrees{60}$, the $C_\mathrm{L}$ changes with $d/c$ are slight, as for $h_\mathrm{m}/c=0.865$.

In summary, for $d/c<10$ there is a post-ceiling peak-lift loss versus the $d/c=10$ case that is most prominent for smaller $h_\mathrm{m}/c$ and $\alpha<\degrees{60}$. The effect is reduced versus the channel cases due to the lower peak forces. For $h_\mathrm{m}/c=0.865$ and $\alpha\ge\degrees{45}$ where the LE is closest, the variations with $d/c$ are smallest, likely related to the poorer performance of the ceiling data after the first circulatory-$C_\mathrm{L}$ maximum. This may also play a role in the lack of a clear relationship between the $C_\mathrm{L}$ behavior at the ceiling end, e.g., a rise or fall there, and afterward. With $h_\mathrm{m}/c=0.865$ and both $\alpha=\degrees{30}$ and $\alpha=\degrees{45}$, for the shorter $d/c=2$ and 4 ceilings the third circulatory-$C_\mathrm{L}$ peak recovers toward the final NO peak value before deceleration, but for $d/c=6$ the run length is insufficient after the ceiling ends to judge recovery.

\vspace{-5 pt}
\subsection{Effect of Vertical Distance from the Ground}\label{sec:grnd_H}

\begin{figure}[t!]
	\centering
	\subfloat{
		\hspace{-5pt}\includegraphics[height=0.175\textheight,keepaspectratio]{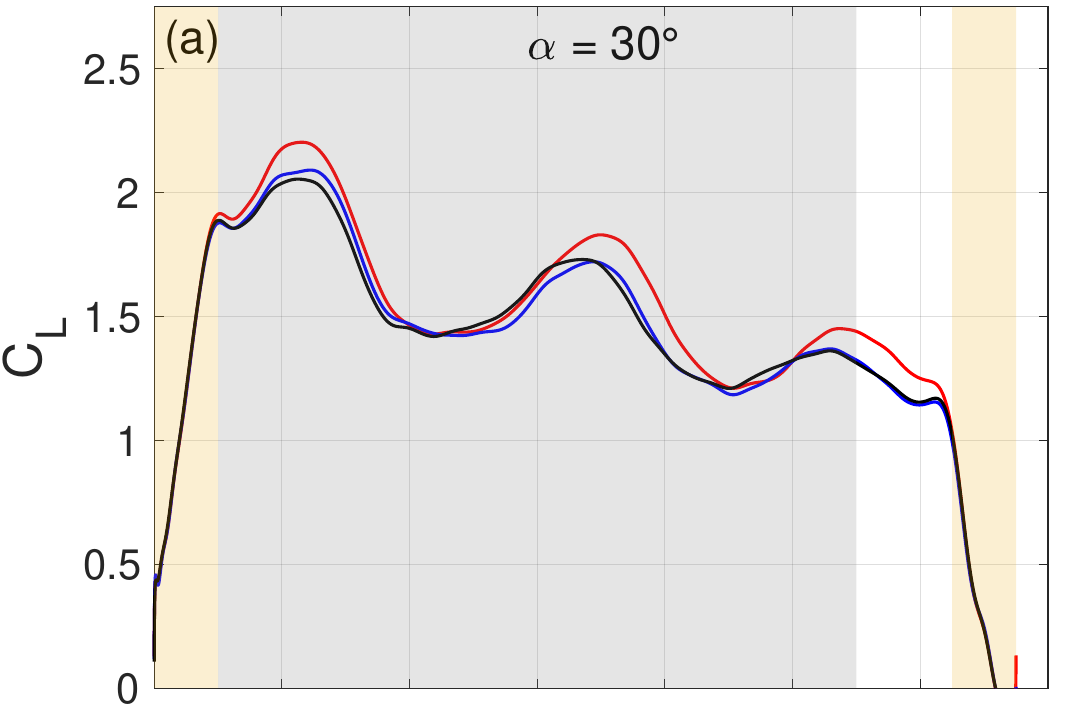}\label{fig:grnd_H_10_30deg}}
	\subfloat{
		\includegraphics[height=0.175\textheight,keepaspectratio]{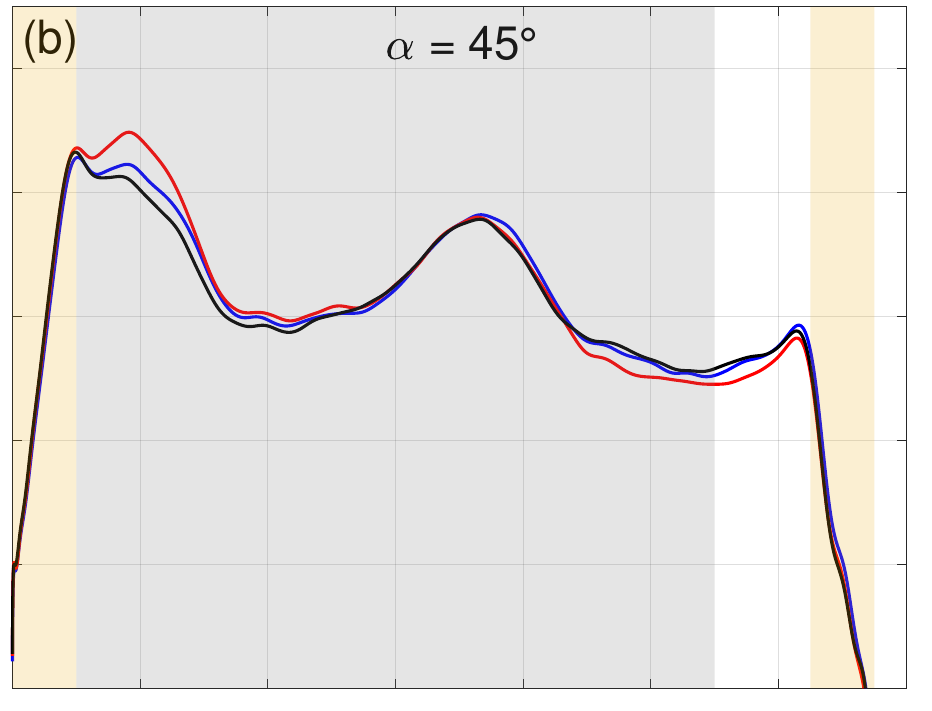}\label{fig:grnd_H_10_45deg}}
	\subfloat{
		\includegraphics[height=0.175\textheight,keepaspectratio]{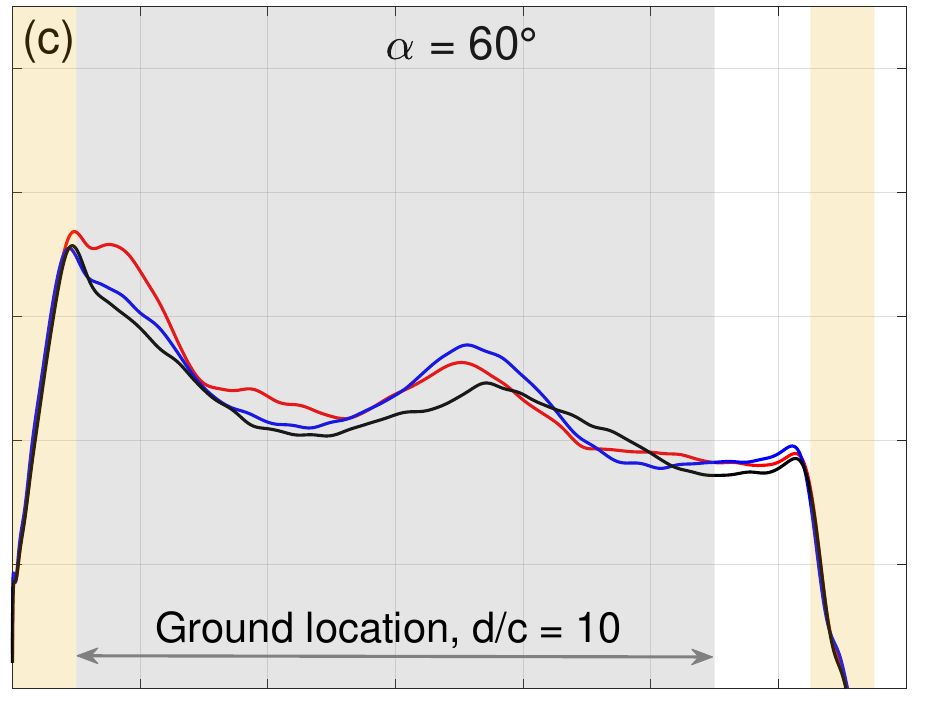}\label{fig:grnd_H_10_60deg}}
	\\ \vspace{-10pt}
	\subfloat{
		\hspace{-5pt}\includegraphics[height=0.19914\textheight,keepaspectratio]{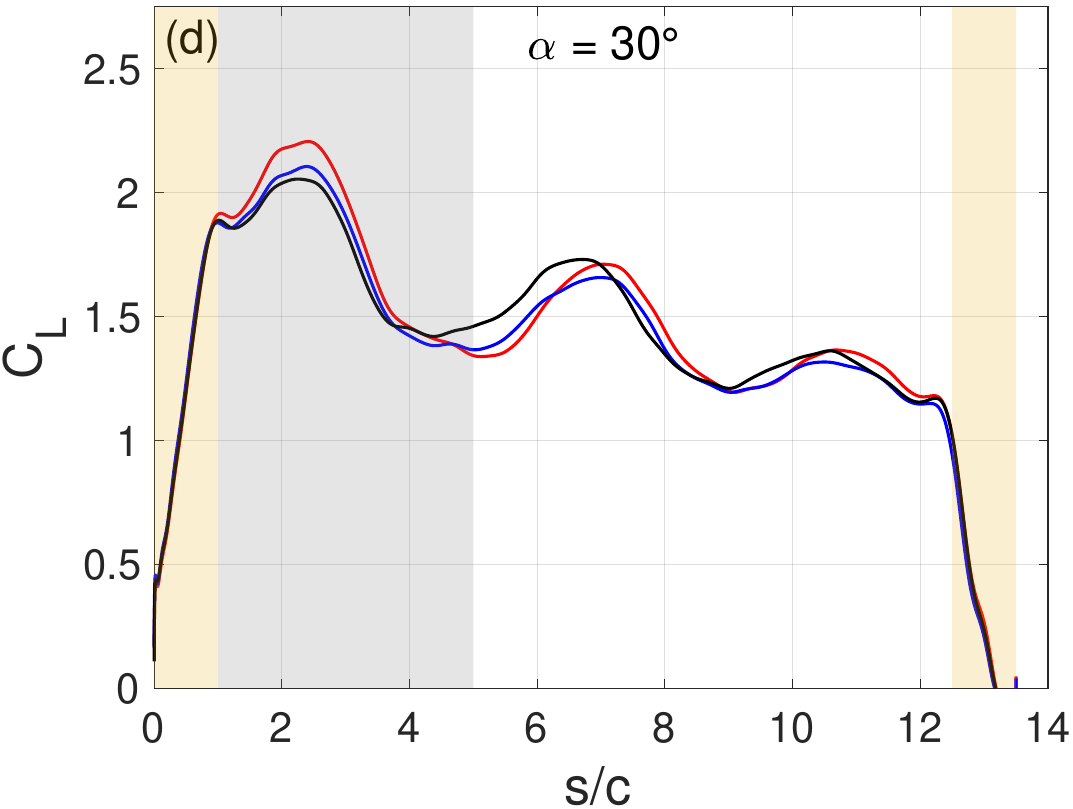}\label{fig:grnd_H_04_30deg}}
	\subfloat{
		\includegraphics[height=0.19914\textheight,keepaspectratio]{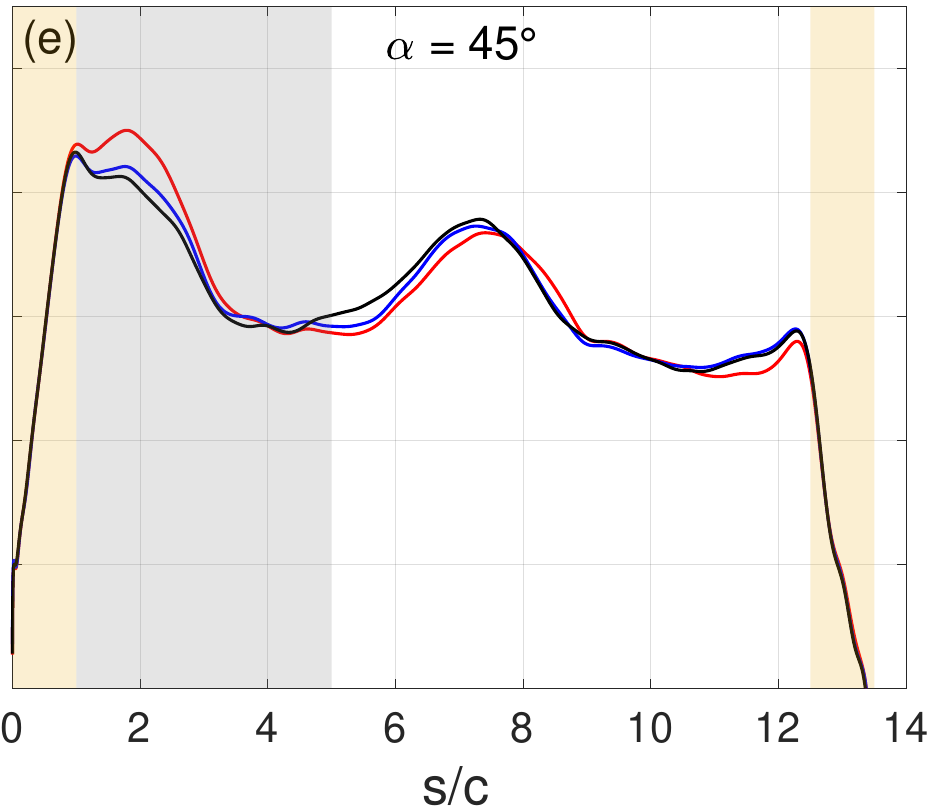}\label{fig:grnd_H_04_45deg}}
	\subfloat{
		\includegraphics[height=0.19914\textheight,keepaspectratio]{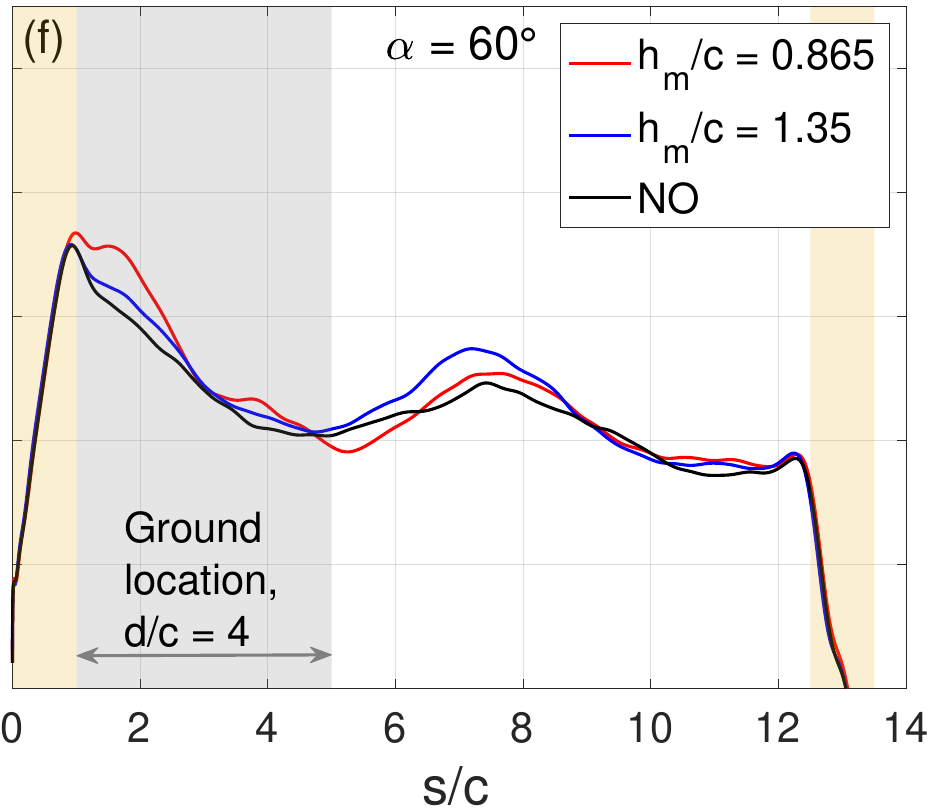}\label{fig:grnd_H_04_60deg}}
	\vspace{-4pt}
	\caption{
	Variations with vertical distance to the ground ($h_\mathrm{m}/c$); $d_\mathrm{a}/c=1$. Top row: $d/c=10$, bottom: $d/c=4$.
	}\label{fig:grnd_H}
\end{figure}

Figure \ref{fig:grnd_H} gives the influence of the wing height from the ground to mid-chord, $h_\mathrm{m}/c$; the gray patch shows the ground position. The vertical gap between the TE and ground is $h_\mathrm{TE}/c$. With $h_\mathrm{m}/c=0.865$, it is $h_\mathrm{TE}/c=0.62$, $0.51$, and $0.43$ for $\alpha=\degrees{30}$, \degrees{45}, and \degrees{60}, respectively, and $h_\mathrm{TE}/c=1.1$, $1.0$, and $0.92$ for $h_\mathrm{m}/c=1.35$.

For $d/c=10$ long grounds (Fig.\ \ref{fig:grnd_H}, top row), at $\alpha=\degrees{30}$ and \degrees{45} the first circulatory-$C_\mathrm{L}$ peak increases with decreasing $h_\mathrm{m}/c$. Similarly, at $\alpha=\degrees{60}$ for the lowest $h_\mathrm{m}/c=0.865$, the $C_\mathrm{L}$ is highest and a circulatory-force only maximum is evident just after the added-mass peak. The $\alpha=\degrees{60}$ NO and $h_\mathrm{m}/c=1.35$ cases have a blended first circulatory-$C_\mathrm{L}$ and added-mass peak as mentioned above, with nearly the same magnitude. At each $\alpha$, the difference in the initial ($s/c<4$) circulatory-$C_\mathrm{L}$ between the NO and ground cases is most evident for $h_\mathrm{m}/c=0.865$, and only slight for 1.35. Therefore, a substantial GE is not observed until $h_\mathrm{TE}/c=0.43$--$0.62$ which depends on $\alpha$.

For nominally-2D flapping wings in hover over a ground that reciprocate in horizontal translation, Gao and Lu \cite{Gao2008} and Lu \etal\ \cite{Lu2014} showed that the stroke-averaged $C_\mathrm{L}$ increases with smaller $h_\mathrm{m}/c$ past the NO value when $h_\mathrm{m}/c$ is below a cutoff; the half-stroke travel of both studies is $2.5c$, encompassing the first circulatory-$C_\mathrm{L}$ peak of the current results. For the CFD of Gao and Lu \cite{Gao2008} at \Rey\ = 100 and experiments of Lu \etal\ \cite{Lu2014} at \Rey\ = 1,000, for a mid-half-stroke $\alpha=\degrees{45}$ the cutoff is $h_\mathrm{m}/c\approx1.5$; Lu \etal\ \cite{Lu2014} also reported cutoffs of $\sim$2.5 and $\sim$1.5 for $\alpha=\degrees{30}$ and \degrees{60}, respectively. This is consistent with the first circulatory-$C_\mathrm{L}$ peak trends shown here, in that a $C_\mathrm{L}$ increase is observed for $h_\mathrm{m}/c\leq1.35$ ($<1.5$) at all $\alpha$, however the flapping-wing studies did not test $h_\mathrm{m}/c<1$. Here, for both $h_\mathrm{m}/c$ this $C_\mathrm{L}$ peak is highest for \degrees{45}, slightly smaller at \degrees{30} but the difference is within the experimental error, and lowest for \degrees{60} (Fig.\ \ref{fig:grnd_H}). Lu \etal\ \cite{Lu2014} found that at $h_\mathrm{m}/c=1$ the mean and peak $C_\mathrm{L}$ are also largest at a mid-half-stroke angle of $\alpha=\degrees{45}$, but instead are lowest at \degrees{30} and slightly reduced at \degrees{60}. The flapping-wing kinematics include periodic reciprocation with pitching, creating wing-wake interactions not present for the simple translating motion here, so differences are expected. In the moving-ground, fixed-wing wind-tunnel experiments of Bleischwitz \etal\ \cite{Bleischwitz2017,Bleischwitz2018}, for rigid $\AR=2$ flat-plate wings at $\alpha=\degrees{15}$ and \degrees{25} there is very little difference in time-averaged $C_\mathrm{L}$ between the no-ground and GE cases for $h_\mathrm{TE}/c=1$, but a substantial increase in $C_\mathrm{L}$ with lower $h_\mathrm{TE}/c$ for $h_\mathrm{TE}/c\le0.5$. This is consistent with the present higher-$\alpha$, larger-\AR\ unsteady data for $s/c<4$, in that for $h_\mathrm{m}/c=1.35$ and $\alpha=\degrees{30}$--\degrees{60} having $h_\mathrm{TE}/c=1.1$--0.92, the GE is slight, whereas for $h_\mathrm{m}/c=0.865$ with $h_\mathrm{TE}/c=0.62$--0.43, there is a clear rise in the first circulatory-$C_\mathrm{L}$ peak.

For the second circulatory-$C_\mathrm{L}$ peak, at $\alpha=\degrees{30}$ the $h_\mathrm{m}/c=1.35$ and NO levels are nearly the same but the $h_\mathrm{m}/c=0.865$ value is larger (Fig.\ \ref{fig:grnd_H}, top row). For $\alpha=\degrees{45}$, both ground-case peaks are very close to the NO level, with the $h_\mathrm{m}/c=1.35$ one slightly higher. At $\alpha=\degrees{60}$ the trend is opposite that of \degrees{30}, with the $h_\mathrm{m}/c=1.35$ peak highest and $h_\mathrm{m}/c=0.865$ peak smaller, both above the NO level. Regarding $h_\mathrm{TE}/c$: for $\alpha=\degrees{30}$ this second circulatory-$C_\mathrm{L}$ peak is larger for $h_\mathrm{TE}/c=0.62$ than $1.1$ (where it is near the NO value), for $\alpha=\degrees{45}$ the peak is similar among the $h_\mathrm{TE}/c=0.51$, 1.0, and NO cases, but for $\degrees{60}$ this peak is higher for $h_\mathrm{TE}/c=0.92$ than 0.43, both larger than the NO result. Flow data will help explain the lack of clear $h_\mathrm{m}/c$ or $h_\mathrm{TE}/c$ trends, which indicates an $\alpha$ effect as the flow evolves. The $C_\mathrm{L}$ peak-timing behavior with $h_\mathrm{m}/c$ for the ground data has differences with the channel and ceiling results. For the second circulatory-$C_\mathrm{L}$ maximum, Fig.\ \ref{fig:grnd_H} ($d/c=10$) shows that at $\alpha=\degrees{30}$ both ground cases are shifted slightly after the NO peak (as is the third peak for $h_\mathrm{m}/c=0.865$), with a smaller shift for $h_\mathrm{m}/c=1.35$. However, for $\alpha=\degrees{45}$ the ground and NO second-circulatory maxima have similar $s/c$ locations. At $\alpha=\degrees{60}$, this peak is instead shifted earlier than the NO case, with $s/c$ shifts ahead of $\sim$$0.3c$ and $\sim$$0.4c$ for $h_\mathrm{m}/c=1.35$ and 0.865, respectively. Therefore, the lift-timing trends also vary with $\alpha$. For the channel and ceiling $d/c=10$ data, for all $\alpha$ the second circulatory-$C_\mathrm{L}$ peak and any afterward are progressively earlier with smaller $h_\mathrm{m}/c$, unlike the $\alpha=\degrees{30}$ and \degrees{45} ground results. Also, in almost all channel and ceiling cases, these peak-timing shifts from the NO data are larger than those of the ground curves. This implies that channel and ceiling obstacles, which affect the LEV flow directly, generally have a greater impact on the circulatory-$C_\mathrm{L}$-peak timing versus a ground. Considering the TEV, as found by Bleischwitz \etal\ \cite{Bleischwitz2017} it should be more parallel to the confining ground rather than the chord. Future flow data will help understand these interactions.

Figure\ \ref{fig:grnd_H}, bottom row, shows $d/c=4$ cases. Up to $s/c\approx3$, $2c$ before the ground ends, the $C_\mathrm{L}$ is very similar to the $d/c=10$ data. For all cases except $\alpha=\degrees{60}$ with $h_\mathrm{m}/c=1.35$, near the ground's end and past it toward the second circulatory-$C_\mathrm{L}$ peak, the $C_\mathrm{L}$ falls below the $d/c=10$ and NO curves with case-dependent timing. This is an adverse effect as the ground distance to the wing suddenly increases. Next, the second circulatory-$C_\mathrm{L}$ peak is somewhat lower than for $d/c=10$ in all cases, with the largest differences for $h_\mathrm{m}/c=0.865$; for $h_\mathrm{m}/c=1.35$ and $\alpha=\degrees{60}$ the change is slight. This shifts the $C_\mathrm{L}$ peak below the NO case for $\alpha=\degrees{30}$ and \degrees{45}. At $\alpha=\degrees{30}$, the third circulatory-$C_\mathrm{L}$ peak is also smaller for each $h_\mathrm{m}/c$, with the $h_\mathrm{m}/c=1.35$ result less than the NO level. In all cases except $\alpha=\degrees{30}$ and $h_\mathrm{m}/c=1.35$, beyond the second circulatory peak the $C_\mathrm{L}$ is overall closer to the NO data than for $d/c=10$, showing a recovery after the detrimental ground-end interaction. Even this $\alpha=\degrees{30}$ case has less than 5\% differences with the NO curve there.

\vspace{-5 pt}
\subsection{Variations with Ground Length}\label{sec:grnd_L}

\begin{figure}[t!]
	\centering
	\subfloat{
		\hspace{-5pt}\includegraphics[height=0.175\textheight,keepaspectratio]{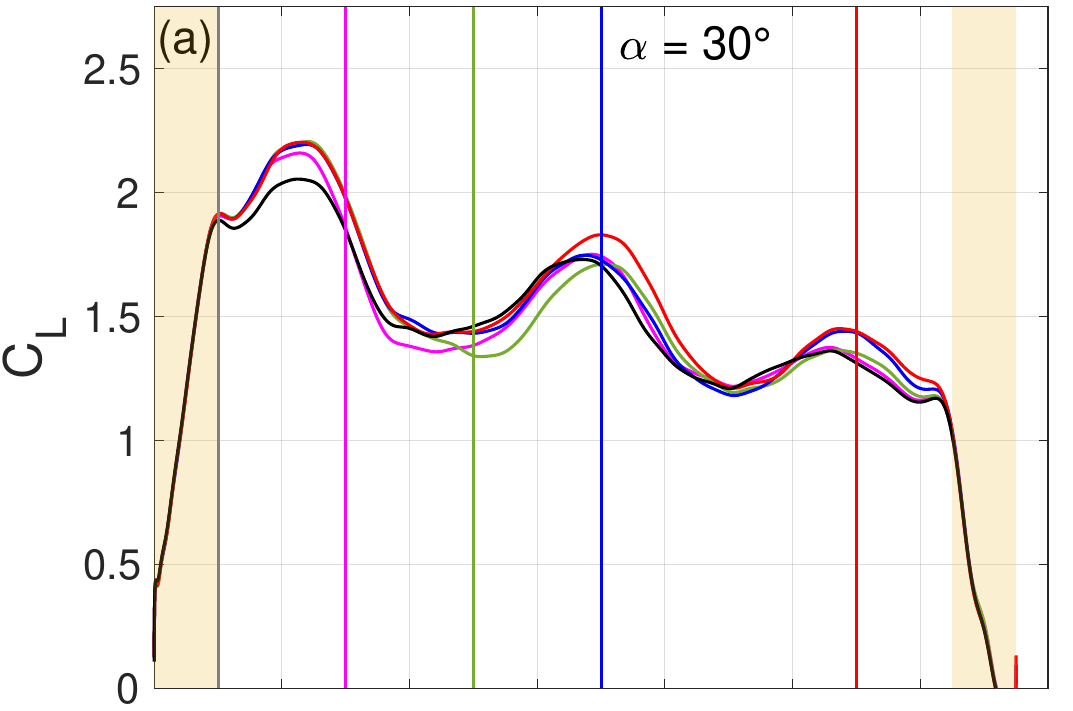}\label{fig:grnd_L_0p865_30deg}}
	\subfloat{
		\includegraphics[height=0.175\textheight,keepaspectratio]{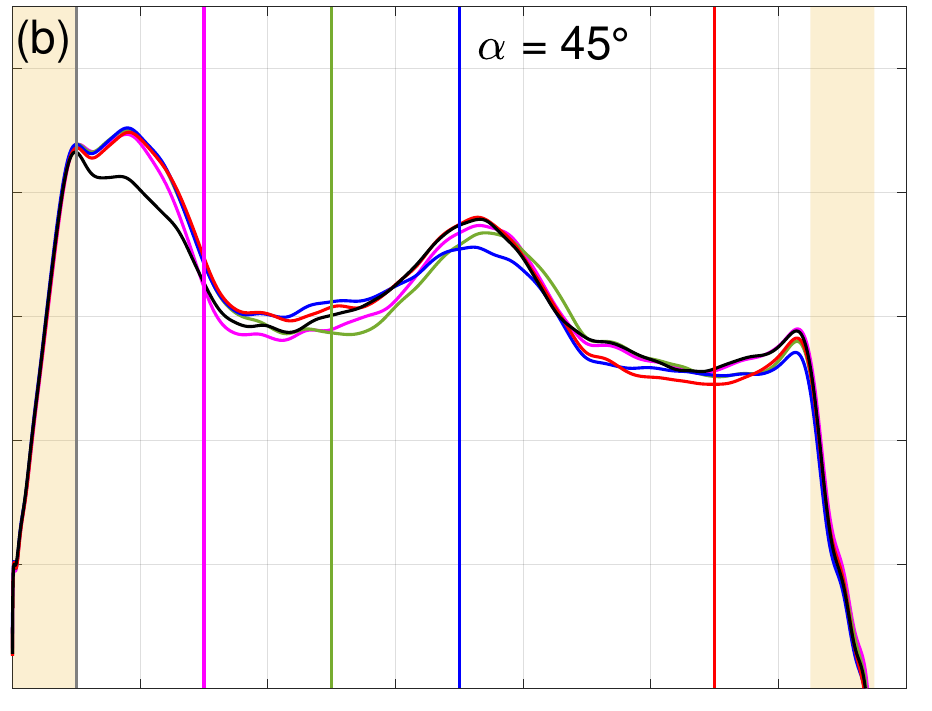}\label{fig:grnd_L_0p865_45deg}}
	\subfloat{
		\includegraphics[height=0.175\textheight,keepaspectratio]{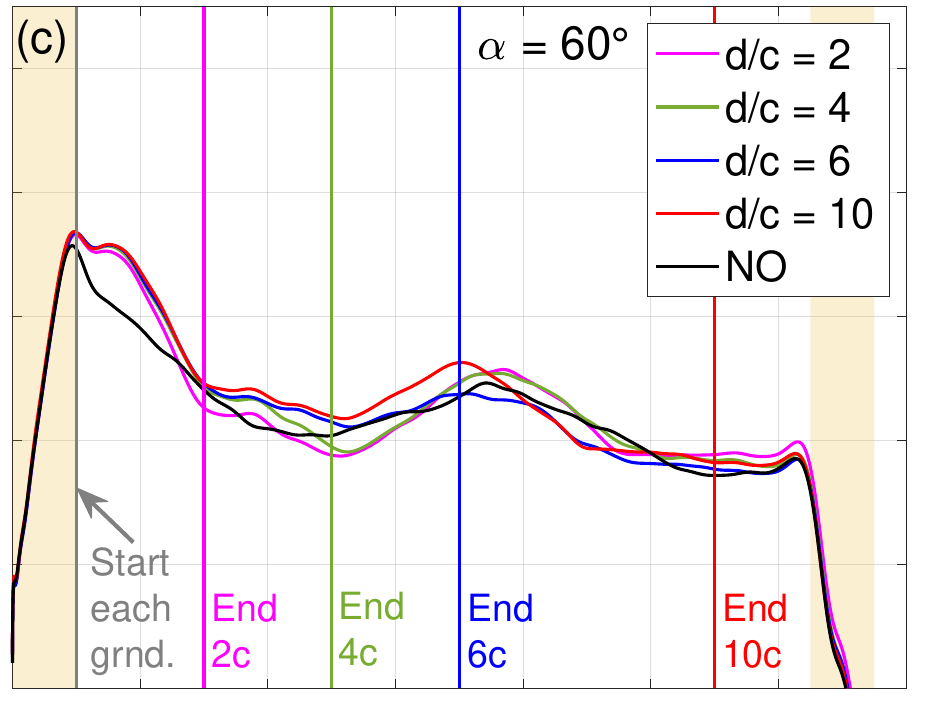}\label{grnd_L_0p865_60deg}}
	\\ \vspace{-10pt}
	\subfloat{
		\hspace{-5pt}\includegraphics[height=0.19914\textheight,keepaspectratio]{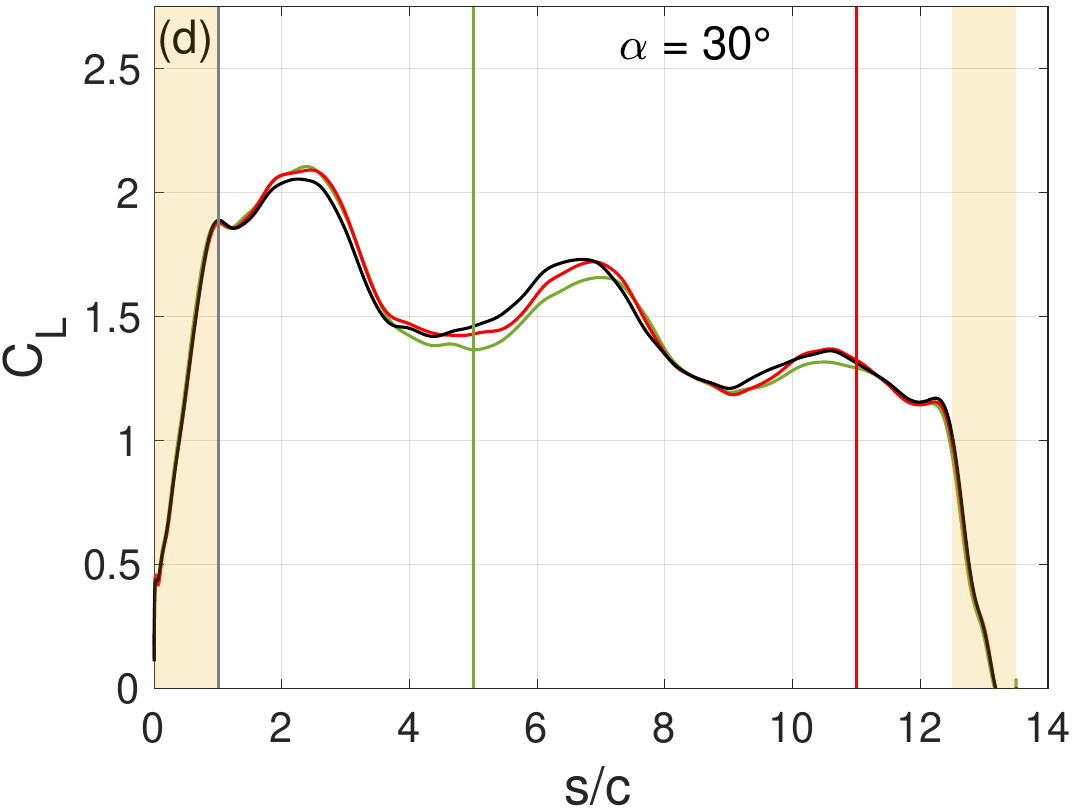}\label{fig:grnd_L_1p35_30deg}}
	\subfloat{
		\includegraphics[height=0.19914\textheight,keepaspectratio]{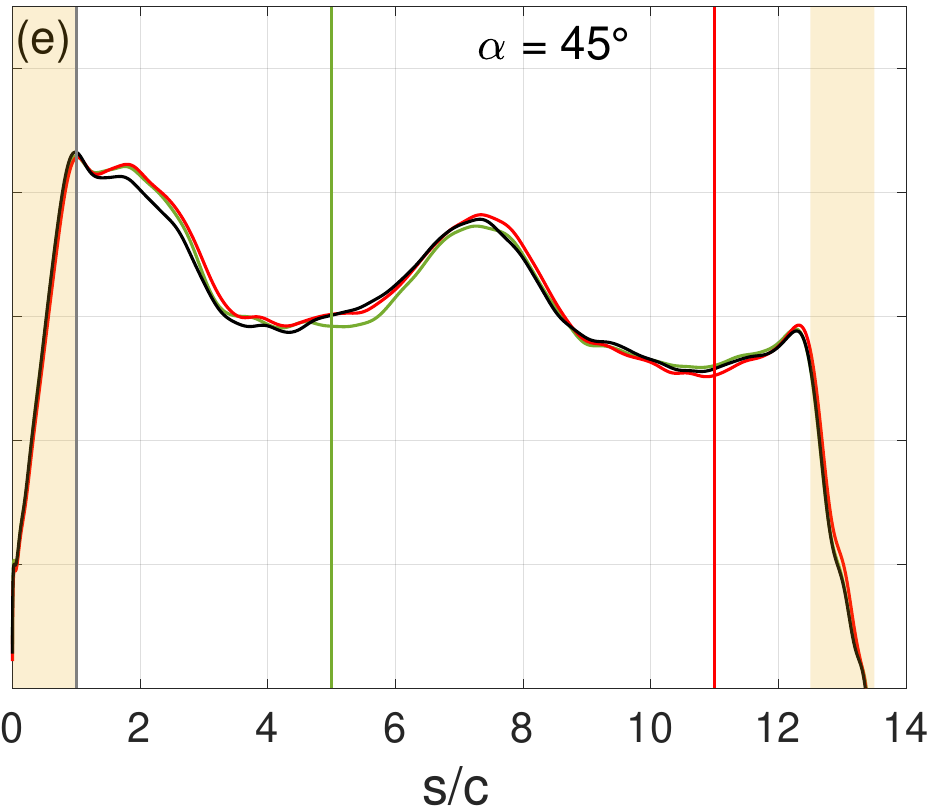}\label{fig:grnd_L_1p35_45deg}}
	\subfloat{
		\includegraphics[height=0.19914\textheight,keepaspectratio]{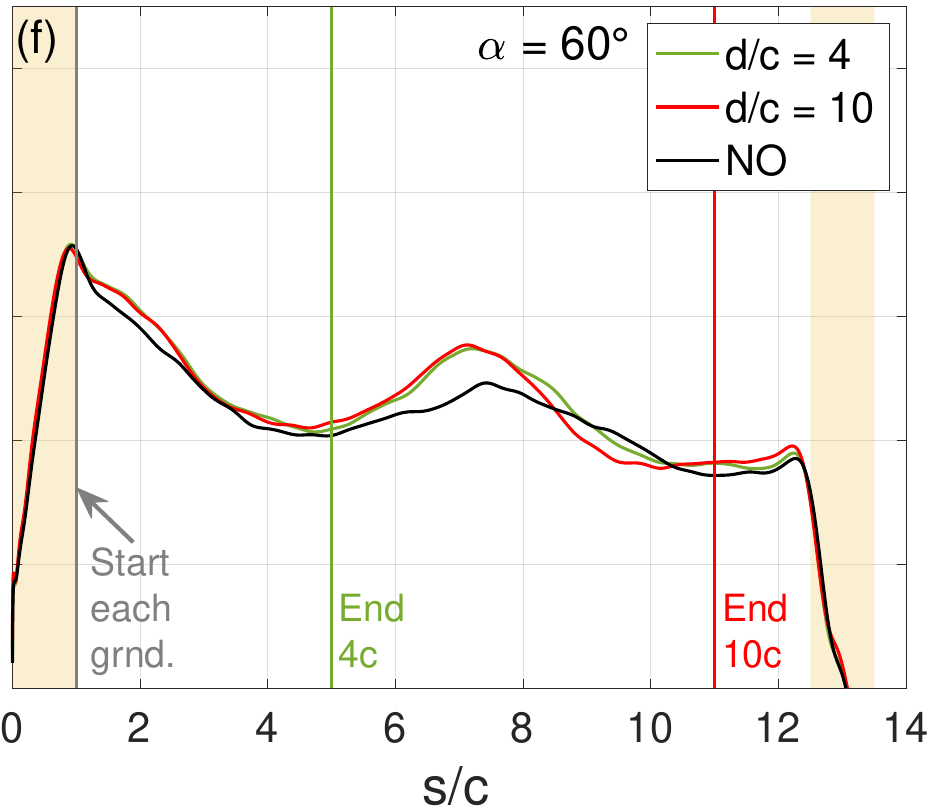}\label{fig:grnd_L_1p35_60deg}}
	\vspace{-4pt}
	\caption{
	Ground length ($d/c$) variations; $d_\mathrm{a}/c=1$. Top row: $h_\mathrm{m}/c=0.865$, bottom row: $h_\mathrm{m}/c=1.35$.
	}\label{fig:grnd_L}
\end{figure}

Ground-length and end effects are further examined in Fig.\ \ref{fig:grnd_L}. The top row with the closest ground, $h_\mathrm{m}/c=0.865$, is discussed first for $d/c=2$--10 lengths. Using $d/c=10$ as a reference, for all $\alpha$ and $d/c<10$, before the ground ends the $C_\mathrm{L}$ falls below the $d/c=10$ curve, as discussed for $d/c=4$ above. In almost all cases this occurs at $\sim$$1c$ or more of travel prior to the ground end; for $\alpha=\degrees{30}$ and $d/c=4$ it is slightly later. This ``warning'' could help detect the ground end in future control studies. In all $d/c<10$ data the $C_\mathrm{L}$ is at least temporarily lower than the NO curve in the range of $1c$ of travel before to $2c$ after the ground ends, so the loss is nontrivial. At all $\alpha$, for $d/c=2$ and 4 the ground end occurs during a $C_\mathrm{L}$ decrease or near a minimum, respectively, but for $d/c=6$ the end is close to the second circulatory-$C_\mathrm{L}$ peak. This timing difference does not appreciably affect the $C_\mathrm{L}$ reduction near the ground end versus $d/c=10$, nor give a clear $C_\mathrm{L}$ loss trend compared to the NO data. This may be related to a lesser influence on the LEV for grounds.

After the ground-end loss, considering first $\alpha=\degrees{30}$ and \degrees{45}, for $d/c=2$ and 4 the $C_\mathrm{L}$ eventually recovers near the NO result (Fig.\ \ref{fig:grnd_L}, top row). The timing shifts between the NO and ground $C_\mathrm{L}$ curves make it difficult to determine a recovery time. For $d/c=2$, for both $\alpha$ values the $C_\mathrm{L}$ curve is very close to the NO result $\sim$$4.8c$ after the ground end, at $s/c\approx7.8$. However, for $\alpha=\degrees{30}$ the $C_\mathrm{L}$ first crosses above the NO curve earlier at $s/c\approx6.6$. With $d/c=4$, for $\alpha=\degrees{30}$ the $C_\mathrm{L}$ curve rises over the NO value at $s/c\approx7.0$ ($2c$ after the end) then lags the NO data but with a similar $C_\mathrm{L}$ magnitude. At \degrees{45}, past the ground end the second circulatory-$C_\mathrm{L}$ peak is below the NO value, then the $C_\mathrm{L}$ crosses above the NO data at $s/c\approx7.7$, after which they overlap some ($s/c\approx9$--10.6). In summary, for $\alpha=\degrees{30}$ and \degrees{45} with $d/c=2$ and 4, recovery behavior is observed roughly within $2c$--$5c$ of travel past the ground end. For $d/c=6$ and $\alpha=\degrees{30}$, after this longer ground ends the third circulatory-$C_\mathrm{L}$ peak instead overshoots the NO value to the $d/c=10$ level, despite its prior peak being closer to the NO case, showing some favorable GE. Conversely, for \degrees{45} the $d/c=6$ ground curve is mainly below the NO value for $s/c>8.8$ (as are the $d/c=10$ data), so more time is needed for a possible recovery.

At $\alpha=\degrees{60}$, for $d/c=2$ and 4 the $C_\mathrm{L}$ curves are similar from $s/c\approx5.3$ (near the $d/c=4$ ground end) onward, and after $s/c=6.4$ they rise above the NO value to a larger second-circulatory peak, then plateau slightly higher than the NO curve. Therefore, despite the temporarily-adverse ground-end influence, overall the beneficial GE persists for longer than at $\alpha=\degrees{30}$ and \degrees{45}, possibly from the smaller $h_\mathrm{TE}/c$. A greater run length is needed to observe when the $C_\mathrm{L}$ relaxes toward the NO case. For $d/c=6$, the ground ends near the second circulatory-$C_\mathrm{L}$ peak, lowering it below the $d/c=10$ value and slightly under the NO level; after $\sim$$3.3c$ past the obstacle end, the $C_\mathrm{L}$ recovers near the NO data.

To summarize the $h_\mathrm{m}/c=0.865$ post-exit behavior, for all $\alpha$ with $d/c=2$ and 4 the $C_\mathrm{L}$ recovers to or above the NO level within $1.5c$--$5c$ of travel past the ground end. For $d/c=6$, the $\alpha=\degrees{30}$ curve crosses over the NO $C_\mathrm{L}$ at $\sim$$4c$ after the end then attains a higher third circulatory peak, while at $\alpha=\degrees{60}$ the ground case becomes similar to the NO result $\sim$$3.3c$ past the end; this has consistency with the $d/c=2$ and 4 results. However, at $\alpha=\degrees{45}$ for $d/c=6$ a sustained recovery is not found in the $5.5c$ travel after the end. More work is needed to study differences with $\alpha$ for $d/c=6$.

Figure \ref{fig:grnd_L}, second row, gives the $h_\mathrm{m}/c=1.35$ results for $d/c=4$ and 10. With the ground farther away, the variations in $C_\mathrm{L}$ peak magnitude and timing with $d/c$ are generally smaller. This includes the adverse ground-end effect present for $\alpha=\degrees{30}$ and \degrees{45} ($d/c=4$), but absent here for $\alpha=\degrees{60}$ although $h_\mathrm{TE}/c$ is the smallest among them. For $\alpha=\degrees{30}$ and $d/c=4$, the second and third circulatory-$C_\mathrm{L}$ peaks (after the ground ends) are reduced further from the NO data than for $h_\mathrm{m}/c=0.865$, giving a worse recovery despite the larger $h_\mathrm{m}/c=1.35$. Flow data will help explain this.

\vspace{-5 pt}
\subsection{Comparison of Channels, Ceilings, and Grounds}

\begin{figure}[t!]
	\centering
	\subfloat{
		\hspace{-5pt}\includegraphics[height=0.19914\textheight,keepaspectratio]{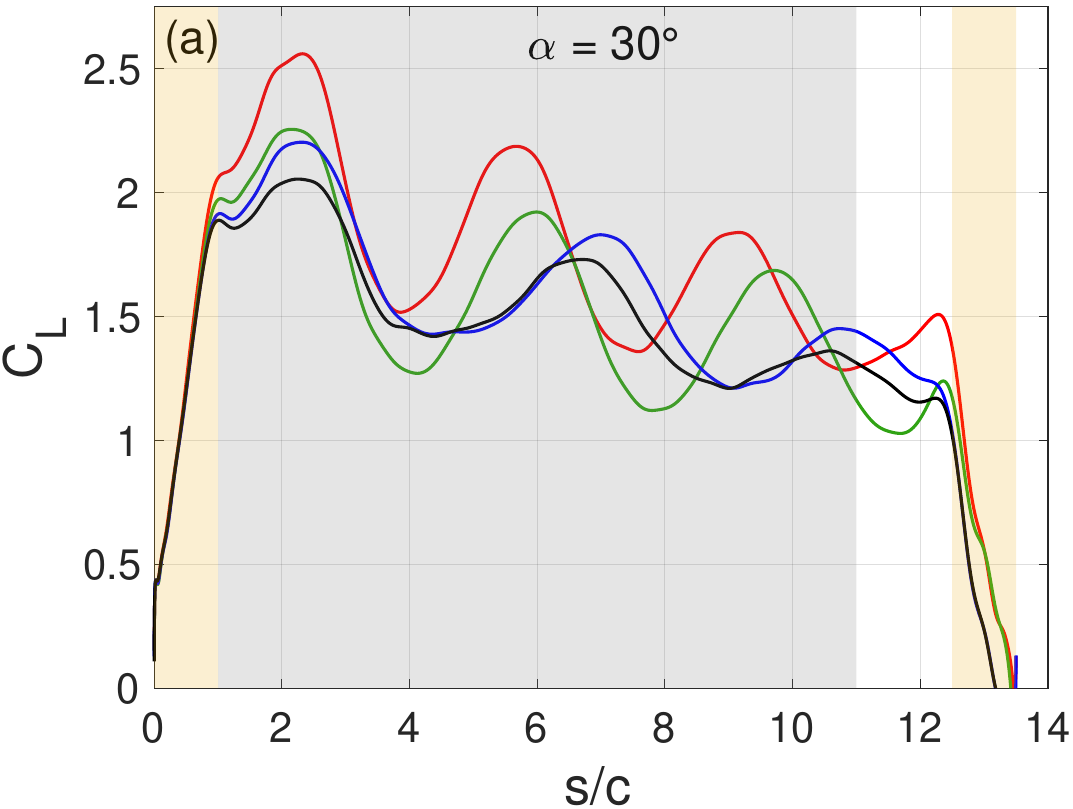}\label{fig:compare_H_10_30deg}}
	\subfloat{
		\includegraphics[height=0.19914\textheight,keepaspectratio]{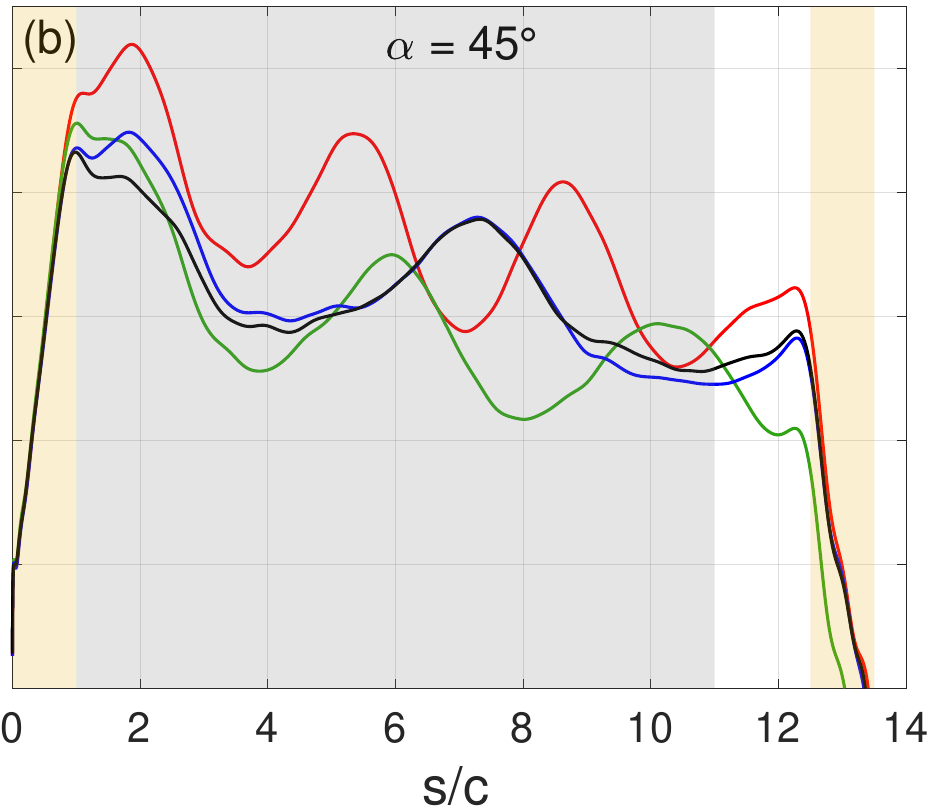}\label{fig:compare_H_10_45deg}}
	\subfloat{
		\includegraphics[height=0.19914\textheight,keepaspectratio]{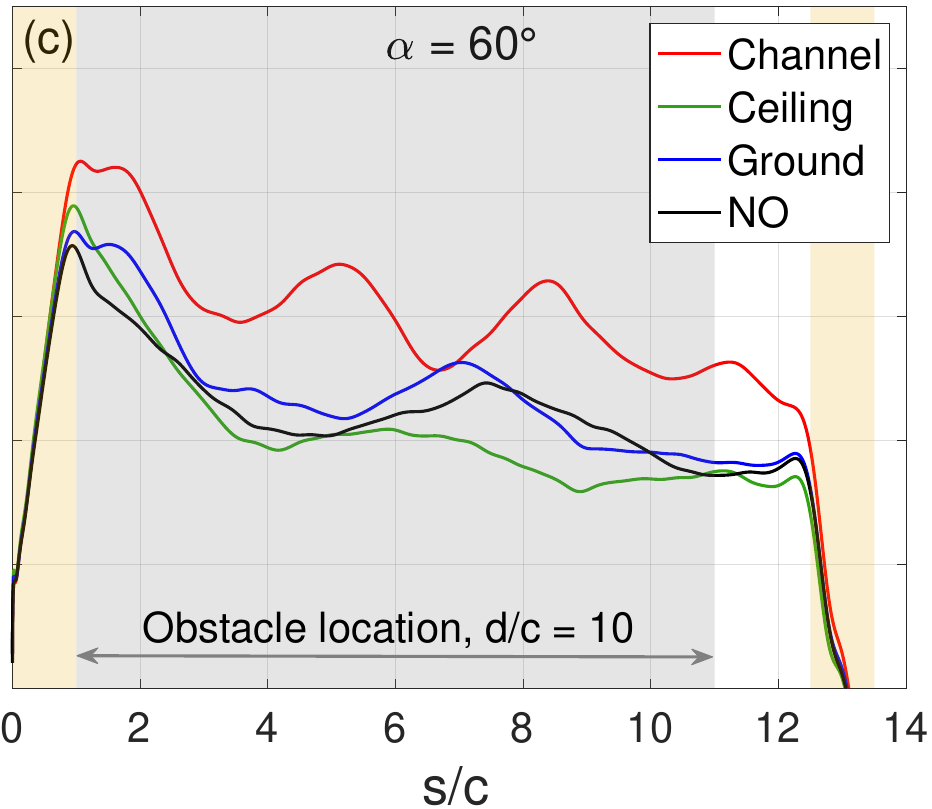}\label{fig:compare_H_10_60deg}}
	\vspace{-4pt}
	\caption{
	Comparison of channel, ceiling, and ground cases for $h_\mathrm{m}/c=0.865$ and $d/c=10$, with $d_\mathrm{a}/c=1$.
	}\label{fig:compare_H}
\end{figure}

Fig.\ \ref{fig:compare_H} compares all obstacles for a $d/c=10$ length and the smallest $0.865c$ mid-chord height. The symmetric confinement of the channel yields the greatest circulatory-$C_\mathrm{L}$ maxima, with the second and third peaks timed the earliest. The ceiling peaks occur later, next the ground and NO ones that are closer; the ceiling cases also have the lowest $C_\mathrm{L}$ minima. For the smallest LE-to-ceiling gaps ($\alpha\ge\degrees{45}$), the second circulatory-$C_\mathrm{L}$ maximum is the least of all the cases.

\vspace{-5 pt}
\subsection{Variations with Approach Distance}

\begin{figure}[t!]
	\centering
	\subfloat{
		\hspace{-5pt}\includegraphics[height=0.19914\textheight,keepaspectratio]{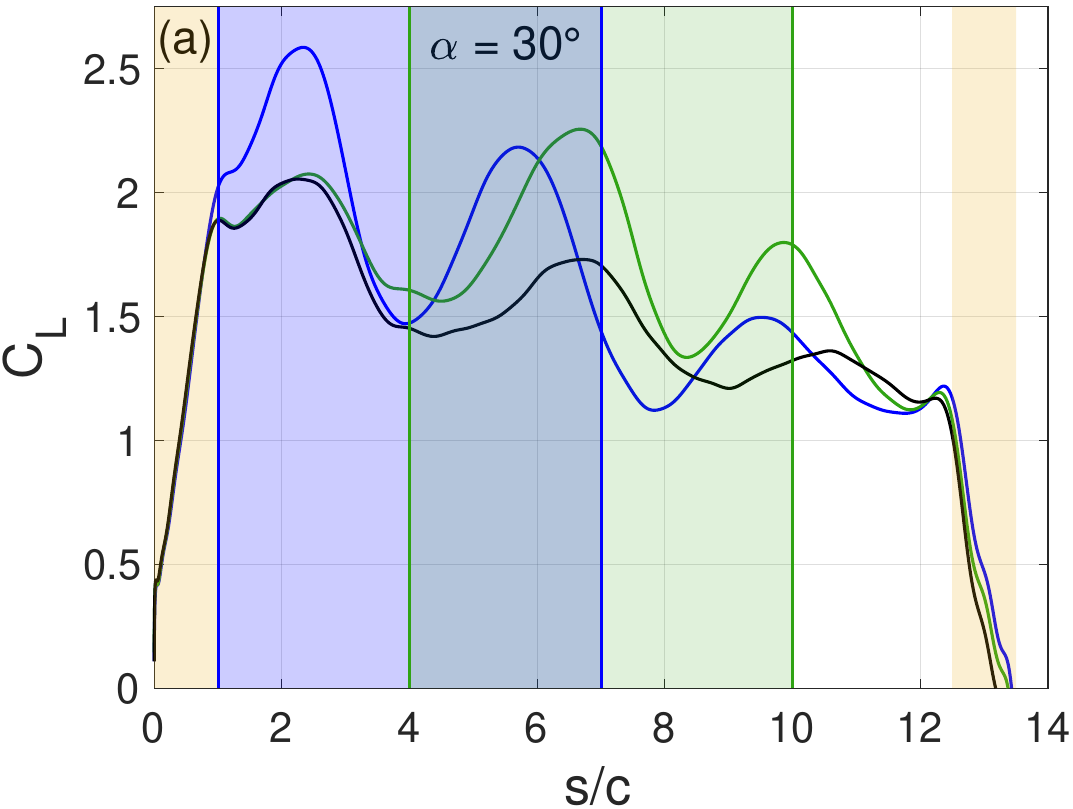}\label{fig:approach_30deg}}
	\subfloat{
		\includegraphics[height=0.19914\textheight,keepaspectratio]{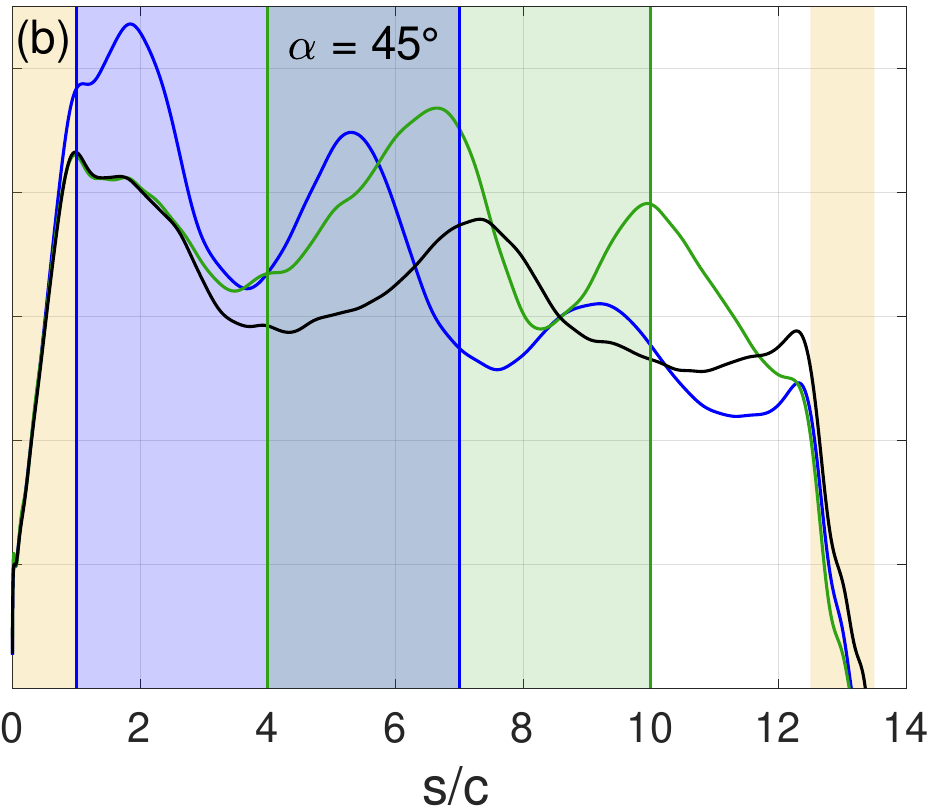}\label{fig:approach_45deg}}
	\subfloat{
		\includegraphics[height=0.19914\textheight,keepaspectratio]{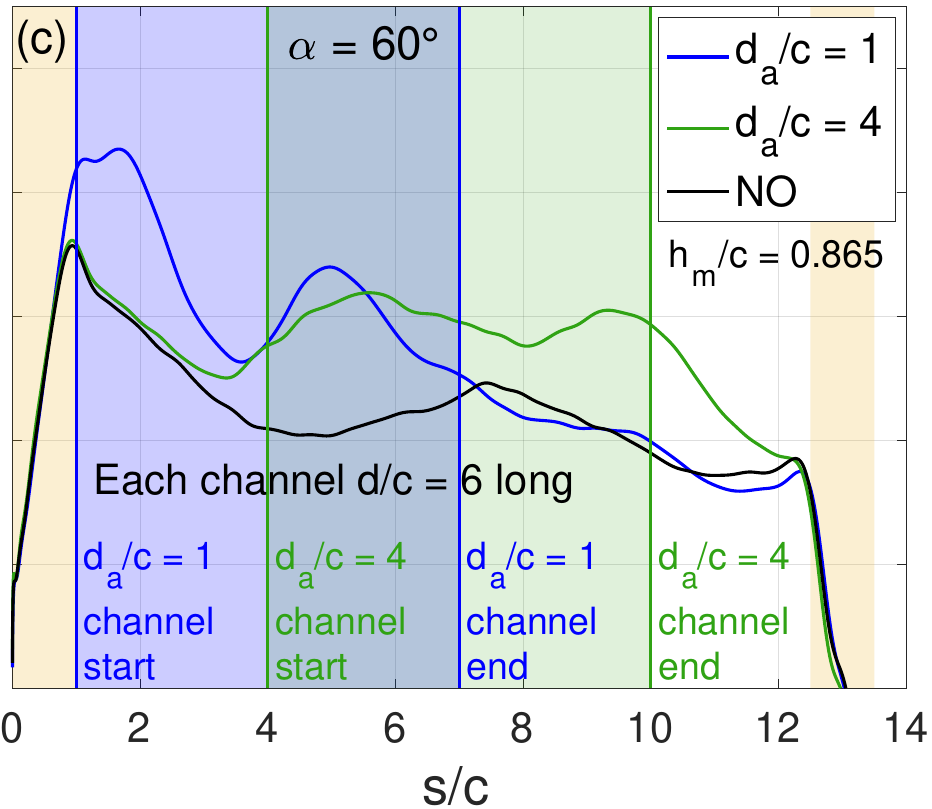}\label{fig:approach_60deg}}
	\vspace{-4pt}
	\caption{
	Variations with approach distance ($d_\mathrm{a}/c$) for $d/c=6$ channels with $h_\mathrm{m}/c=0.865$.
        }\label{fig:approach}
\end{figure}

Figure \ref{fig:approach} shows changes with approach distance, $d_\mathrm{a}/c=1$ and 4, from the wing LE start to a channel with $d/c=6$ and $h_\mathrm{m}/c=0.865$; all results above have $d_\mathrm{a}/c=1$. The $d_\mathrm{a}/c=4$ case (green) has the highest second circulatory-$C_\mathrm{L}$ peak for $\alpha=\degrees{30}$ and \degrees{45}, which occurs as the wing reaches $\sim$2.7$c$ past the channel entrance ($s/c\approx6.7$). However, for $d_\mathrm{a}/c=1$ (blue) its second peak is lower, despite its first circulatory peak being much larger and the wing having moved farther in its channel (4.7$c$ and 4.3$c$ past the entrance for $\alpha=\degrees{30}$ and \degrees{45}, respectively). With $d_\mathrm{a}/c=1$, the wing reaches the channel as the first LEV is growing, since $C_\mathrm{L}$ is increasing toward the first circulatory maximum. Conversely, for $d_\mathrm{a}/c=4$ the channel entrance is near the next $C_\mathrm{L}$ minimum, which should correspond to the first LEV shedding or close to it and being much larger \cite{Mulleners2017,Stevens2017}. The higher second circulatory-$C_\mathrm{L}$ peak for $d_\mathrm{a}/c=4$ may be related to the large first LEV interacting with the channel start and creating more favorable conditions for the second LEV forming in the channel, versus $d_\mathrm{a}/c=1$ for which its first LEV sheds in the confined channel followed by the second. Future flow data will help explain this. The smaller second peak for $d_\mathrm{a}/c=1$ is unlikely from its channel ending earlier at $s/c=7$, versus $s/c=10$ for $d_\mathrm{a}/c=4$ (both have $d/c=6$ channels), since this peak is nearly the same when $d/c$ is increased to $10$ (Fig.\ \ref{fig:chan_L}). For $\alpha=\degrees{60}$, this maximum is instead larger for $d_\mathrm{a}/c=1$ than 4, which may be related to the higher blockage.

For all $\alpha$, the second circulatory-$C_\mathrm{L}$ peak occurs earlier for $d_\mathrm{a}/c=1$, for which the wing has moved relatively farther in its channel than for $d_\mathrm{a}/c=4$ at the same chords traveled. With $d_\mathrm{a}/c=4$ the $C_\mathrm{L}$ curve departs from the NO data $\sim$$1c$ ahead of the channel entrance for each $\alpha$, versus less than $0.5c$ for $d_\mathrm{a}/c=1$. This may be due to the larger $d_\mathrm{a}/c=4$ LEV formed before the wing reaches the channel creating a stronger interaction. Therefore the ``warning'' that the obstacle is approaching likely depends on the LEV dynamics, but more study is needed.

\vspace{-5 pt}
\section{Conclusions}

Force measurements in a water towing tank are used to understand the time-varying lift produced by a rectangular, flat-plate wing at high $\alpha$ moving through or past finite-length obstacles. The obstacles are simple rectangular channels, ceilings, or grounds, made from walls in the tank. The wing has $\AR_\mathrm{eff}=8$, translates at fixed $\alpha$, and accelerates from rest over $1c$ to constant velocity at $\Rey=7{,}000$; $\alpha=\degrees{30}$, \degrees{45}, and \degrees{60} are tested. The parameters studied are the height from the wing mid-chord to the streamwise obstacle surface ($0.865c$ and $1.35c$), obstacle streamwise length ($2c$--$10c$), and approach distance from the wing LE starting position to the obstacle; most cases have a $1c$ approach. Based on prior NO studies, the initial $C_\mathrm{L}$ peak during acceleration is from combined added-mass and circulatory forces, with acceleration ending at the obstacle start, followed by circulatory-lift peaks associated mainly with LEV formation and shedding.

\vspace{-2 pt}

For channels the wing is centered vertically inside, and with reduced channel gap-height the circulatory-$C_\mathrm{L}$ peaks increase in magnitude. Starting with the second such peak, these maxima occur progressively earlier as the gap narrows, compared to the NO case. These behaviors are probably due to the wing's blockage effect, which for smaller gaps will increase the flow speed around the wing to give larger $C_\mathrm{L}$ and faster flow evolution. In most cases, before the wing exits the channel there is a lift loss, in all cases this occurs afterward. For the narrowest channel the performance is poorest past the exit if the exiting coincides with a $C_\mathrm{L}$ peak, but if not, the post-exit $C_\mathrm{L}$ values are often above the NO case. Therefore, channel length and exit effects should be related to the timing of the LEV formation versus the channel end.

The ceiling cases show some similar trends. As the LE-to-ceiling gap is reduced, the first circulatory-$C_\mathrm{L}$ maximum increases, and the peaks afterward are more advanced in time versus the NO data. However, when the gap is $\sim$$0.5c$ or lower the second circulatory-$C_\mathrm{L}$ peak is below the NO value, and the post-obstacle ceiling-length effect on $C_\mathrm{L}$ is smaller than for channels. After the ceiling, the $2c$--$4c$ long cases recover near the NO data within the run length. For larger gaps this means a $C_\mathrm{L}$ decline, but the smaller gap with $\alpha=\degrees{45}$ has a weakly-adverse ceiling-end effect and then rises.

For grounds, as the wing height decreases the first circulatory-$C_\mathrm{L}$ peak is larger, but the trends for the second circulatory maximum depend on $\alpha$. Unlike the channels and ceilings, this peak's timing is closer to the NO data, since the ground's effect on the LEV is likely smaller. As the wing nears and passes the ground end, in almost all cases there is a lift loss versus if the ground persisted. At the closest mid-chord height, for ground lengths $2c$--$6c$ the lift falls below the NO result close to the ground end, then most cases show $C_\mathrm{L}$ recovery at or over the NO value within $1.5c$--$5c$ of travel.

When the approach distance is changed from $1c$ to $4c$ for the narrowest channel, for $\alpha\le\degrees{45}$ the latter yields a larger second circulatory-$C_\mathrm{L}$ peak in the channel. This may be related to the first LEV being shed before the wing enters the channel, versus being inside it for the $1c$ approach, creating different formation conditions for the second LEV and peak.

For the closest channel and ceiling cases, there is a clear $C_\mathrm{L}$ rise above the NO curve prior to the obstacle start, and for the channel and ground obstacles, in many cases a $C_\mathrm{L}$ loss occurs before the end is reached. With channels, the data indicate that the timing of the LEV formation with respect to the start and end can substantially affect the $C_\mathrm{L}$. These results could inform work on obstacle sensing. Overall, future flow data will help explain the features reported above.

\section*{Acknowledgments}

The authors wish to thank Dr.\ Juhi Chowdhury for assistance with the experimental setup. This work was supported in part by the Air Force Office of Scientific Research, award no.\ FA9550-23-1-0170, supervised by Dr.\ Gregg Abate.

\bibliography{references}

\end{document}